\documentclass{aa}

\usepackage{graphicx}
\usepackage{txfonts}
\usepackage{xcolor}
\usepackage{comment}
\usepackage{multirow}
\usepackage{hhline}
\usepackage[shortlabels]{enumitem}
\usepackage{amsmath}
\usepackage{placeins}
\usepackage{multicol}

\usepackage{hyperref} 

\makeatletter
  \renewcommand*\aa@pageof{, page \thepage{} of \pageref*{LastPage}}
\makeatother

\usepackage{booktabs}

\newcommand\doublerule{\toprule\specialrule{\heavyrulewidth}{\doublerulesep}{0.95em}}

\newcommand \msun{\mathrm{M}_{\odot}}

\newcommand \rcorecorr{r_{\mathrm{core,corr}}}
\usepackage{tikz}
\usepackage{scalerel}

\usepackage{orcidlink}

\begin{document}
\def\linefilter{$\mathrm{F680N_{line}}\,$}

\title{Clump formation in ram-pressure stripped galaxies: evidence from mass function}

   \author{Eric Giunchi\thanks{E-mail: \email{eric.giunchi2@unibo.it}}
          \orcidlink{0000-0002-3818-1746}\,
          \inst{1,3,4}
          \and
          Claudia Scarlata
          \orcidlink{0000-0002-9136-8876}\,
          \inst{2}
          \and
          Ariel Werle
          \orcidlink{0000-0002-4382-8081}\,
          \inst{3}
          \and
          Bianca M. Poggianti
          \orcidlink{0000-0001-8751-8360}\,
          \inst{3}
          \and
          Alessia Moretti
          \orcidlink{0000-0002-1688-482X}\,
          \inst{3}
          \and
          Marco Gullieuszik
          \orcidlink{0000-0002-7296-9780}\,
          \inst{3}
          \and
          Benedetta Vulcani
          \orcidlink{0000-0003-0980-1499}\,
          \inst{3}
          \and
          Alessandro Ignesti
          \orcidlink{0000-0003-1581-0092}\,
          \inst{3}
          \and
          Antonino Marasco
          \orcidlink{0000-0002-5655-6054}\,
          \inst{3}
          \and
          Anita Zanella
          \orcidlink{0000-0001-8600-7008}\,
          \inst{3}
          \and
          Anna Wolter
          \orcidlink{0000-0001-5840-9835}\,
          \inst{4}
          }

   \institute{Dipartimento di Fisica e Astronomia "Augusto Righi", Universit\`a di Bologna, via Piero Gobetti 93/2, I-40129 Bologna, Italy
   \and
       Minnesota Institute for Astrophysics, School of Physics and Astronomy, University of Minnesota, 316 Church Street SE, Minneapolis, MN 55455, USA
    \and
        INAF-Osservatorio Astronomico di Padova, Vicolo Osservatorio 5, 35122 Padova, Italy
    \and
        Dipartimento di Fisica e Astronomia, Universit\`a di Padova, Vicolo Osservatorio 3, 35122 Padova, Italy
    \and
        INAF - Osservatorio Astronomico di Brera, via Brera, 28, 20121, Milano, Italy
         }

   \date{\today}

 
  \abstract
   {The mass function (MF) of young ($\mathrm{age\lesssim 200\,Myr}$) stellar clumps is an indicator of the mechanism driving the collapse of the interstellar medium (ISM) into the giant molecular clouds. Typically, the MF of clumps in main-sequence galaxies is described by a power law ($dN/dM_*\propto M_*^{-\alpha}$) with slope $\alpha=2$, hinting that the collapse is driven by turbulence.}
   {To understand whether the local environment affects star formation, we have modelled the clump MF of six cluster galaxies, from the GASP survey, undergoing strong ram-pressure stripping. This phenomenon, exerted by the hot and high-pressure intra-cluster medium (ICM), has produced long tails of stripped ISM, where clumps form far away from the galactic disk and are surrounded by the ICM itself.}
   {Clumps were selected from \textit{HST}-UVIS/WFC3 images, covering from near-UV to red-optical bands and including H$\alpha$ emission-line maps. The catalogue comprises 398 H$\alpha$ clumps (188 in tails, 210 in disk outskirts, the so-called extraplanar region) and 1270 UV clumps (593 tail, 677 extraplanar). Using mock images, we quantified the mass completeness and bias of our sample. Keeping these two effects into account, we adopted a Bayesian approach to fit the clump mass catalogue to a suitable MF.}
   {The resulting MFs are steeper than the expected value $2$. In the tails, the H$\alpha$ clumps have slope $\alpha=2.31\pm 0.12$, while the UV slope is larger ($2.60\pm 0.09$), in agreement with ageing effects. Similar results are found in the extraplanar region, with H$\alpha$ slope $\alpha=2.45^{+0.20}_{-0.16}$ and UV slope $\alpha=2.63^{+0.20}_{-0.18}$, even though in this case they are consistent within uncertainties.}
   {We suggest that the steepening results from the higher-than-usual turbulent environment, arising from the interaction between the ISM and the ICM. As shown by recent works, this process can favour the fragmentation of the largest ISM clouds, inhibiting the formation of very massive clumps.}


   \authorrunning{Eric Giunchi et al.}
   \maketitle

\section{Introduction}\label{sec:intro}
Nowadays it is widely accepted that the formation of new stars in galaxies occurs in the density peaks of the cold molecular gas, in the cores of the giant molecular clouds \citep{Lada2003,Bressert2010}.
These dense and cold regions represent the final stage of a complex process that involves turbulence, gas cooling, growth of non-linear perturbation, which drives the condensation of atomic gas (mainly neutral hydrogen, $\mathrm{H\,I}$) from galactic to sub-kpc scales down to the pc/sub-pc dense cores of cold molecular gas \citep{Kennicutt2012}.
In this scenario, $\gtrsim10$ pc-scaled star-forming clusters and compact groups of stellar clusters (the so-called clumps), with masses $\gtrsim10^4\,\msun$ \citep{Portegies2010}, represent the connecting piece between the galactic- and core-scale regimes.
Indeed these stellar clusters are the smallest structures in galaxies containing enough stars to fully sample the stellar initial mass function (IMF). Therefore they represent suitable sites to study in a statistical way the star-formation mechanism as a function of the clump environment, like the global properties of the host galaxy, whether the host galaxy itself is in the field or in a cluster and the local conditions of the medium in which the stellar clumps and the molecular gas clouds are embedded.

Models describing the fragmentation of star-forming regions as a scale-free, turbulence-driven process predict that the mass function $\mathrm{MF}\equiv dN/dM_*$ of these regions (i.e. the number of objects per mass bin) is a power law in the form

\begin{equation}\label{eq:mf}
\mathrm{MF}(M_*)\propto M_*^{-\alpha},
\end{equation}

\noindent with $\alpha=2$ \citep{Elmegreen2002,Elmegreen2006}. Similar results are found in simulations of neutral atomic and molecular medium clump formation \citep{Renaud2024}. Other theoretical works find slightly flatter distributions ($\alpha = 1.7-1.8$)  \citep{Hennebelle2007,Audit2010,Fensch2023}, especially when increasing the gas fraction \citep{Renaud2024}.
As time proceeds after the clump formation, the slope of the clump mass function is thought to change, as a consequence of many processes, including the clump dynamical evolution, relaxation and the interactions within the surrounding environment. However, whether the net effect is the steepening or the flattening the mass function \citep{Gieles2009,Fujii2015} is still debated, since many observational effects make the comparison with models difficult.

The launch of the \textit{Hubble Space Telescope} (\textit{HST}) has given the possibility to perform unprecedented high-resolution observational surveys of low-redshift galaxies (LEGUS, \citealt{Calzetti2015}; DYNAMO, \citealt{Fisher2017}; LARS, \citealt{Messa2019}; PHANGS-HST, \citealt{Lee2022}), which brought a great improvement to our knowledge about star-forming clumps. These surveys covered galaxies in very different regimes and environments, from normal spirals (like in LEGUS and PHANGS) to turbulent disks (DYNAMO), to starburst triggered by merging and peculiar morphology (LARS). 

Thanks to these and other observational studies of young star-forming regions, further confirmations have shown that their MFs are well described by a power law with slopes consistent with the turbulent cascade model ($\alpha=2$, \citealt{Zhang1999,Hunter2003,deGrijs2003,Gieles2006,Adamo2017,Messa2017,Messa2018b}, see also \citealt{Krumholz2019} and references therein), even in tidal debris \citep{Rodruck2023} and $z\sim1-1.5$ lensed galaxies \citep{Livermore2012}. Very often a cut-off at high masses ($>10^5\,\mathrm{M_\odot}$) is observed \citep{Haas2008,Adamo2017,Messa2017,Livermore2012}. In some cases the distribution is found to be shallower \citep{deGrijs2006}, or affected by the local environment, with a steeper distribution in the inter-arm regions of spiral galaxies \citep{Haas2008,Messa2018b,Messa2019}.

However, yet the full understanding of how the environment affects the formation and the final properties of the clumps is not achieved. In this context, the study of star-forming clumps in jellyfish galaxies can put strong constraints on how gas compression and changes in the properties of the surrounding gaseous medium can affect the clumps themselves.

Jellyfish galaxies are peculiar objects found in galaxy clusters (and groups), typically recently accreted and therefore experiencing for the first time the effects of the hot and high-pressure intracluster medium (ICM). The interplay between the infalling galaxy and the ICM can trigger episodes of strong ram-pressure stripping (RPS, \citealt{Gunn1972,Boselli2022review}), with the interstellar medium (ISM) pushed far away from the galactic disk in form of long tails up to more than 100 kpc long. Interestingly, being a pure hydro-dynamical interaction, the stellar disk is left almost undisturbed \citep{Poggianti2017b}. Even though the loss of the ISM ultimately results in the quenching of the star formation in the galaxy \citep{Vulcani2020a,Cortese2021}, several previous works find short-term star-formation enhancement in RPS galaxies (\citealt{Vulcani2018,Vulcani2019,Vulcani2020} for observational results, \citealt{Goller2023} for evidence from TNG simulation) and the presence of in-situ star formation in compact knots of gas stripped out of the galactic disks \citep{Yoshida2008,Smith2010,Merluzzi2013,Abramson2014,Kenney2015,Fossati2016,Consolandi2017,Jachym2019,Cramer2019}.

GASP (GAs Stripping Phenomena in galaxies with MUSE, \citealt{Poggianti2017b}) is an ESO Large Program that studies how different gas removal processes affect the properties of galaxies. The survey includes cluster galaxies observed at different RPS stages. First observations were carried out with MUSE on the VLT, in order to investigate the properties of both the stellar and the ionized gaseous components, both in the disks and in the stripped tails. The sample includes 114 galaxies in the mass range $10^9-10^{11.5}\,\msun$ and at redshift $0.04<z<0.07$, 64 of which are cluster galaxies observed at different RPS stages, 12 are cluster galaxies (both passive and star-forming) serving as control sample, 38 are in groups, filaments and isolated (Poggianti et al. in prep.). RPS candidates were chosen from the \cite{Poggianti2016} galaxy catalogue for showing long tails in B-band optical images, suggestive of gas-only removal. GASP led to the observation of in-situ star formation occurring in compact knots in the tails of many individual jellyfish galaxies of the sample \citep{Bellhouse2017,Gullieuszik2017,Moretti2018b,Moretti2020,Poggianti2019b,Poggianti2019a}.

In order to observe these knots with a better spatial resolution, six GASP galaxies have been observed with the UVIS/WFC3 mounted onboard of \textit{HST} \citep{Gullieuszik2023}, improving the PSF angular size by a factor $\sim 14$ with respect to the seeing-limited MUSE images (from $1\arcsec$ to $\sim0.07\arcsec$).
Photometric observations cover a spectral range from UV- to I-band rest-frame, including the H$\alpha$ emission line.

This work is part of a series of papers studying the high-resolution sample of star-forming clumps detected from the aforementioned \textit{HST} images: 1) \cite{Gullieuszik2023} presented the dataset and the visual properties of the galaxies thanks to the improved spatial resolution; 2) \cite{Giunchi2023a} focused on the sample of star-forming clumps and complexes, studied their luminosity and size distribution functions and their luminosity-size relation; 3) \cite{Giunchi2023b} studied the morphology of the clumps and how RPS affects it; 4) finally, \cite{Werle2024} fitted the Spectral Energy Distributions (SEDs) of the clumps to get their masses, ages and star-formation histories (SFHs), studying how these properties correlate with the morphology and spatial distribution of the clumps.

The present work combines the results obtained in \cite{Giunchi2023a} and \cite{Werle2024}, aiming at studying the mass distribution of the clumps observed in these galaxies, by means of a robust Bayesian technique.
This paper is structured as follows: Sect. \ref{sec:dataset} introduces the dataset, the sample of clumps and the techniques adopted to retrieve their masses and ages; Sect. \ref{sec:mass_function} describes how we took into account the effects that can lead to wrong estimates of the MF properties (i.e. mass completeness and bias); Sect. \ref{sec:res} shows the results of our fits to the MFs; Sect. \ref{sec:conclusions} compares our results with previous studies from the literature and discuss how the peculiar formation environment of our stellar clumps can affect their MF.

This work adopts standard cosmology parameters $H_0=70\,\mathrm{km\,s^{-1}\,Mpc^{-1}}$, $\Omega_M=0.3$ and $\Omega_\Lambda=0.7$. All the coordinates are reported in the J2000 epoch. The parameter space of a given likelihood is explored by means of a Markov-chain Monte Carlo (MCMC), computationally performed using the python software package \textsc{emcee} \citep{Foreman2013}.

\section{Dataset}\label{sec:dataset}

\begin{table*}[t!]
\fontsize{10pt}{10pt}\selectfont
\setlength{\tabcolsep}{4pt}
\renewcommand{\arraystretch}{1.4} 
\centering
\caption{Main properties of our target galaxies and of their clusters.}
\begin{tabular}{cccccccccc}
\doublerule
ID$_\mathrm{P16}$ & RA & Dec & $M_*$ & $R_e$ & $z_{gal}$ & cluster & $z_{clus}$ & $\sigma_{clus}$ & References\\
& (J2000) & (J2000) & $[\times\mathrm{10^{10}\,M_{\odot}}]$ & [kpc] &&&&$[\mathrm{km/s}]$&\\
$(1)$ & $(2)$ & $(3)$ & $(4)$ & $(5)$ & $(6)$ & $(7)$ & $(8)$ & $(9)$ & $(10)$\\\hline
JO$175$ & 20:51:17.593 & -52:49:22.34 & $3.2\pm0.5$ & $3.24^{+0.37}_{-0.34}$ & 0.0468 & A3716 & 0.0457 & $753^{+36}_{-38}$ & (4,10,14)\\
\multirow{2}{*}{JO$201$} & \multirow{2}{*}{00:41:30.295} & \multirow{2}{*}{-09:15:45.98} & \multirow{2}{*}{$6.2\pm0.8$} & \multirow{2}{*}{$7.73^{+0.60}_{-0.79}$} & \multirow{2}{*}{0.0446} & \multirow{2}{*}{A85} & \multirow{2}{*}{0.0559} & \multirow{2}{*}{$859^{+42}_{-44}$} & (3,4,5,6,8,9,10,\\ 
&&&&&&&&& 13,14,15,16,18)\\
JO$204$ & 10:13:46.842 & -00:54:51.27 & $4.1\pm0.6$ & $4.58^{+0.22}_{-0.26}$ & 0.0424 & A957 & 0.0451 & $631^{+43}_{-40}$ & (2,4,6,12,18)\\ 
JO$206$ & 21:13:47.410 & +02:28:35.50 & $9.1\pm0.9$ & $8.94^{+1.08}_{-1.14}$ & 0.0513 & IIZW108 & 0.0486 & $575^{+33}_{-31}$ & (1,4,6,10,13,18,19)\\ 
JW$39$ & 13:04:07.719 & +19:12:38.41 & $17.0\pm3.0$ & $10.16^{+1.23}_{-1.34}$ & 0.0650 & A1668 & 0.0634 & 654\tablefootmark{a} & (14,18,19,21,22)\\
\multirow{2}{*}{JW$100$} & \multirow{2}{*}{23:36:25.054} & \multirow{2}{*}{+21:09:02.64} & \multirow{2}{*}{$29.0\pm7.0$} & \multirow{2}{*}{$7.35^{+0.43}_{-0.74}$} & \multirow{2}{*}{0.0602} & \multirow{2}{*}{A2626} & \multirow{2}{*}{0.0548} & \multirow{2}{*}{$650^{+53}_{-49}$} & (4,6,7,10,11,\\
&&&&&&&&& 15,17,18,19,20)\\
\bottomrule
\end{tabular}
\tablefoot{Columns are: GASP ID of the galaxy as in \citealt{Poggianti2016} (1), RA and Dec of the galaxy (2 and 3), galaxy stellar mass with $1\sigma$ uncertainties (4), effective radius with $1\sigma$ uncertainties (5), galaxy redshift (6), ID of the host cluster (7), cluster redshift (8), cluster velocity dispersion with $68\%$ upper and lower uncertainties (9), references (10). Masses are taken from \cite{Vulcani2018} and derived
with our spectrophotometric code \textsc{SINOPSIS} \citep{Fritz2017}. Effective radii are 
computed as the half-light radii by fitting I-band (computed integrating MUSE data) isophotes to ellipses and taken from \cite{Franchetto2020} and converted to kpc. Cluster redshifts and velocity dispersions are computed from the spectroscopic redshifts of the cluster members and taken from \cite{Biviano2017} and \cite{Cava2009}.}
\tablebib{ 1) \cite{Poggianti2017a}, 2) \cite{Gullieuszik2017}, 3) \cite{Bellhouse2017}, 4) \cite{Poggianti2017b}, 5) \cite{George2018}, 6) \cite{Moretti2018b}, 7) \cite{Poggianti2019b}, 8) \cite{Bellhouse2019}, 9) \cite{George2019}, 10) \cite{Radovich2019}, 11) \cite{Moretti2020}, 12) \cite{Deb2020}, 13) \cite{Ramatsoku2020}, 14) \cite{Bellhouse2021}, 15) \cite{Tomicic2021a}, 16) \cite{Campitiello2021}, 17) \cite{Ignesti2022a}, 18) \cite{Tomicic2021b}, 19) \cite{Ignesti2022b}, 20) \cite{Sun2021}, 21) \cite{Peluso2022}, 22) \cite{George2023}.\\
\tablefoottext{a}{Uncertainties not available.}}
\label{tab:targets}
\end{table*}

The six galaxies \citep{Gullieuszik2023} were observed with the UVIS channel of the WFC3, mounted on board of Hubble Space Telescope (\textit{HST}).
The targets are jellyfish galaxies belonging to the GASP sample and therefore were previously observed with MUSE, showing particularly long H$\alpha$ emitting tails (between 40 and 120 kpc) and a large number of H$\alpha$ clumps ($\sim 250$), compared to the other galaxies of the sample (\citealt{Poggianti2019a}, see also \citealt{Vulcani2020}). The properties of the galaxies and of their host clusters are summarized in Table \ref{tab:targets}.

\begin{figure*}[t!]
\centering
\resizebox{0.88\textwidth}{!}{\includegraphics[height=1cm]{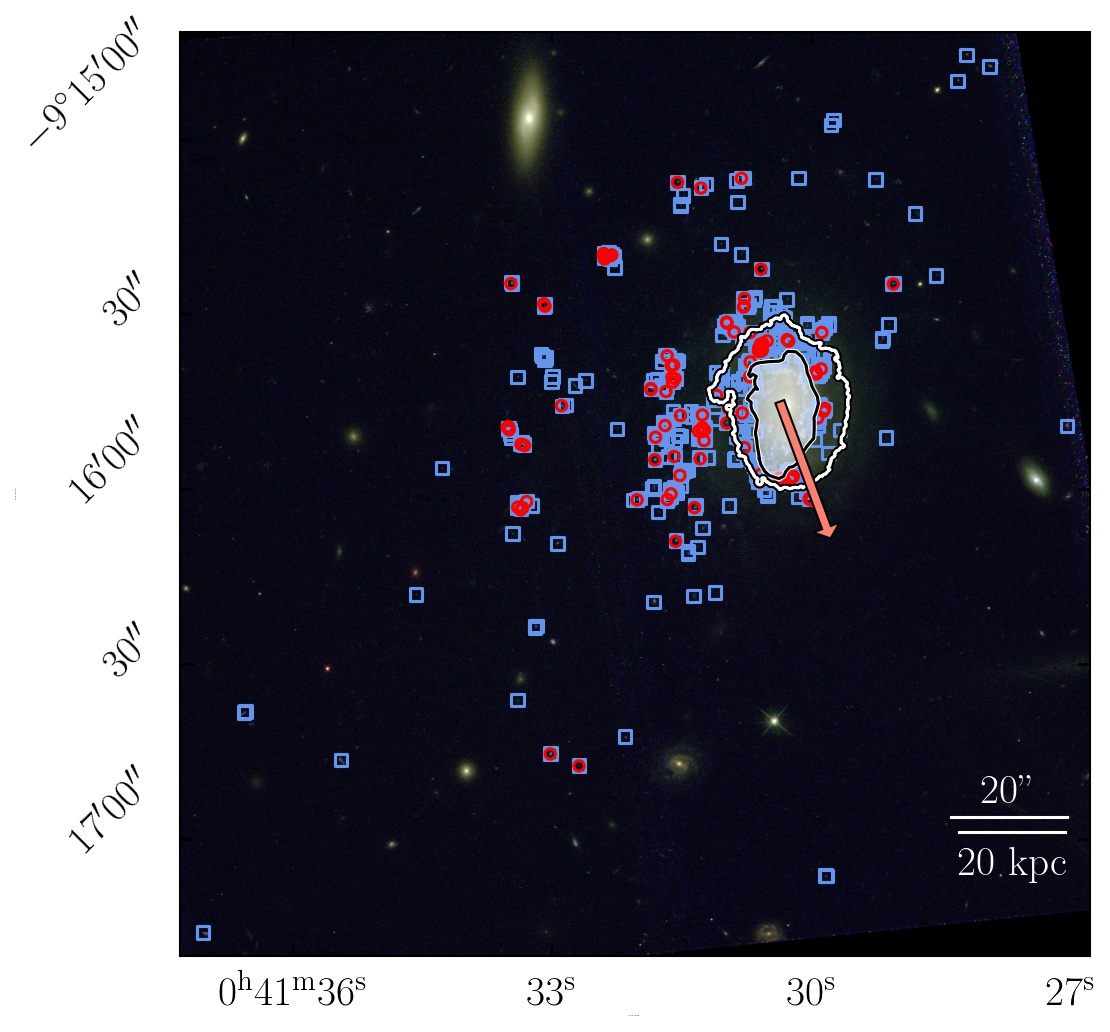}\hspace{-0.04cm}
    \includegraphics[height=1cm]{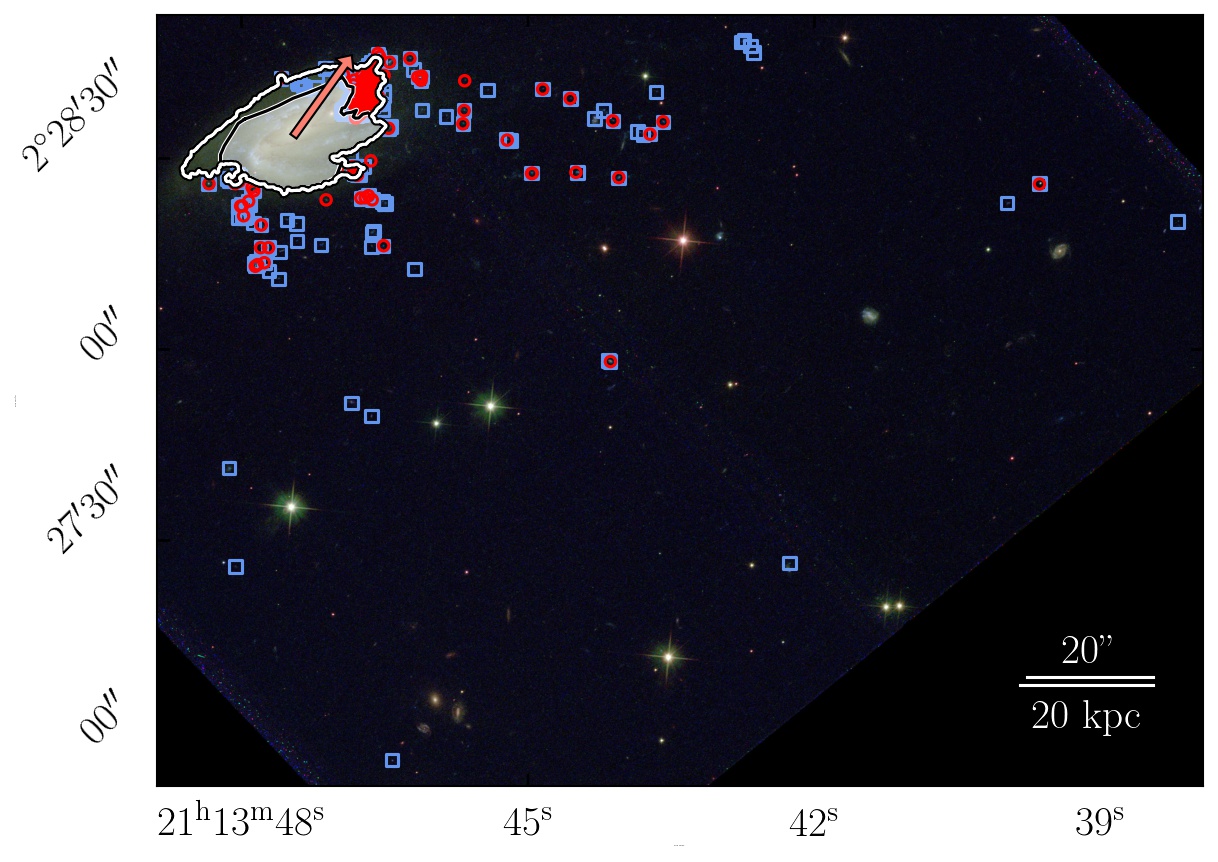}}\vspace{-0.2cm}\\
    \resizebox{0.88\textwidth}{!}{\includegraphics[height=1cm]{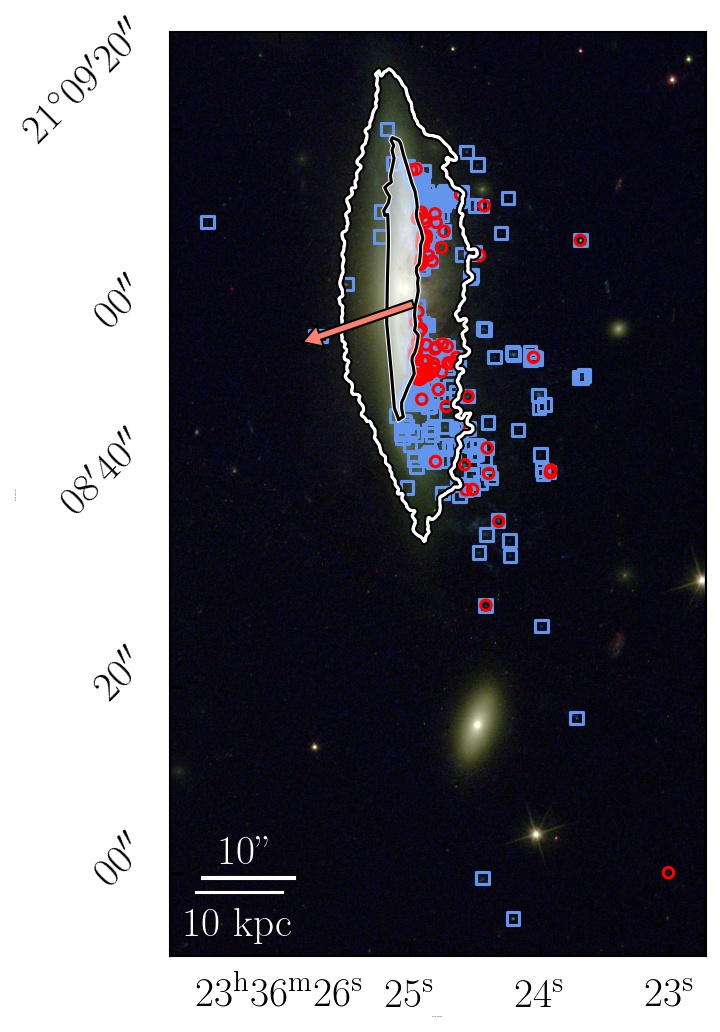}\hspace{-0.04cm}
    \includegraphics[height=1cm]{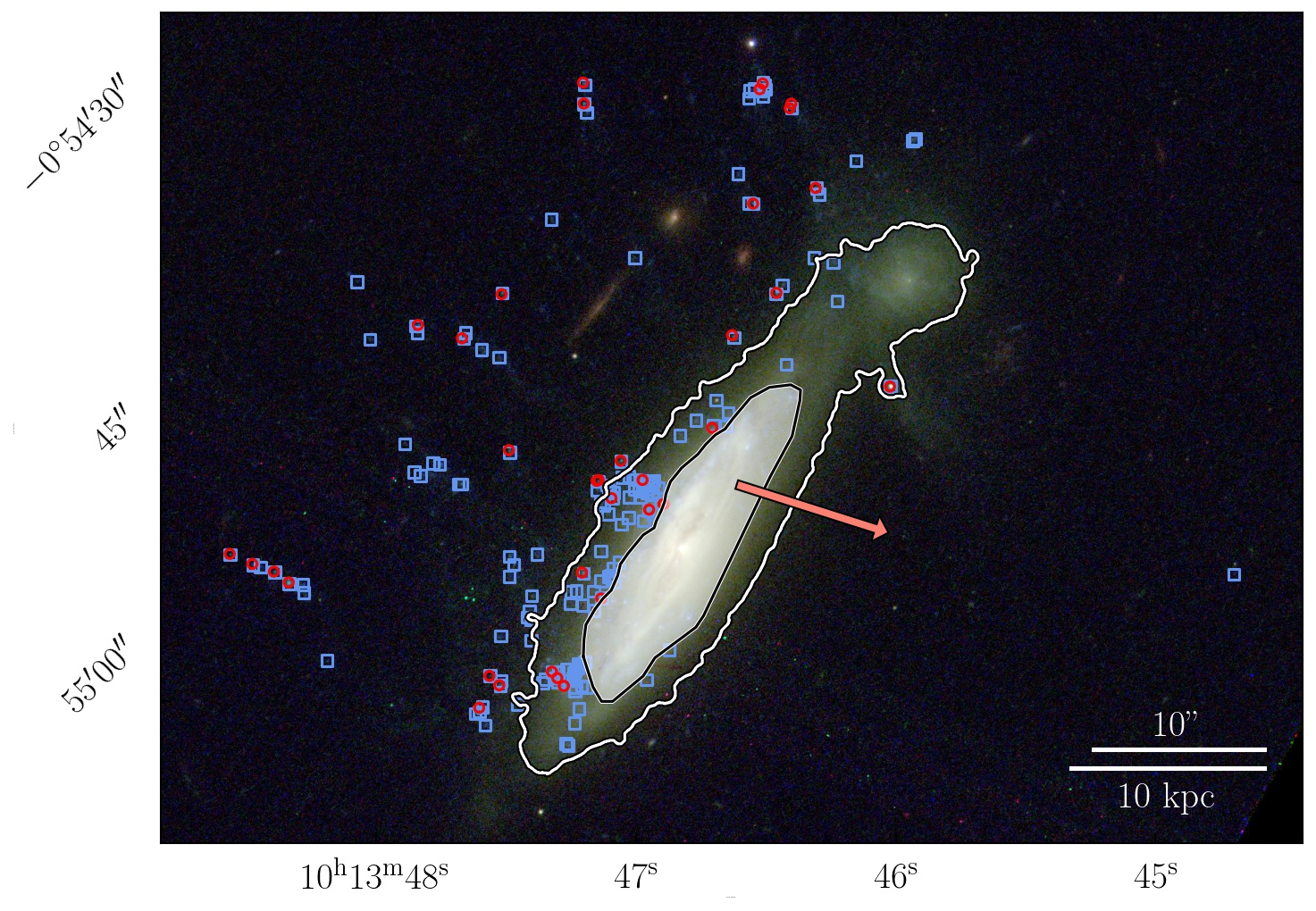}}\vspace{-0.2cm}\\
    \resizebox{0.88\textwidth}{!}{\includegraphics[height=1cm]{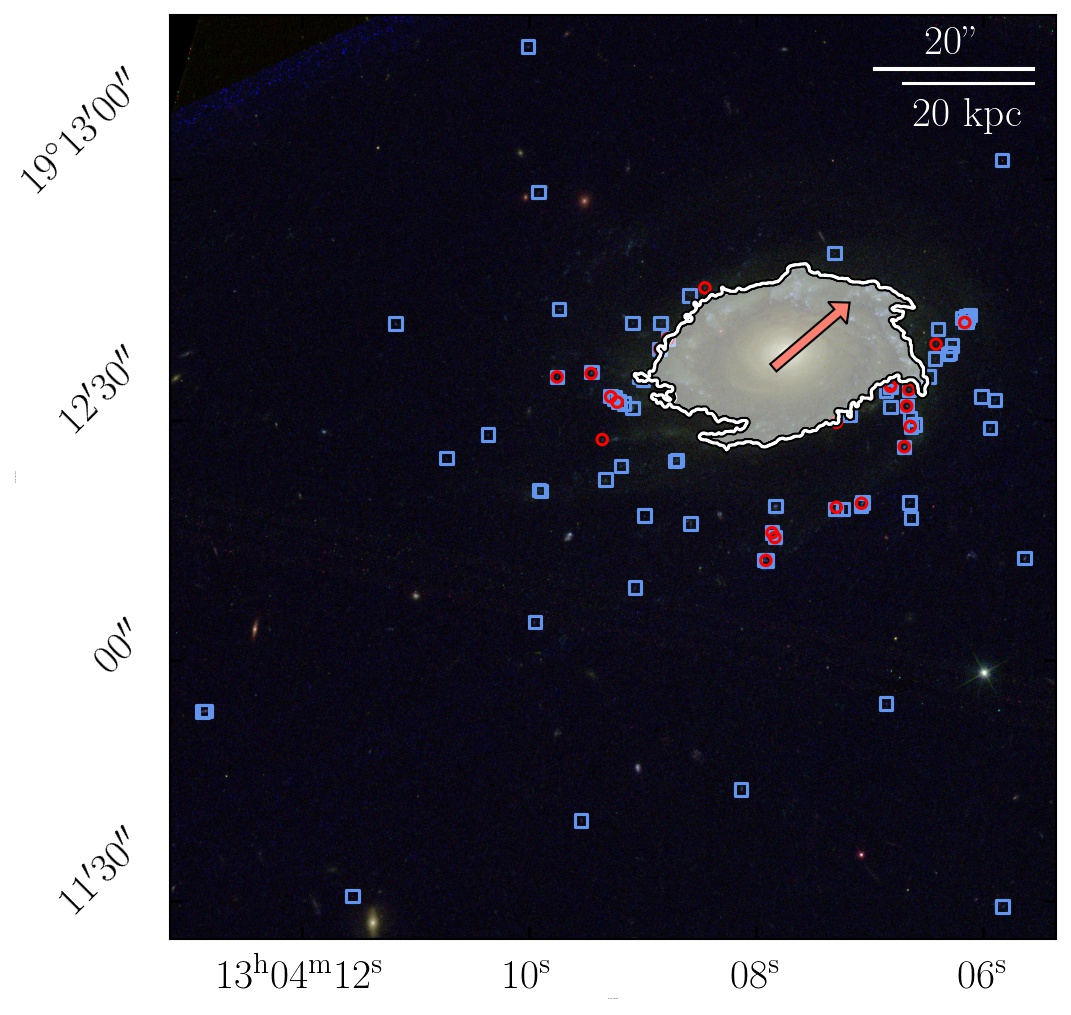}\hspace{-0.04cm}
    \includegraphics[height=1cm]{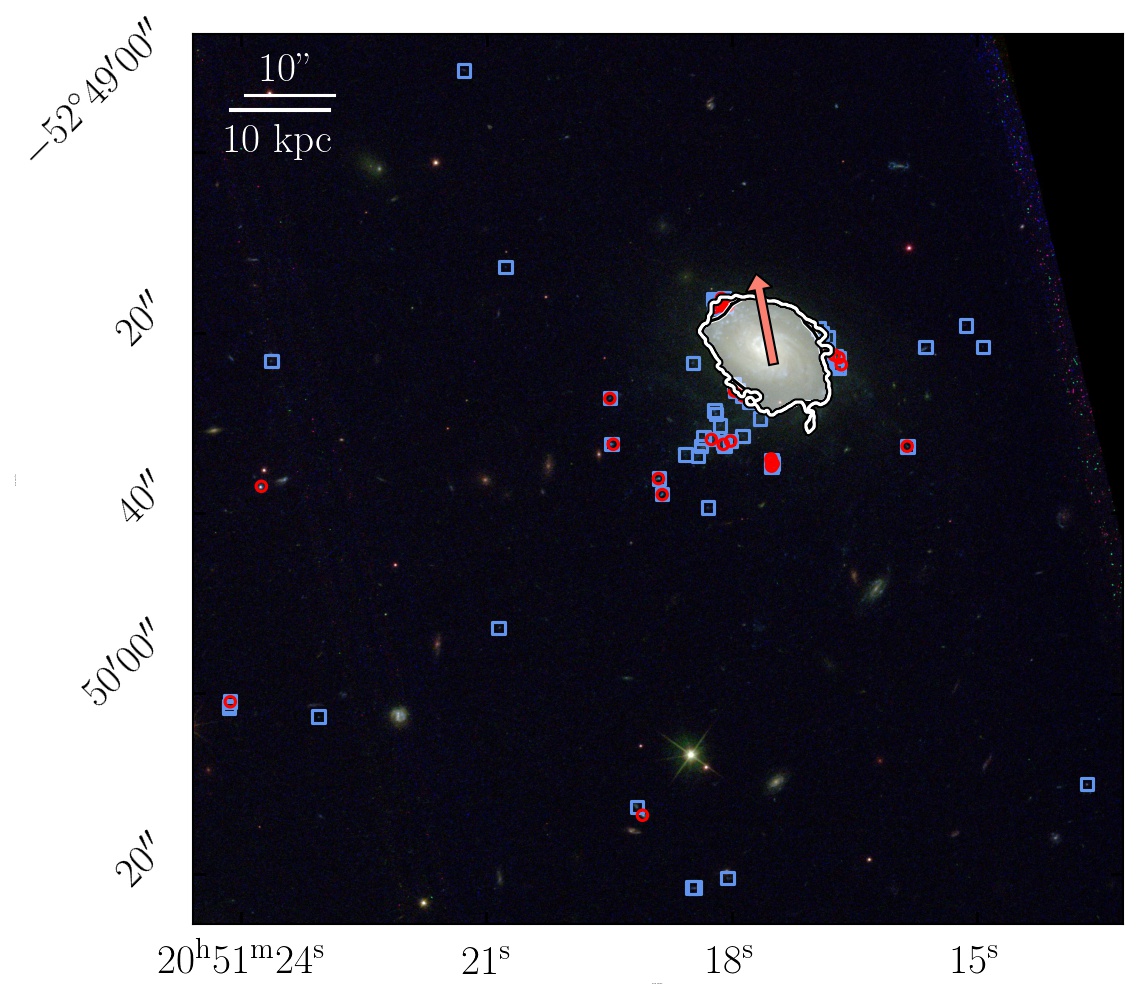}}
\caption{RGB images of the six studied galaxies. The galaxies are: JO201 and JO206 (top row), JW100 and JO204 (middle row), JW39 and JO175 (bottom row).
The images are obtained following the prescription given in \citep{Gullieuszik2023} and summarized in Sect. \ref{sec:dataset}. White and black contours separate the tail, extraplanar and disk regions, as described in Sect. \ref{sec:spatcat}. The white shaded area highlights the disk region, which is not considered throughout this work. We point out that the two contours almost coincide in JW39 and JO175 (bottom left and bottom right panels, respectively).
Blue squares and red round dots mark the position of UV and H$\alpha$ leaf clumps with robust mass estimates \citep{Giunchi2023a,Werle2024}, summarized in Sects. \ref{sec:clumps} and \ref{sec:method_mass}. The pink arrow points toward the centre of the galaxy cluster (taken from \citealt{Cava2009} and \citealt{Biviano2017}).}
\label{fig:gals}
\end{figure*}

The \textit{HST} dataset, together with its reduction and analysis, is described in detail in \cite{Gullieuszik2023}, and here briefly summarized.
The pixel angular size of UVIS/WFC3 is $0.04\arcsec$, and the Point-Spread Function (PSF) is almost constant in all the filters, with a FWHM of $\sim 0.07\arcsec$\footnote{https://hst-docs.stsci.edu/wfc3ihb/chapter-6-uvis-imaging-with-wfc3/6-6-uvis-optical-performance}, granting $\sim 70\,\mathrm{pc}$ physical resolution.
The targets were observed in 4 broad-band photometric filters (F275W, F336W, F606W, F814W) and one narrow-band filter, F680N. The broad-band filters cover the UV- to I-band rest-frame spectral interval, while the narrow-band filter collects the H$\alpha$ emission line at the redshift of the targets (column 5 in Table \ref{tab:targets}).
The images are available in the Mikulski Archive for Space Telescopes (MAST): \href{http://dx.doi.org/10.17909/tms2-9250}{10.17909/tms2-9250}. The ID of the project is GO-16223; PI Gullieuszik.
The RGB images of the targets are shown in Fig. \ref{fig:gals}, where one can appreciate the high level of details achieved by \textit{HST} and the large field of view covering both the disk and the tails of the galaxies. The RGB images are obtained following the same prescription of \cite{Gullieuszik2023}, using the F814W and F606W data for the R and G channels, respectively; for the B channel a pseudo B-band image $F_\mathrm{B}$ is created as $F_\mathrm{B}=0.25\,(F_\mathrm{F275W}+F_\mathrm{F336W})+0.5\,F_\mathrm{F606W}$, where $F_\mathrm{i}$ is the flux in the i$th$ filter.

The continuum emission in the F680N filter is interpolated from the adjacent filters (F606W and F814W) and subtracted in order to extract the intensity of the H$\alpha$ line (either in emission or absorption).

\subsection{Distinction of disk, extraplanar and tail regions}\label{sec:spatcat}

As described in detail in \cite{Giunchi2023a}, for each galaxy we defined three spatial categories to be studied separately: tail, extraplanar and disk. The tail region is the area outside the most external layer of the optical stellar disk of the galaxy, defined as the $2\sigma$ (where $\sigma$ is the background noise, computed as described in \citealt{Gullieuszik2023}) contour of the reddest photometric band available (F814W).
Inside the optical stellar disk, we visually inspected the UV (F275W+F336W) images of the galaxies to define an inner contour separating clumps looking mostly undisturbed (disk clumps, inside the contour) from those already showing clear signs of RPS, like elongated and filamentary structures aligned with the stripping direction (extraplanar clumps, outside the contour).
The overlap of these clumps with the disk is most likely due to the combined effects of projection and their formation at very early stages of RPS, when the gas had just been stripped and was still very close to the galaxy.

In Fig. \ref{fig:gals} we show the contours defining the three spatial categories as white and black contours: the former traces the optical galactic disk, the latter divides clumps affects by RPS from those looking undisturbed.The tail region includes everything outside the white contour, the extraplanar region is in between the white and black contours and the disk region is inside the black contour. We point out that, in some galaxies (namely JW39 and JO175, both in the bottom row of Fig. \ref{fig:gals}), the two contours are almost coincident, since most of the clumps inside the optical galactic disk do not show signs of RPS.

\subsection{Catalog of star-forming clumps}\label{sec:clumps}

In this section we summarize how the star-forming clumps and complexes were detected and selected, together with the main properties of the samples.
For a complete description we refer to \cite{Giunchi2023a}, in which the luminosity and size distribution functions and the luminosity-size relation of the clumps are studied.

The clumps are detected and selected independently in continuum-subtracted H$\alpha$-line and UV (F275W filter, $\sim 260$ nm rest-frame at $z=0.05$) images. Both these tracers probe recently formed stars. H$\alpha$ emission is due to gas ionized by very massive stars, therefore probes star formation on timescales $\sim 10$ Myr, while UV emission is associated to the continuum of OBA stars and therefore probes star formation on timescales $\sim 200$ Myr \citep{Kennicutt1998a,Kennicutt2012,Haydon2020}. Using both tracers allows us to study how star formation in clumps changes both in time and space.

Clumps are selected using \textsc{Astrodendro}\footnote{https://dendrograms.readthedocs.io/en/stable/index.html}, a Python software package able to detect bright sources and, where present, also brighter sub-structures and secondary peaks within larger sources. The input parameters to be set include the minimum number of pixels that a clump candidate must possess, the minimum flux of each pixel and the flux step used to search for bright sub-structures. In our case, we required a clump candidate to have at least 5 pixels (the minimum number of pixels needed to sample the PSF of \textit{HST}), with a minimum flux of $2\sigma$ (where $\sigma$ is the single-pixel detection limit, computed as described in \citealt{Gullieuszik2023}) and a flux step equal to $5\sigma$.
The resulting collection of structures is then catalogued as a hierarchical tree made of trunks and then nested into branches, and finally reaching the top of the structures, the so-called leaves.

From the source sample detected by \textsc{Astrodendro}, only candidates with total signal-to-noise ratio $\mathrm{SNR}>2$ in at least three out of five photometric bands are kept.
We used MUSE observations to confirm candidates in the tails and exclude background objects (see \citealt{Poggianti2019a}), by getting local redshift measurements from the corresponding spaxel, to be compared with the one of the galaxy. Candidates outside of the MUSE field of view are confirmed if they have $\mathrm{SNR}>2$ in all the five photometric bands and positive H$\alpha$ emission.

As described in detail in \cite{Giunchi2023a}, for every clump a variety of physical parameters can be computed. The fluxes of the clumps in the five photometric filters are computed by integrating within their area, while their sizes are defined as twice the PSF-corrected core radius ($\rcorecorr$). The core radius is the geometric mean of the semi-major and semi-minor axes of the clumps, which are the standard deviations of the surface brightness distribution of the clumps, computed along the direction of maximum elongation and the perpendicular direction, respectively.
The photometric samples include 2406 (1708 disk, 375 extraplanar, 323 tail) H$\alpha$-selected clumps and 3745 (2021 disk, 825 extraplanar, 899 tail) UV-selected clumps.
As previously stated, throughout this paper we will focus on the tail and extraplanar clumps.

\subsection{Estimate of masses and ages of the clumps}\label{sec:method_mass}
The aim of this work is the study of the mass function of the H$\alpha$- and UV-selected clumps observed in the tail and extraplanar regions of these galaxies, therefore we need masses and (as we will see in Sect. \ref{sec:sed}) ages of the clumps in the photometric catalogues described in the previous section.
The procedure adopted to get stellar masses $M_*$ and mass-weighted ages $\mathrm{age_{mw}}$ of the clumps is extensively described in \cite{Werle2024} and here briefly summarized.

Both those quantities are obtained using \textsc{Bagpipes} (Bayesian Analysis of Galaxies for Physical Inference and Parameter EStimation, \citealt{Carnall2018}), a code modelling the observed Spectral Energy Distribution (SED), in our cases represented by the flux density in the five photometric filters.

First of all only leaf clumps with $\mathrm{SNR}>2$ in all the filters are selected.
The code uses the $2016$ update of the simple stellar population (SSP) models from \cite{Bruzual2003} with a \cite{Kroupa2001} IMF; the solar metallicity is assumed to be $Z_\odot = 0.017$.
Adopting a Bayesian fitting technique, \textsc{Bagpipes} returns posterior probability distributions (PDFs) for each fitted quantity; in the following, the reference value of a physical property is computed as the median of its PDF.
Clumps star formation histories (SFHs) are modelled as a single delayed exponential \citep{Carnall2019}.

In case of stellar populations younger than 20 Myr, the code includes also emission lines and nebular continuum emission from \textsc{Cloudy} \citep{Ferland2017}.
Dust is modelled after the Milky Way extinction curve from \cite{Cardelli1989}, assuming $R_V=3.1$ and two foreground dust screens with different $V$-band extinction $A_V$, for stellar populations older and younger than 20 Myr, respectively. The young-to-old $A_V$ ratio is kept between 1 and $2.5$.
The total assembled stellar mass is left to vary from 0 to $10^{10}M_\odot$ and the ages between 0 and 500 Myr.

The resulting fits underwent a series of quality-check controls.
To evaluate how the underlying stellar populations belonging to the galaxy disk might affect our results (especially for extraplanar clumps), a star formation history with two components was tested, to model both the young and old stellar populations.
If the addition of a second component led to unconstrained mass and age, or if the old component was found to be more than $100$ times more massive than the young one, the clump was discarded, as in both cases the properties of the young stellar populations could not be constrained well enough. This happens $\sim 7\%$ ($\sim 16\%$) of the H$\alpha$ (UV) clumps (mostly in the extraplanar region).
For $\sim 8\%$ of the H$\alpha$ clumps and $\sim 22\%$ of the UV clumps, the model fluxes are outside the 2$\sigma$ error bars of the observations for one or more filters, thus the fit cannot be considered satisfactory and is therefore discarded.

\begin{figure*}[t!]
\includegraphics[width=0.49\textwidth]{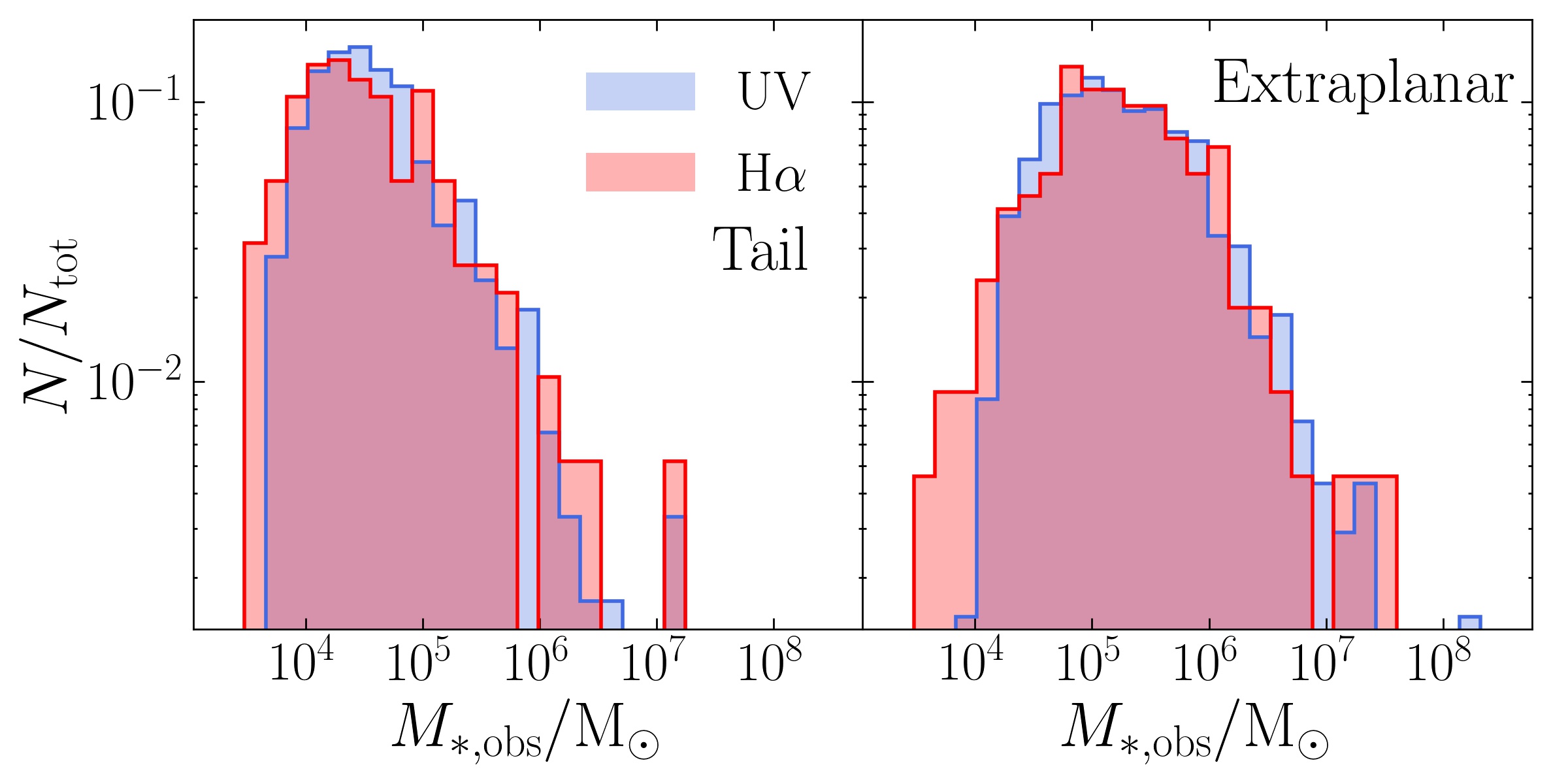}
\includegraphics[width=0.49\textwidth]{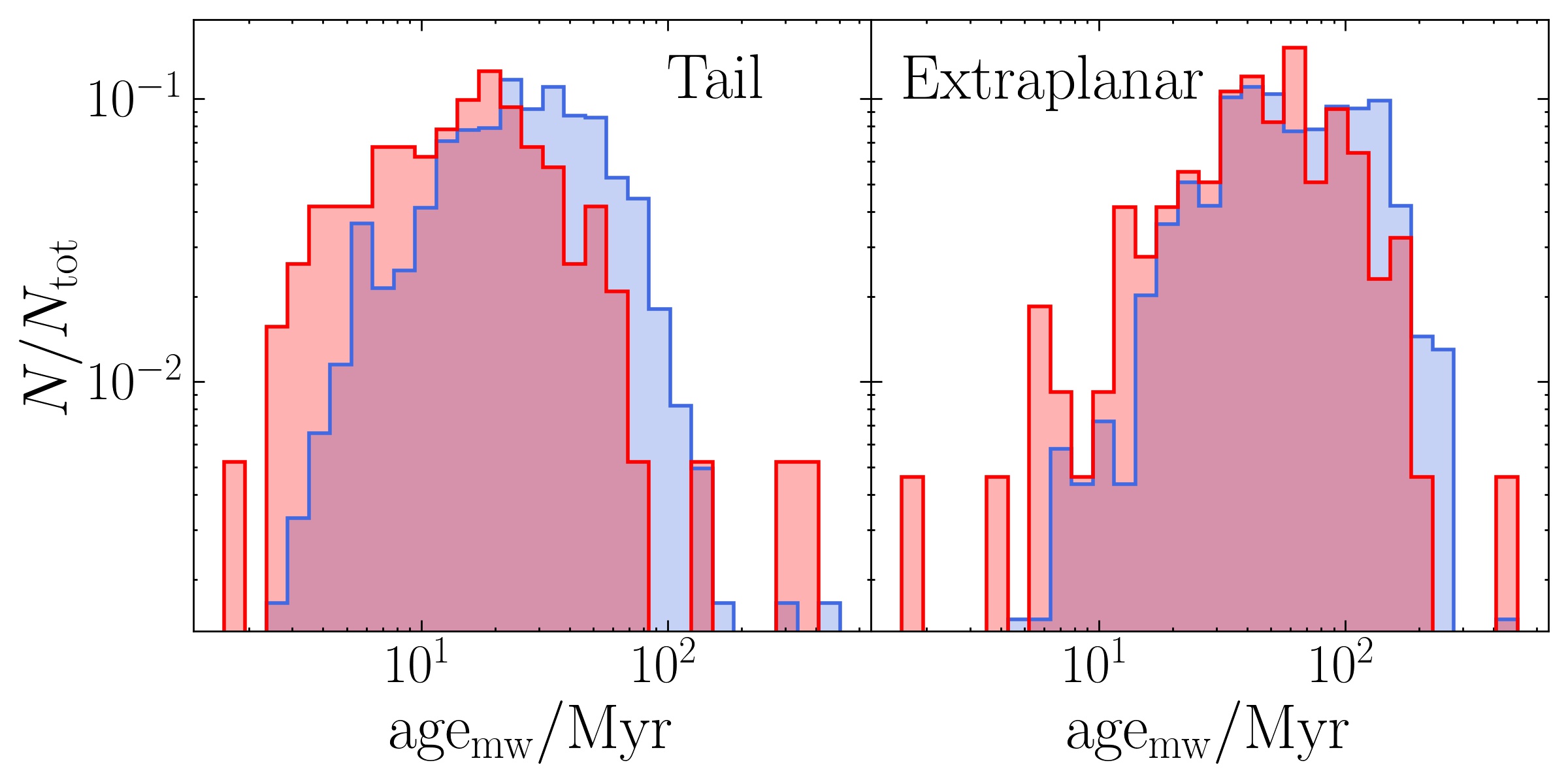}
\caption{Stellar mass (top row) and mass-weighted age (bottom row) distributions of H$\alpha$ (red histograms) and UV (blue histograms) clumps. Clumps are divided in tail (left panels) and extraplanar (right panels) accordingly to Sect. \ref{sec:spatcat}, and the sample is selected following the procedure described in Sect. \ref{sec:method_mass}.}
\label{fig:mass_age_distr}
\end{figure*}

After these quality-check cuts, our final sample consists of 398 H$\alpha$-selected clumps (188 tail, 210 extraplanar) and 1270 UV-selected clumps (593 tail, 677 extraplanar). The clump positions are plotted in Fig. \ref{fig:gals} as red rounds and blue squares, for the H$\alpha$ and UV clumps, respectively.
The distributions of the stellar mass and mass-weighted ages of H$\alpha$ and UV clumps, divided also in tail and extraplanar regions, are shown in Fig. \ref{fig:mass_age_distr}. Our masses span a large interval going from a few $10^3\,\msun$ up to $10^8\,\msun$. The most massive clumps of the sample are located in the extraplanar region, while the low-mass regime is populated by the H$\alpha$ clumps, whether in tails or extraplanar regions. The ages go from a few Myr to hundreds of Myr, with the youngest clumps preferentially selected in H$\alpha$ and the oldest ones in UV. A detailed study of these mass and age distributions was performed by \cite{Werle2024}.

\section{Modelling the mass function}\label{sec:mass_function}
Before fitting a chosen function to the clumps mass distribution, we need to quantify the main sources of bias and uncertainty that could lead to incorrect results.
We considered two effects: 1) the completeness of the sample as a function of the clump mass; 2) the bias and uncertainty on the estimate of the intrinsic mass of a clump, which highly depends on how the fluxes are computed and on the model adopted for the SED fitting.

One way to quantify these effects is to study images of mock clumps with given mass and age as they were real images, comparing the input quantities with those retrieved as output, similarly to what has been done for other surveys of extended objects \citep{Villegas2010,Sun2024}. In order to do that, we modelled the morphology and SED of the clumps as a function of their mass and age and, based on these models, generate the library of mock images. The age of the clumps determines the shape of the SED and the mass-to-light ratio, while the mass of the clumps constrains the bolometric luminosity (fixed the mass-to-light ratio) and the size.
Further details about how this library is created is given in the Appendix, from \ref{sec:stacking} to \ref{sec:re_obs}, and here briefly summarized:

\begin{enumerate}
    \item We stacked a subsample of resolved clumps, characterized by smooth, round, non-peculiar morphology and properly normalized in size and flux in order to be comparable with each other. The stacked image is then fitted to two 2D Moffat functions, producing the stacked model, fully normalized both in size and flux. This was done independently for H$\alpha$ and UV clumps;
    \item Clump masses are extracted from a flat distribution (in order to evenly sample and study all the mass regimes), while the ages follow the observed age distribution derived as in Sect. \ref{sec:method_mass};
    \item The physical size of the clump is derived for a given mass from the clump size-mass relation (fitted as described in Sect. \ref{sec:size_mass});
    \item The shape of the SED is determined by the age of the clumps, which determines for example how bright the H$\alpha$ line or the UV part of the spectrum are. In particular, we adopted the resulting SEDs obtained from the fits performed with \textsc{bagpipes} in \cite{Werle2024};
    \item Finally, for a given age, the mass-to-light ratio determines the flux normalization of the SED.
\end{enumerate}

We point out that the steps listed above are performed independently for the four subsamples of H$\alpha$ tail, H$\alpha$ extraplanar, UV tail and UV extraplanar clumps, that can possibly have different size-mass relations, age distributions and mass-to-light ratios. The only exception is step n. 1 (i.e. the generation of the stacked clump model), which is performed separately for H$\alpha$ and UV clumps, but with no further divisions related to the spatial category.

The subsequent step is described in detail in Sect. \ref{sec:re_obs} and consists in adding the mock clumps to the real images, on which we performed the same procedure adopted to detect the real clumps (\citealt{Giunchi2023a}, summarized in Sect. \ref{sec:clumps}) and obtained their masses (\citealt{Werle2024}, summarized in Sect. \ref{sec:method_mass}). This process is iterated many times, in order to reach a sufficient statistics ($\sim 1000$ mock clumps per galaxy).
The number of clumps added to the tail and extraplanar regions at each iteration is set by the need not to change the crowding of the real underlying clump population of the studied galaxy.
Whether the mock clump is re-detected or not and (if it is) the comparison between input and output properties quantify the aforementioned sources of uncertainty.

In the following sections, we will show in details the results for what concerns the completeness (Sect. \ref{sec:mass_completeness}) and the mass bias (Sect. \ref{sec:in_out_mass}).

\subsection{Completeness}\label{sec:mass_completeness}
We define the completeness as the fraction of re-detected clumps per intrinsic mass.
We modelled it as a logistic function characterized by an \textit{S}-shaped curve, monotonically increasing from an asymptotic value (for $x \rightarrow -\infty$) to another (for $x \rightarrow +\infty$). The equation describing the logistic function \citep{McCullagh1989,Peng2002} is

\begin{equation}\label{eq:completeness}
    C(\log m_\mathrm{in})=\cfrac{k}{[1+e^{-p(\log m_\mathrm{in}-\log m_\mathrm{in,0})}]},
\end{equation}

\noindent where $m_\mathrm{in}$ is the intrinsic mass, $k$ is the value of the asymptotic value for $m_\mathrm{in}\rightarrow +\infty$, $p$ regulates the sharpness of the increase of the function and $m_\mathrm{in,0}$ is the scale intrinsic mass at which $C(\log m_\mathrm{in,0})=k/(1+e)$.

The space of the free parameters $(k,p,\log m_\mathrm{in,0})$ is explored according to the log-likelihood function of the logistic regression  \citep{McCullagh1989,Peng2002}

\begin{equation}\label{eq:logL_compl}    \log\mathcal{L}_C=\sum_{i=1}^{N_\mathrm{det}}\log C(\log m_{\mathrm{in},i})+\sum_{j=1}^{N_\mathrm{non-det}}\log[1-C(\log m_{\mathrm{in},j})],
\end{equation}

\noindent where $N_\mathrm{det}$ and $N_\mathrm{non-det}$ are the numbers of re-detected and not detected mock clumps added to the mock observations, respectively, while $m_{\mathrm{in},i}$ and $m_{\mathrm{in},j}$ are the input stellar masses of the $i-$th re-detected and $j-$th not detected mock clumps, respectively.
We compute the best-fitting parameters as the median of the PDFs obtained by exploring the parameter space to maximize $\mathcal{L}_C$ and listed in the rows $(2)-$to$-(4)$ of Table \ref{tab:completeness_discrepancy}. We point out that this fitting procedure we followed (Eq. \ref{eq:logL_compl}) is binning independent.

\begin{table*}[t!]
\fontsize{10pt}{10pt}\selectfont
\setlength{\tabcolsep}{4pt}
\renewcommand{\arraystretch}{1.4} 
\centering
\caption{Best-fitting parameters of the completeness and mass bias functions for H$\alpha$ tail (first row), H$\alpha$ extraplanar (second row), UV tail (third row) and UV extraplanar (fourth row) mock clumps.}
\begin{tabular}{ll|ccc|ccccc}
\bottomrule
\bottomrule
Filter & region & $k$ & $p$ & $\log m_\mathrm{in,0}$ & $a$ & $b$ & $c$ & $d$ & $e$\\
& & $[\times 10^{-1}]$ &  &  &  & $[\times 10^{-1}]$ & $[\times 10^{-3}]$ & $[\times 10^{-1}]$ & $[\times 10^{-2}]$\\
$(0)$ & $(1)$ & $(2)$ & $(3)$ & $(4)$ & $(5)$ & $(6)$ & $(7)$ & $(8)$ & $(9)$\\
\hline
\multirow{2}{*}{H$\alpha$} & tail & $9.91\pm 0.02$ & $4.85\pm0.16$ & $4.77\pm 0.01$ & $2.63\pm 0.15$ & $3.8\pm 0.6$ & $33\pm 5$ & $2.6\pm 0.2$ & $-0.9\pm 0.3$ \\
& extra. & $9.29\pm 0.08$ & $2.95\pm0.09$ & $5.49\pm 0.02$ & $2.0\pm 0.5$ & $6.6\pm 1.5$ & $8\pm 13$ & $8.3\pm 0.3$ & $-7.3\pm 0.8$\\ \hline
\multirow{2}{*}{UV} & tail & $9.87\pm 0.04$ & $3.88\pm 0.13$ & $5.02\pm 0.01$ & $1.77\pm 0.20$ & $7.6\pm 0.7$ & $-5.2\pm 6.0$ & $5.3\pm 0.2$ & $-6.1\pm 0.3$\\
& extra. & $7.84\pm 0.13$ & $2.38\pm 0.08$ & $5.80\pm 0.02$ & $1.8\pm 0.6$ & $6.7\pm 2.0$ & $13\pm 15$ & $10.6\pm 0.6$ & $-11\pm 1$\\
\hline
\end{tabular}
\tablefoot{The completeness and mass bias functions are modelled as described in Sects. \ref{sec:mass_completeness} and \ref{sec:in_out_mass}. 
Column $(0)$ and $(1)$ refer to the mock clumps reference filter and region, respectively. Columns $(2)$ to $(4)$: free parameters of the completeness $(k,p,\log m_\mathrm{in,0})$ as described in Eq. \ref{eq:completeness} (i.e. the maximum value) the steepness of the rise and the input mass at which the rise starts. Columns $(5)$ to $(9)$: free parameters of the mass bias function as in Eq. \ref{eq:discrepancy_function}, where $(a,b,c)$ describe the input-output mass correlation and $(d,e)$ the intrinsic scatter of the correlation.}
\label{tab:completeness_discrepancy}
\end{table*}

Now, correcting for the non-detected clumps when the completeness is below $0.5$, where more than half of the clumps are lost, may not bring to robust estimates. Therefore we force the completeness to $0$ for input masses below the value $m_{\mathrm{in},1/2}$ at which $C(\log m_{\mathrm{in},1/2})=0.5$. The operative completeness $\widetilde{C}(\log m_\mathrm{in})$ becomes

\begin{equation}\label{eq:completeness_effective}
    \begin{cases}
        0\quad\quad&\mathrm{for}\quad\log m_\mathrm{in}<\log m_{\mathrm{in},1/2};\\
        C(\log m_\mathrm{in})\quad\quad&\mathrm{for}\quad\log m_\mathrm{in}\geq\log m_{\mathrm{in},1/2}.
    \end{cases}
\end{equation}

\begin{figure*}[t!]
\resizebox{\textwidth}{!}{\includegraphics[height=1cm]{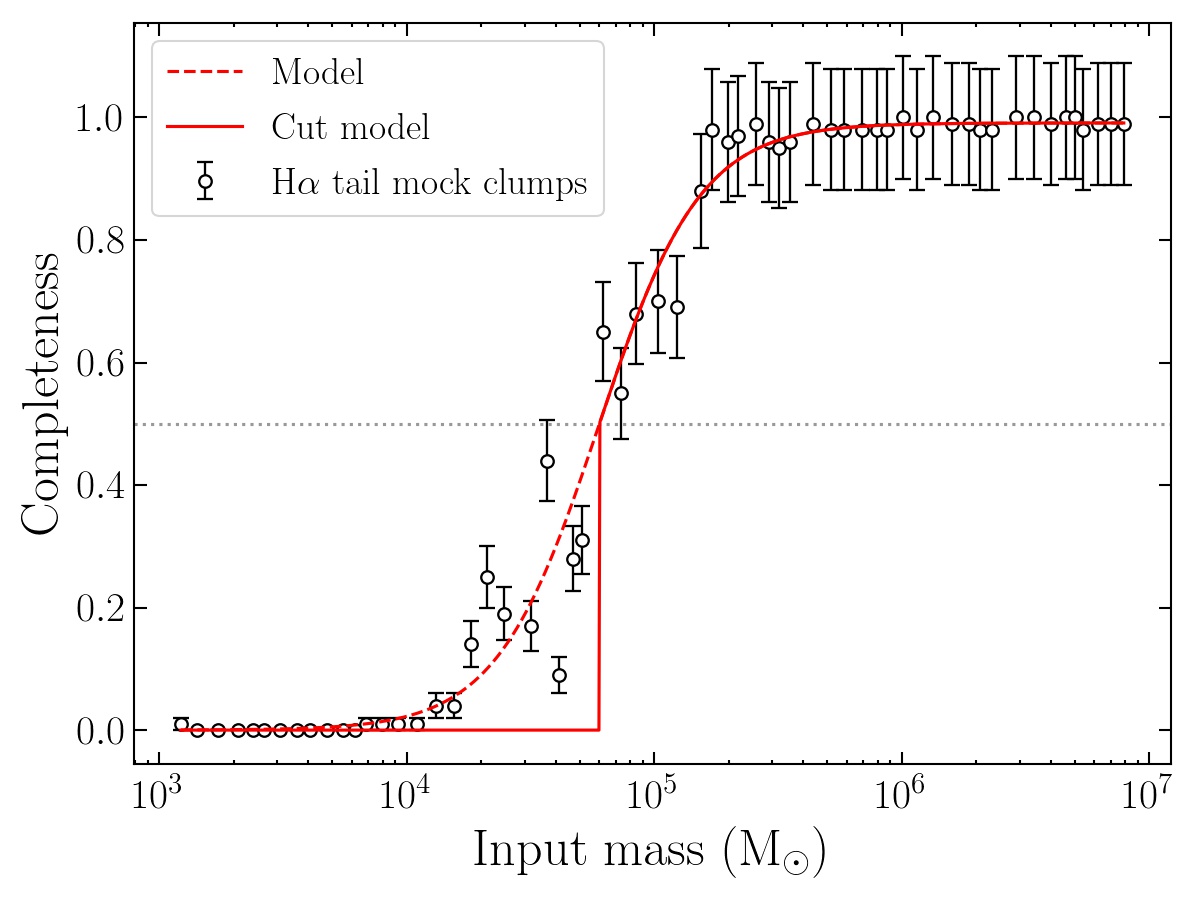}\hspace{-0.1cm}
    \includegraphics[height=1cm]{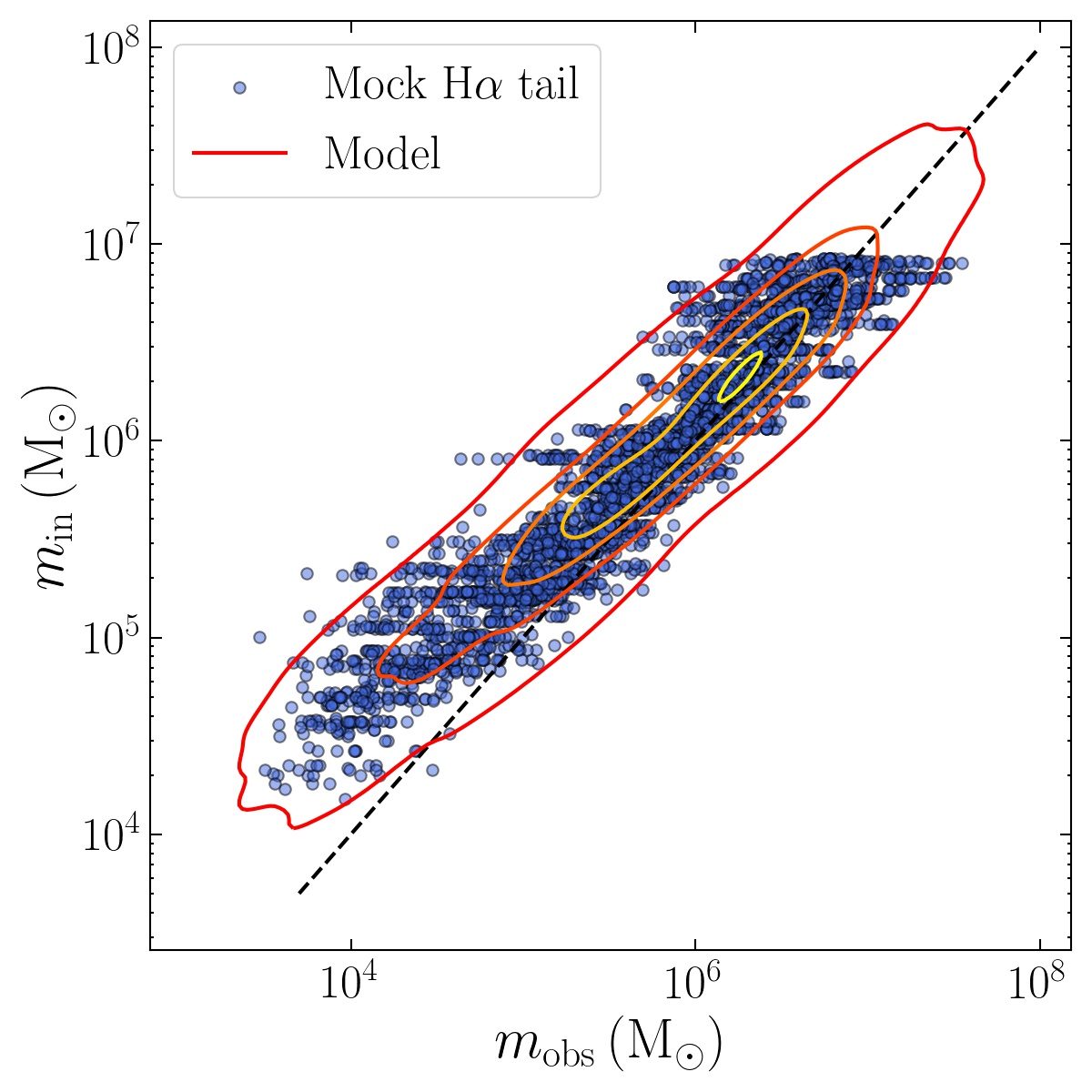}}
\caption{Completeness (left) and mass estimation bias (right) for H$\alpha$ tail clumps. Left: the red solid line is the best-fitting model computed as in Eq. \ref{eq:completeness_effective}, cut to 0 for values below $0.5$ (Sect. \ref{sec:mass_completeness}), while the red dashed line shows the uncut function. The threshold value of $0.5$, below which the function is forced to 0, is plotted as a grey dotted line. The black empty circles show the binned mass completeness derived from the mock clumps (with uncertainties equal to the Poissonian error), for a qualitative visualization of the goodness of the fit (the fitting technique does not depend on the binning). Right: relationship between observed ($x$-axis) and intrinsic ($y$-axis) mass for H$\alpha$ tail re-detected mock clumps. Each clump is plotted as a blue dot. The black dashed line is the $1:1$ relation. The red contours are computed from the best-fitting model describing the clumps distribution (Eq. \ref{eq:discrepancy_function}).}
\label{fig:ha_tail_params}
\end{figure*}

As an example, on the left panel of Fig. \ref{fig:ha_tail_params} we show the H$\alpha$ tail completeness function (where the dashed line is the profile of $C(\log m_\mathrm{in})$, before forcing it to 0). For a visual comparison, we over-plotted the observed input mass completeness; this is computed by binning the mock clumps in input mass, assigning to the $k-$th bin a value $C_k=N_{\mathrm{det},k}/N_k$, where $N_k$ and $N_{\mathrm{det},k}$ are the total number and the number of re-detected clumps in the $k-$th bin, respectively.
The same plots for the other filters and regions are shown in the left panels of Fig. \ref{fig:uv_tail_params}, showing that the best-fitting functions well reproduce the results from the mock observations, with low-mass clumps more likely to be lost, as expected.

\subsection{Mass estimate bias}\label{sec:in_out_mass}
Our capability of recovering the intrinsic mass of a star-forming clump is affected by many factors, like the noise of the image, which especially affects the faint regions of the clump, the definition of clump flux \citep{Giunchi2023a} and the models used to infer the mass \citep{Werle2024}. The comparison between the intrinsic mass of the mock clumps and the observed value can quantify this effect and unveil the presence of systematics.

In the right panel of Fig. \ref{fig:ha_tail_params} we show the comparison between observed and intrinsic mass of the re-detected H$\alpha$ tail mock clumps. Note that for high masses there is a good agreement between them, while at low masses the observed mass is systematically smaller than the intrinsic one. Similar plots for the different filters and regions (UV tail, H$\alpha$ extraplanar and UV extraplanar, respectively) are shown in Fig. \ref{fig:uv_tail_params}, where one can see how the underlying disk highly increases the uncertainties on the mass estimate in the extraplanar region.

We modelled the distribution of $(\log m_\mathrm{obs},\log m_\mathrm{in})$ as a quadratic relation $F$ with linearly decreasing scatter $\sigma$ depending on the observed mass $\log m_\mathrm{obs}$:

\begin{equation}\label{eq:discrepancy_function}
\begin{cases}
F(\log m_\mathrm{obs}) =& a+b\cdot\log m_\mathrm{obs}+c\cdot(\log m_\mathrm{obs})^2+\\
& +g[0,\sigma(\log m_\mathrm{obs})] \\
\sigma(\log m_\mathrm{obs}) =& d+e\cdot\log m_\mathrm{obs}
\end{cases}
\end{equation}

The best-fitting values of the free parameters $(a,b,c,d,e)$ are the median values of the PDFs obtained sampling the parameter space of the log-likelihood \citep{Ponomareva2017,Posti2018,Bacchini2019}

\begin{equation}\label{eq:logL_inout}
    \log \mathcal{L}_\mathrm{bias}=-\cfrac{1}{2}\sum_{i=1}^{N_\mathrm{cl}}\left[ \cfrac{d_i^2}{\sigma_\mathrm{tot}^2} + \ln (2\pi \sigma_\mathrm{tot}^2) \right],
\end{equation}

\noindent where $N_\mathrm{cl}$ is the number of re-detected mock clumps, $d_i^2=\log m_{\mathrm{in},i}^2-[a+b\cdot\log m_{\mathrm{obs},i}+c\cdot(\log m_\mathrm{obs})^2]^2$ is the distance between the $i-$th intrinsic mass and the one inferred from the model, and $\sigma_\mathrm{tot}=\sigma(\log m_{\mathrm{obs},i})$ is the intrinsic scatter at the $i-$th observed mass. The uncertainty on the mass estimate given by \textsc{bagpipes} is neglected a posteriori since it is much smaller than the intrinsic scatter obtained by this fit.
The best-fitting parameters are listed in the last five columns of Table \ref{tab:completeness_discrepancy} and the model distributions are shown as contours in Fig. \ref{fig:ha_tail_params} (and for the other filters and regions in Fig. \ref{fig:uv_tail_params}), where one can see that they well describe the mock clumps distributions.

Thanks to this model, at a given $j-$th clump observed mass $\log m_{\mathrm{obs},j}$ we define a Gaussian distribution $G_j(\log m_\mathrm{in})$ with mean and standard deviation given by Eq. \ref{eq:discrepancy_function}, which describes the distribution of possible intrinsic masses that can result in the observed mass $\log m_{\mathrm{obs},j}$.

\subsection{Modelling the mass function}\label{sec:mf_fit}
The fitting to the mass function (MF) must take into account the effects that we have quantified in the previous sections; the number of clumps with a certain mass is affected by incompleteness and a clump with a given observed mass may have a different input mass, which is the real value on which the MF depends.
The likelihood we used to explore the parameter space is taken from \citep{Vasiliev2019}, which was applied in a different context (fitting a sample of globular clusters to a distribution function) but with the same ingredients we have considered. The functional form of the log-likelihood is

\begin{equation}\label{eq:logL_mf}
\begin{split}
    &\ln\mathcal{L}_\mathrm{MF}(\pmb{\theta})=N_\mathrm{cl}\,\ln\pi(\pmb{\theta})+\\
    &+\sum_{i=1}^{N_\mathrm{cl}}\ln\frac{\int_{\log m_{\mathrm{in},1/2}}^{+\infty}\mathrm{MF}(\omega,\pmb{\theta})\,\widetilde{C}(\omega)\,G_i(\omega)\,d\omega}{\int_{\log m_{\mathrm{in},1/2}}^{+\infty}\mathrm{MF}(\omega,\pmb{\theta})\,\widetilde{C}(\omega)\,d\omega},
\end{split}
\end{equation}

\noindent where $N_\mathrm{cl}$ is the number of real UV or H$\alpha$ clumps with reliable mass estimates (Sect. \ref{sec:method_mass}), $\omega=\log m_\mathrm{in}$ (the clump intrinsic mass), $\mathrm{MF}(\omega,\pmb{\theta})$ is the mass function with free parameters $\pmb{\theta}$ (the functional form will be described later on), $\pi(\pmb{\theta})$ is the prior of the parameters (which depends on the parameters of the MF), $G_i(\omega)$ is the input mass distribution of the $i-$th observed mass $\log m_{\mathrm{obs},i}$ (Sect. \ref{sec:in_out_mass}) and $\widetilde{C}(\omega)$ is the completeness (Sect. \ref{sec:mass_completeness}). Note that the integral over input masses is done only for values larger than $\log m_{\mathrm{in},1/2}$, avoiding a mass regime in which the fit is highly affected by incompleteness.

\section{Results}\label{sec:res}
As stated in Sect. \ref{sec:intro}, whether the clump mass functions present a high-mass cut-off or not, and what drives the value of this mass cut-off, is still highly debated in literature. In order to test the putative presence of a cut-off mass in our sample, in the following sections we have fitted our clump mass datasets both to a single power law and to a Schechter function.

\subsection{Fit of the mass function to a single power-law}\label{sec:pl_test}
\begin{figure*}[t!]
\centering
\includegraphics[width=0.25\textwidth]{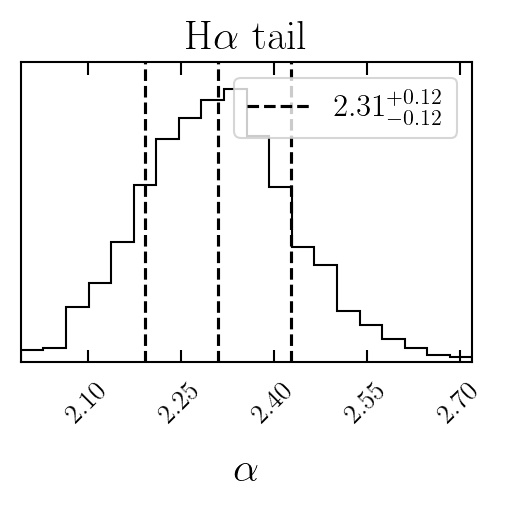}\hspace{-0.25cm}
\includegraphics[width=0.25\textwidth]{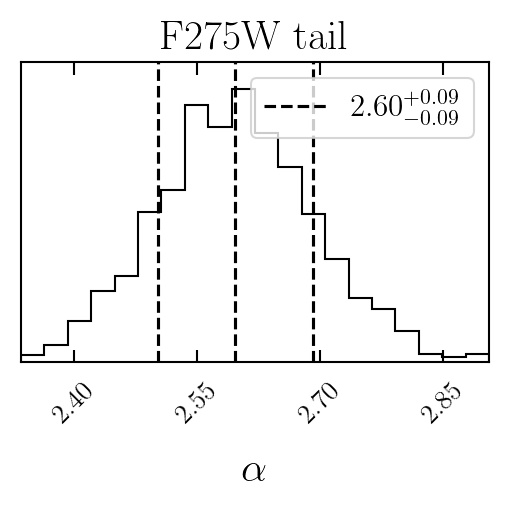}\hspace{-0.25cm}
\includegraphics[width=0.25\textwidth]{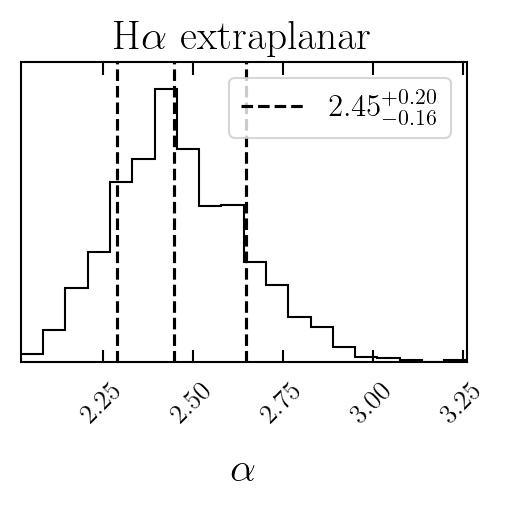}\hspace{-0.25cm}
\includegraphics[width=0.25\textwidth]{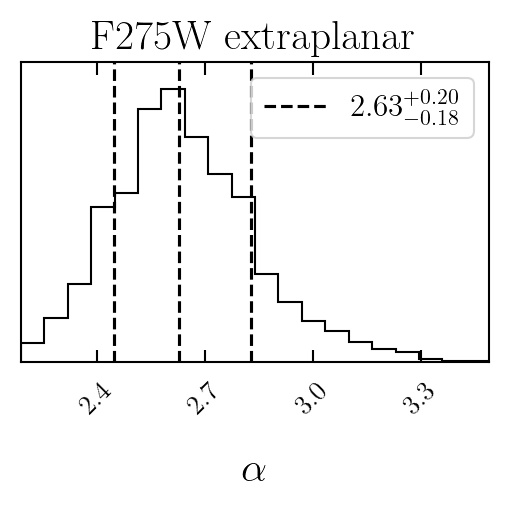}\\
\vspace{-0.3cm}
\caption{Posterior distribution of the slope parameters $\alpha$, inferred fitting a power law to the mass distribution of tail (first two panels) and extraplanar (last two panels) clumps (H$\alpha$ on the left panel, UV on the right panel). The vertical dashed black lines are the 16th, 50th and 84th percentiles, reported also in the legends.}
\label{fig:pl_hists}
\end{figure*}

We fitted the observed data to a MF modelled as a single power law (Eq. \ref{eq:mf}), with only the slope $\alpha$ as a free parameter.
The best-fitting slopes are computed as the median values of the parameter distributions, while the uncertainties are the 84th percentile minus the median and the median minus the 16th percentile (upper and lower uncertainty, respectively).
For the tail clumps, the retrieved slopes are $2.31\pm 0.12$ in H$\alpha$ and $2.60\pm 0.09$ in UV. In the extraplanar region, the best-fitting slopes are $2.45^{+0.20}_{-0.16}$ and $2.63^{+0.20}_{-0.18}$ in H$\alpha$ and UV, respectively. The posterior distributions are shown in Fig. \ref{fig:pl_hists}.

Our slopes are all more than $1\sigma$ larger than the typical value found in literature ($\alpha=2.0\pm 0.2$, \citealt{Krumholz2019}), with the only exception of the tail H$\alpha$ clumps. Such large slopes are usually associated to environmental effects and variations in the SFR surface density $\Sigma_\mathrm{SFR}$. Indeed, the environment in which jellyfish galaxies are located and the effects of RPS could halt the formation of massive clumps and explain the steepening.
In Sect. \ref{sec:schechter_test} we will test our results adopting different MF functional models, while in Sect. \ref{sec:conclusions} we will further discuss the possible causes of the steepness of the observed mass function.

The slope of the tail UV mass function is steeper (at $2\sigma$ confidence level) and therefore compatible with models predicting this evolution of the slope as a function of the ageing and dynamical evolution of the clumps \citep{Fujii2015}.

In the extraplanar region, H$\alpha$ clumps show a steeper slope than in the tail regions (yet consistent within $1\sigma$), hinting that the formation of massive clumps is even more halted in this conditions, possibly as a consequence of ram pressure, which is causing both gas compression and stripping, which both are at very early stages in this region, where the gas is still very close to the disk. Similarly to tail clumps, also in the extraplanar region the UV mass function is steeper than the H$\alpha$, even though the slope is identical to the one in the tail. However, we point out that they are consistent within $1\sigma$.
In the following section we will discuss in more detail the behaviour and the observational caveats that are present when working in the extraplanar region. As a consequence of that, the values of the slopes for the extraplanar clumps should be taken with caution.

\subsection{Fit of the mass function to a Schechter}\label{sec:schechter_test}
We explored the case of the Schechter function fitting, instead of the simple power law.
The Schechter function is defined as \citep{Schechter1976}

\begin{equation}\label{eq:schechter}
    \mathrm{MF}(M_*)\propto M_*^{-\alpha}\,e^{-M_*/M_c},
\end{equation}

\noindent with the slope $\alpha$ and the cut-off mass $M_c$. As described in Appendix \ref{sec:mock}, our fitting procedure is able to rule out or not the presence of a cut-off mass even when a small sample (about $100$ clumps) is fitted. Therefore this is a good test to confirm whether the observed mass function is a power law.

\begin{figure*}[t!]
\includegraphics[width=0.47\textwidth]{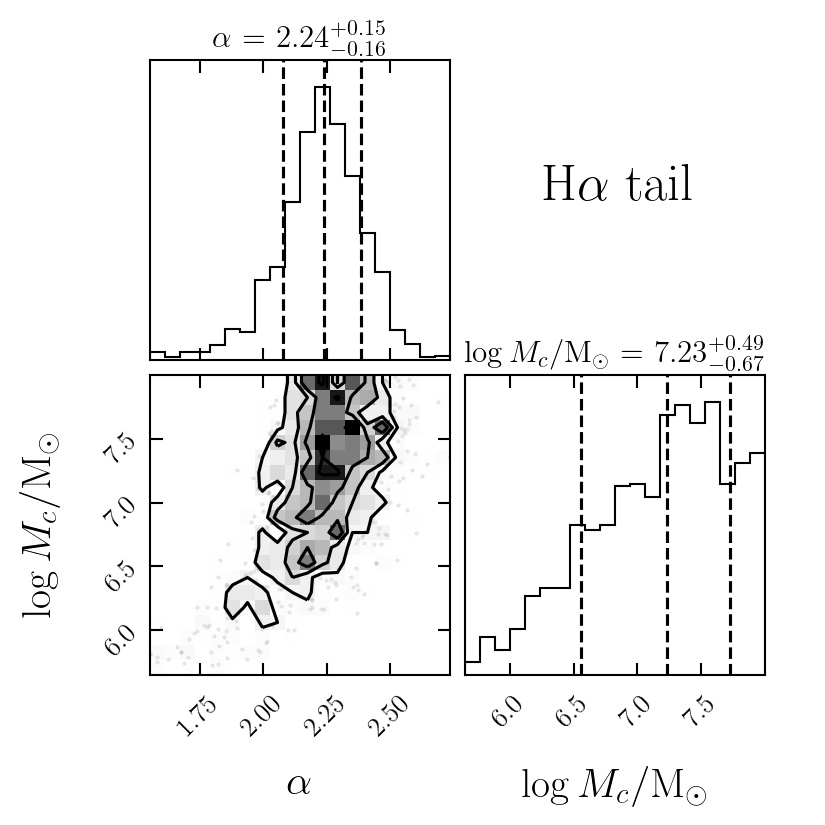}\hspace{-0.3cm}
\includegraphics[width=0.47\textwidth]{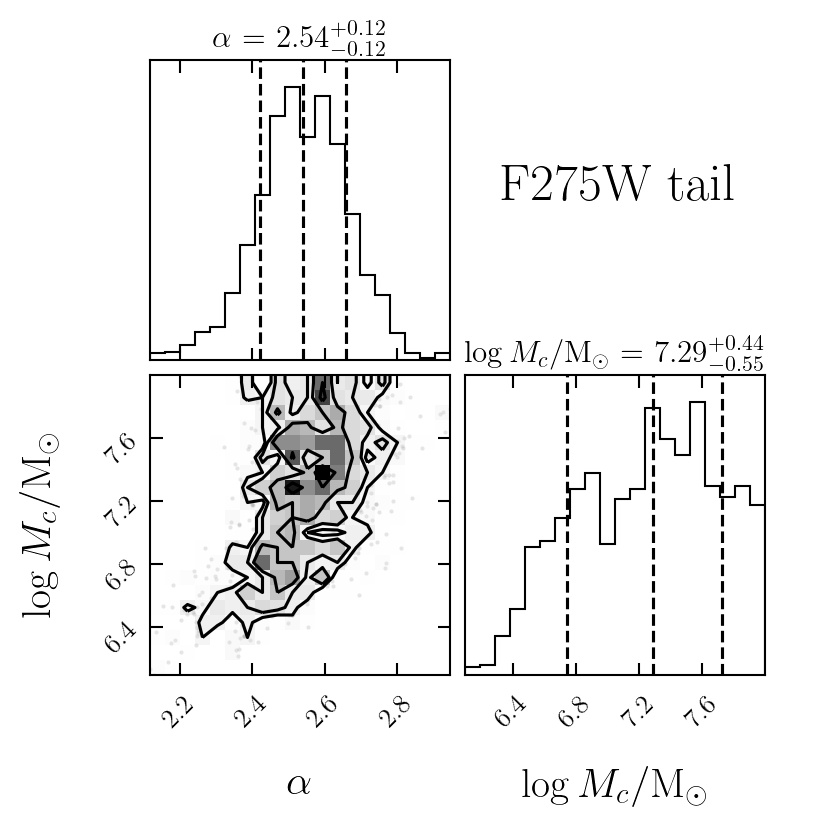}\vspace{-0.2cm}\\
\includegraphics[width=0.47\textwidth]{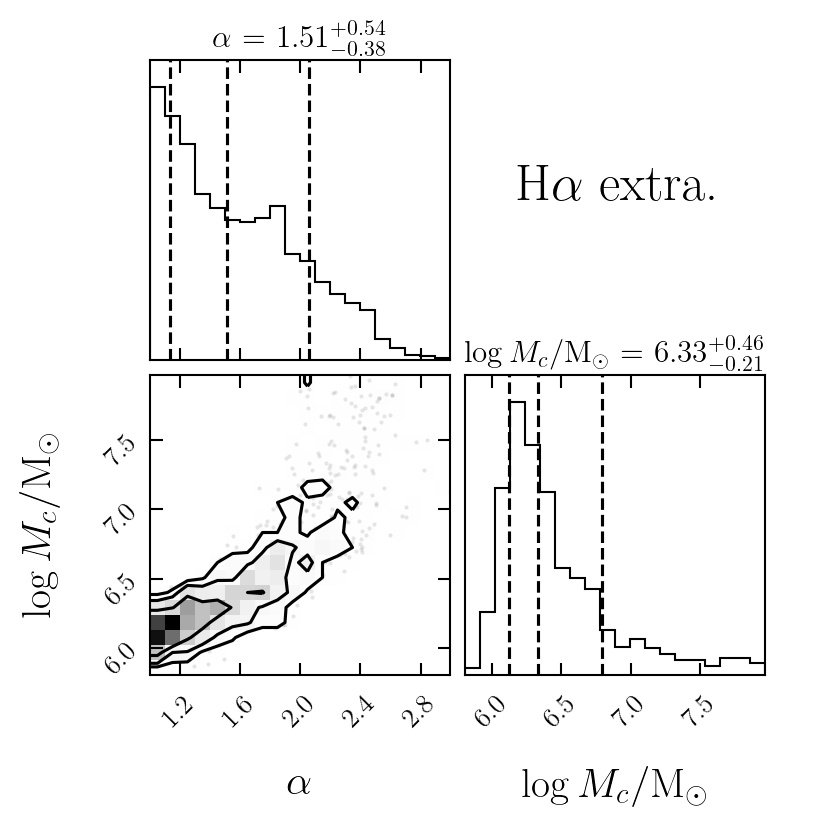}\hspace{-0.3cm}
\includegraphics[width=0.47\textwidth]{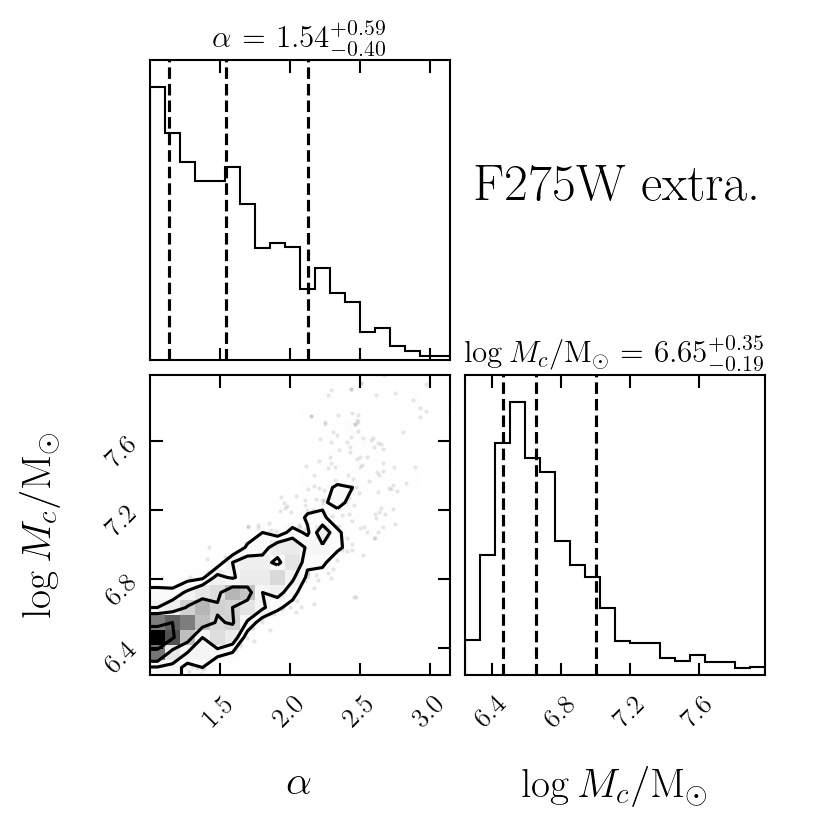}
\caption{Corner plots of the parameter space $(\alpha,\log M_c)$ (the slope and the cut-off mass, respectively), obtained fitting a Schechter mass function (Eq. \ref{eq:schechter}) to the H$\alpha$ and UV tail (top row) and extraplanar (bottom row) clumps mass distribution. Values on top of each histogram are the median of each distribution for the corresponding parameter, with uncertainties given by the 84th percentile minus the median and the median minus the 16th percentile (upper and lower uncertainty, respectively). The same values are also reported as black dashed vertical lines in the histograms.}
\label{fig:schechter_corners}
\end{figure*}

The parameter space is explored using the same procedure described in Sect. \ref{sec:mf_fit} and just changing the functional form of the mass function. The results are shown in corner plots of Fig. \ref{fig:schechter_corners} for the tail (top row) and the extraplanar (bottom row) clumps (either H$\alpha$ and UV).

In the tails the cut-off mass is unconstrained and peaks at masses larger than $10^7\,\msun$, while the best-fitting MF slopes are $2.24^{+0.15}_{-0.16}$ and $2.54\pm 0.12$ for H$\alpha$ and UV clumps, respectively. Both slopes are consistent with the values obtained when fitting a single power law (Sect. \ref{sec:pl_test}), even though systematically smaller. On the other hand, in the extraplanar region both the slopes of the H$\alpha$ and UV clumps have unconstrained results, for which we can only set an upper limit $\alpha \lesssim 2$, while the cut-off masses are constrained with $\log M_c/\msun=6.33^{+0.46}_{-0.21}$ and $\log M_c/\msun=6.65^{+0.35}_{-0.19}$ for the H$\alpha$ and the UV samples, respectively.
By comparing the shape of the contours and the constraints on the parameters, for the tail clumps (either H$\alpha$ or UV) we can rule out the presence of a cut-off mass, for which only a lower limit can be set and at values larger than the most massive clumps of the sample. That suggests that the most correct shape for the tail mass function is a simple power law.

For what concerns the extraplanar regions, the fit suggests the presence of a cut-off, which is well constrained and at values comparable with the fitted mass interval. However, no constraints can be put on the slope, for which the fit sets an upper limit. Furthermore, we performed the Bayesian information criterion (BIC) test \citep{Wit2012} to quantify whether the addition of a cut-off mass concretely improves the goodness of the fit. The BIC of a model is defined as \citep{Wit2012}

\begin{equation}\label{eq:bic}
    \mathrm{BIC}=k\ln(n)-2\mathcal{L},
\end{equation}

\noindent where $k$ is the number of free parameters of the model, $n$ is the number of data points and $\mathcal{L}$ is the likelihood of the model. In our case, the BIC differences $\Delta \mathrm{BIC}$ between the power law and the Schechter cases are always smaller than $2$, therefore there is no strong evidence that the Schechter function is a better model than the power law.

The difficulty in constraining on the parameters of the extraplanar clumps is most likely due to a series of caveats which become negligible when moving to the tails. The underlying old stellar disk and the large blending rate (even in the outskirts of disks) make it not trivial to correctly estimate the stellar mass of the clumps, which are characterized by larger uncertainties and a poor completeness. Both these effects have been extensively discussed and quantified in the previous sections (Sects. \ref{sec:in_out_mass} and \ref{sec:mass_completeness}), but deeper observations would be needed to improve the fit in the extraplanar region by extending the range of fitted masses, including also the power-law regime.
Considering the BIC test ruled out the need of the Schechter function, we presented in the previous section the values of the slopes obtained from the fit to a power law. Still, the resulting slopes for extraplanar clumps are susceptible to the fitted mass range and to the aforementioned observational caveats.

\section{Discussion and conclusion}\label{sec:conclusions}
Throughout this paper we have shown our results for what concerns the mass distribution functions of a sample of star-forming clumps in a set of six galaxies undergoing strong ram-pressure stripping.
The clumps are selected both in H$\alpha$ and UV, and divided in two spatial categories: tail, if observed far away from the galactic plane, and extraplanar, if the clumps is in the outskirts of the galactic disk but already showing the signs of ram-pressure stripping.

The fit of the mass functions of these sub-samples of clumps has been performed by means of a Bayesian algorithm, which can let us fit the distribution in a way that is binning-independent. Modelling our clumps and performing a set of mock observations, we have also quantified the mass completeness and bias of the sub-samples, showing how these observational effects can highly affect the steepness of the observed mass function. Not taking into account these effects tend to lead to shallower mass distributions, bringing to completely different conclusions and interpretations.
Therefore we highly improved our results with respect to our previous study about the luminosity function of these objects \citep{Giunchi2023a}, finding that:

\begin{enumerate}

\item the mass function of the tail clumps is better described by a power law, rather than a Schechter function. Having ruled out the presence of a cut-off mass, our slopes are steeper than the value $\alpha=2$ typically obtained from observations \citep{Krumholz2019} and simulations \citep{Renaud2024} of main-sequence, isolated galaxies, and equal to $2.31\pm 0.12$ in H$\alpha$ and $2.60\pm 0.09$ in UV.
This difference is remarkable: in \cite{Giunchi2023a} it was shown that clumps in jellyfish galaxies are likely to be experiencing an enhancement in star-formation rate surface density ($\Sigma_\mathrm{SFR}$), similarly to star-forming clumps in starburst galaxies \citep{Fisher2017}, as hinted by the observed enhancement in H$\alpha$ luminosity. Previous works have shown that the mass function of young clumps in such environments (including also spiral arms, for instance) is shallower than in main-sequence isolated galaxies or in the inter-arm regions of spiral galaxies and therefore the formation of massive clumps is favoured.
Nonetheless, our sample of clumps, although enhanced in $\Sigma_\mathrm{SFR}$, show large slopes, usually associated to regions with mild star formation, like the inter-arm regions of spiral-armed galaxies \citep{Haas2008,Messa2018b,Messa2019};

\item in the extraplanar region the resulting slopes are $2.45^{+0.20}_{-0.16}$ in H$\alpha$ and $2.63^{+0.20}_{-0.18}$ in UV. The extraplanar H$\alpha$ mass function is steeper than the tail one. Furthermore, similarly to the tail region, also for the extraplanar clumps the UV mass function is steeper than the H$\alpha$ one. However, the uncertainties on the extraplanar slopes are such that they are all consistent within $1\sigma$, making it difficult to drive strong conclusions.
Further tests on the clumps of this region suggest that a cut-off mass, which could bias the inferred value of the slope, is likely to the present, even if the depth of the sample and the observational caveats in this regions makes it difficult to put strong constraints on both parameters and statistical tests rule out the need of a second parameter.
\end{enumerate}

The unpredicted steepening of our mass function has never been observed in galaxies with enhanced $\Sigma_\mathrm{SFR}$ \citep{Livermore2012} and not even in the tidal tails of merging galaxies \citep{Rodruck2023}, therefore it must be tightly connected to the peculiar environment experienced by the clumps in the tails of these jellyfish galaxies, far away from the galactic disk and embedded in the hot and high-pressure ICM. In this context, recent works by \cite{Li2023} and \cite{Ignesti2024} studied the mixing between stripped ISM and the surrounding ICM by means of the velocity structure function (VSF), which is a two-point correlation function that quantifies the kinetic energy fluctuations as a function of the scale l in a velocity field. The former focused on the RPS galaxy ESO $137-001$, the latter studied ten galaxies from the GASP survey, extending the study of the ICM turbulence to galaxies of different masses and RPS stage and including also five of the six galaxies of this work (with the only exception of JO201).
In both works, the VSFs suggest that the mixing of these two gas phases triggers the Kelvin–Helmholtz instability that induces turbulent motions down to scale as small as $0.2$ kpc, regardless of the studied galaxies.
The presence of turbulent motions on scales larger than 1 kpc implies that the collapse of large and massive clumps is hindered by the contribution of turbulence to the gas internal pressure. Furthermore, \cite{Li2023} have found that turbulent motions triggered by the ICM increase with the distance from the disk. Therefore the formation of massive clumps is even more halted in the tails, explaining why the most massive clumps of our samples are located in the extraplanar region (Sect. \ref{sec:method_mass} and Fig. \ref{fig:mass_age_distr}).

It is worth stressing that the interaction between ISM and ICM involves also other processes \citep{Ge2021} not addressed by the afore-mentioned studies, like thermal heating (because of mixing, thermal conduction and radiative heating, \citealt{Campitiello2021}) and magnetic fields, which can either prevent or favour the gas collapse \citep{Mueller2021,Mueller2021b}. These phenomena can affect the gas collapse and the properties of the clumps, however their net effects is still debated due to the complexity of the processes at play.

Our results suggest that star formation is highly sensitive to the environment in which it occurs. Increasing the available statistics of highly resolved clumps, in combination with multi-wavelength observations focusing on the gaseous components, would improve our understanding of the star-formation process.

\begin{acknowledgements}
    We thank the anonymous referee for their useful comments that helped to improve this script.
    EG would like to thank the GASP team, Raffaele Pascale, Carlo Nipoti and Alessandro Della Croce for the useful discussions and comments.
    The research activities described in this paper have been co-funded by the European Union – NextGenerationEU within PRIN 2022 project n.20229YBSAN – Globular clusters in cosmological simulations and in lensed fields: from their birth to the present epoch.
    This paper is based on observations made with the NASA/ESA Hubble Space Telescope obtained from the Space Telescope Science Institute, which is operated by the Association of Universities for Research in Astronomy, Inc., under NASA contract NAS 5-26555. These observations are under the programme GO-16223. All the \textit{HST} data used in this paper can be found in MAST: \href{http://dx.doi.org/10.17909/tms2-9250}{10.17909/tms2-9250}.
    This paper used also observations collected at the European Organization for Astronomical Research in the Southern Hemisphere associated with the ESO programme 196.B-0578.
    This research made use of Astropy, a community developed core Python package for Astronomy by the Astropy Collaboration (\citeyear{Astropy2018}). This project has received funding from the European Research Council (ERC) under the European Union’s Horizon 2020 research and innovation programme (grant agreement No. 833824).
    We acknowledge support from the INAF Minigrant "Clumps at cosmological distance: revealing their formation, nature, and evolution" (Ob. Fu. 1.05.23.04.01).
\end{acknowledgements}

\bibliography{biblio}{}
\bibliographystyle{aa}

\begin{appendix}

\onecolumn
\section{Characterization of the clumps morphology}\label{sec:stacking}

In order to generate a set of mock clumps, we need to characterize the typical morphology of H$\alpha$ and UV clumps, respectively.
Therefore we visually selected a set of 7(13) H$\alpha$-(UV-)resolved clumps, with $\mathrm{SNR}>2$ in all five filters, isolated and with smooth and symmetric surface brightness profiles. For each selected clump we defined the largest possible square region including the clump as the only source.
Then each clump undergoes the following steps:

\begin{enumerate}
    \item{the local background is fitted with a 2D polynomial function of 2nd degree and subtracted to the image. In order to exclude the clump, we masked out a region as large as the total area of the clump defined by \textsc{Astrodendro} \citep{Giunchi2023a} dilated by $3$ pixels using the \textsc{scipy} function \textsc{binary\_dilation}\footnote{\url{https://docs.scipy.org/doc/scipy/reference/generated/scipy.ndimage.binary_dilation.html}}.}
    \item{The image is rotated to align the major axis of the clump to the $x-$axis and the minor axis to the $y-$axis \citep{Giunchi2023b}.}
    \item{The physical scale of the image is normalized for the major axis along the $x-$axis and the minor axis along the $y-$axis. Since the size scales with the luminosity \citep{Giunchi2023a} this normalization let us compare clumps of different sizes and axial ratios, also stretching the image to make the clumps as round as possible.}
    \item{The image is converted from flux to surface brightness and resampled with smaller-size pixels.}
    \item{The image is normalized for an effective surface brightness $\Sigma_\mathrm{eff}=F_\mathrm{bag}/\mathrm{A_{exact}}$, where $F_\mathrm{bag}$ is the flux derived from the \textsc{Bagpipes} best-fitting model of the clump\footnote{We chose to use \textsc{Bagpipes} fluxes, instead of the observed ones, as this choice let us better correlate the SED with the best-fitting mass and age of the clump, which are derived by \textsc{Bagpipes} itself.} and $\mathrm{A_{exact}}$ is the total area of the clump defined by \textsc{Astrodendro} (prior to the pixel dilation). Also this normalization is performed in order to make clumps of different luminosities comparable.}
\end{enumerate}

\begin{figure}[t!]
\centering
\includegraphics[width=0.49\textwidth]{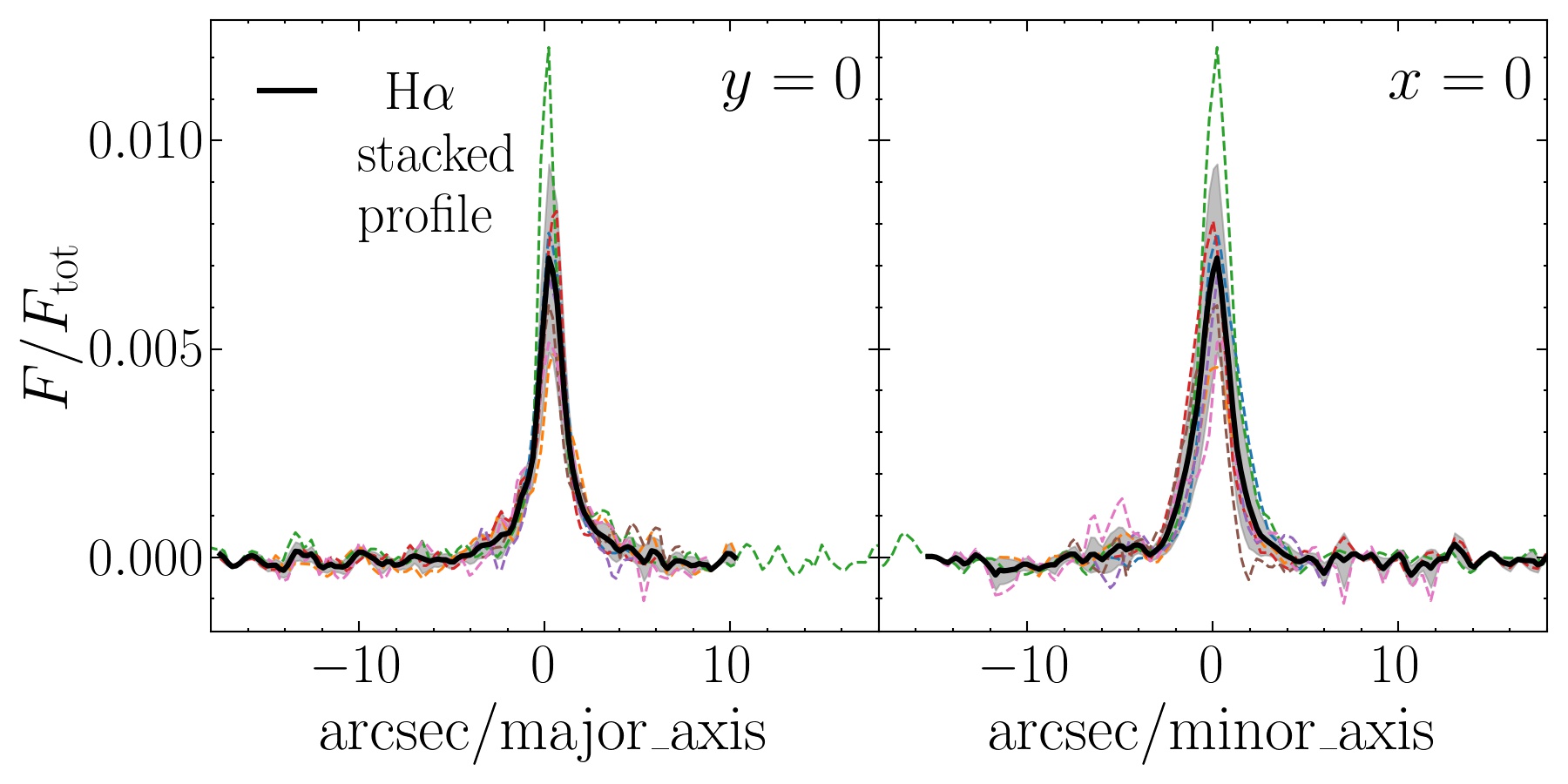}
\includegraphics[width=0.49\textwidth]{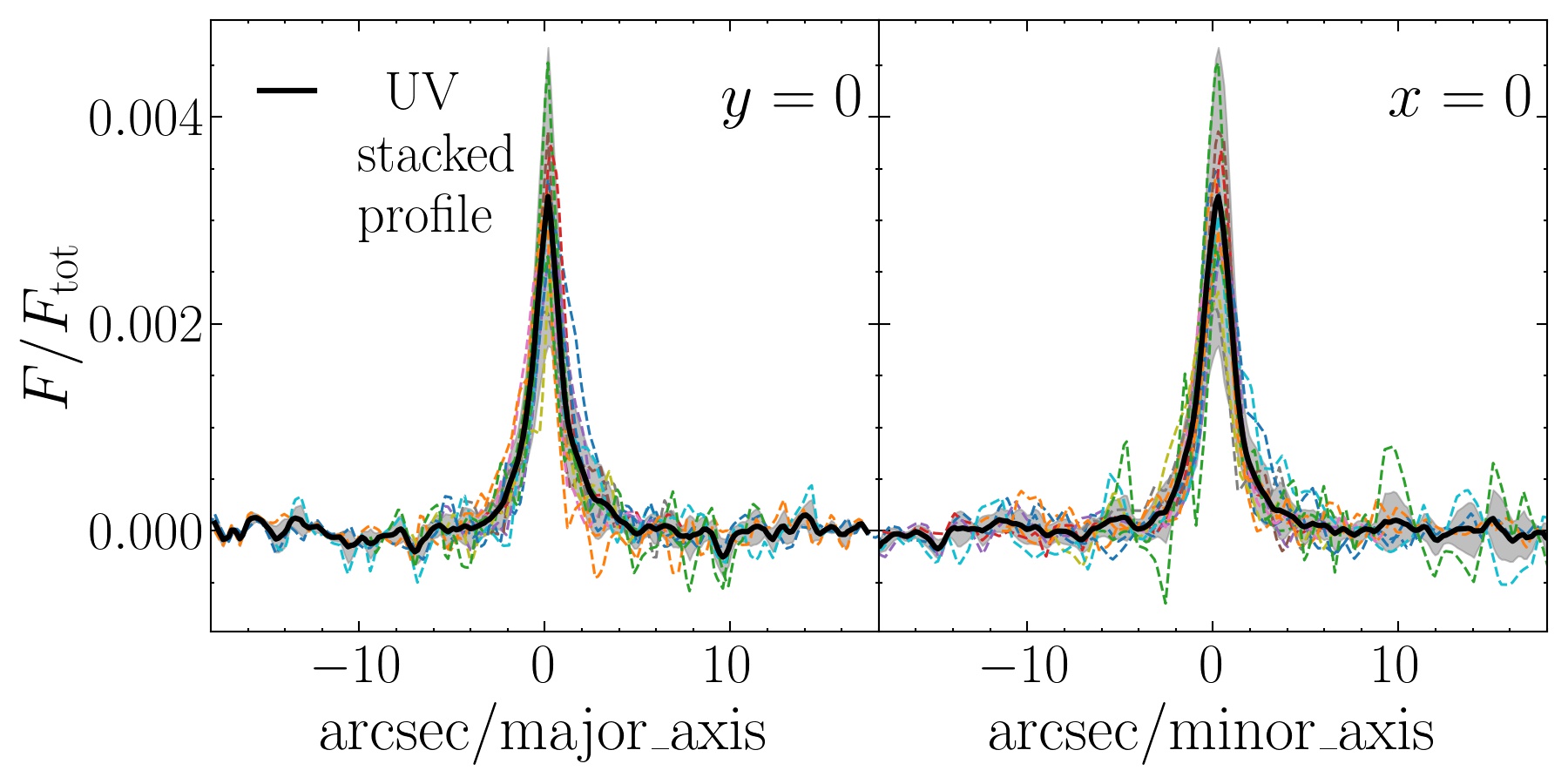}
\caption{Normalized fluxes at $y=0$ (left panels) and $x=0$ (right panels) axis of the H$\alpha$-resolved (top row) and UV-resolved (bottom row) clumps selected for the stacking (dashed lines of varying colours) and the corresponding stacked clump (black solid line). The gray shaded area is the $1\sigma$ uncertainty of the stacked profile. The axes are normalized for scale quantities as described in Sect. \ref{sec:stacking}.}
\label{fig:clump_stack}
\end{figure}

In this way, the rotated and normalized clumps are put in the same reference system, centred at the coordinates of the brightest pixel, and the stacked image is computed as the mean of all the images (where at least two images overlap). In Fig. \ref{fig:clump_stack} we show the stacked normalized flux at $y=0$ and $x=0$ both in H$\alpha$ and UV, compared to the stacked observed clumps. Even though a few outliers are present, the agreement is pretty good and the stacked profile described quite well the average behaviour of the chosen clumps.

The stacked image is then fitted with \textsc{galfit}\footnote{\url{https://users.obs.carnegiescience.edu/peng/work/galfit/galfit.html}}, a tool fitting simultaneously one or more sources in an image adopting a large variety of models, adopting as a model the sum of two 2D Moffat functions\footnote{We compared the two-2D Moffat model with others, including a single 2D Moffat, two 2D Gaussians, Moffat+Gaussian and a single Gaussian, verifying that the combination of two Moffat functions is the best choice to reproduce the stacked clump.}, convolved for the WFC3 PSF of the proper filter (either F680N or F275W), available from \textsc{tinytim} \citep{Krist2011}.
The surface brightness $\Sigma$ of a 2D Moffat function is described by

\begin{equation}\label{eq:moffat}
    \begin{cases}
    \begin{aligned}
        \Sigma (r)&=\Sigma_0\left[1+\left(\cfrac{r}{r_d}\right)^2\right]^{-\gamma_\mathrm{M}}\\
        \Sigma_0&=\cfrac{F_\mathrm{M}(\gamma_\mathrm{M}-1)}{\pi r_d^2}\\
        r_d&=\cfrac{\mathrm{FWHM_\mathrm{M}}}{2\sqrt{2^{1/\gamma_\mathrm{M}}-1}}
    \end{aligned}
    \end{cases}
\end{equation}

\noindent where $\Sigma_0$ and $r_d$ are the scale surface brightness and radius, respectively, defined by the total flux of the Moffat $F_\mathrm{M}$, the full-width at half-maximum $\mathrm{FWHM_{M}}$ and the concentration index $\gamma_\mathrm{M}$.
Now, \textsc{galfit} is not able to fit normalized images in non-physical units, therefore we scaled the stacked image to the average flux and radius of the clumps selected to build the model. This choice will let us correctly evaluate the effects of the PSF on the shape of the clumps, which would be too strongly influenced by it (in case of clumps with the smallest flux and radius) or completely unaffected (in case of clumps with the largest flux and radius).

\begin{table}[t!]
\fontsize{10pt}{10pt}\selectfont
\setlength{\tabcolsep}{4pt}
\renewcommand{\arraystretch}{1.4} 
\centering
\caption{Best-fitting parameters of the two 2D-Moffat functions used to model the stacked clumps.}
\begin{tabular}{ll|ccc}
\bottomrule
\bottomrule
Filter & Function & $F_\mathrm{M}/F_\mathrm{tot}$ & $\mathrm{FWHM_{norm,M}}$ & $\gamma_\mathrm{M}$\\
\hline
\multirow{2}{*}{H$\alpha$} & Moffat $1$ & $0.35\pm 0.16$ & $35\arcsec\pm 28\arcsec$ & $1.39\pm 0.31$\\
& Moffat $2$ & $0.65\pm 0.16$ & $131\arcsec\pm 169\arcsec$ & $10\pm 21$\\\hline
\multirow{2}{*}{UV} & Moffat $1$ & $0.40\pm 0.06$ & $71\arcsec\pm 32\arcsec$ & $1.21\pm 0.22$\\
& Moffat $2$ & $0.60\pm 0.06$ & $175\arcsec\pm 55\arcsec$ & $4.7\pm 2.0$\\
\hline
\end{tabular}
\tablefoot{From left to right: selection filter of the sample stacked to build the model (Filter), the Moffat function (Function), the total flux relatively to the sum of the two Moffat functions ($F_\mathrm{M}/F_\mathrm{tot}$), the full-width at half-maximum normalized as described in Sect. \ref{sec:stacking} ($\mathrm{FWHM_{norm,M}}$), the concentration index ($\gamma_\mathrm{M}$).}
\label{tab:moffat2d}
\end{table}

\begin{figure}[t!]
\centering
\includegraphics[width=0.48\textwidth]{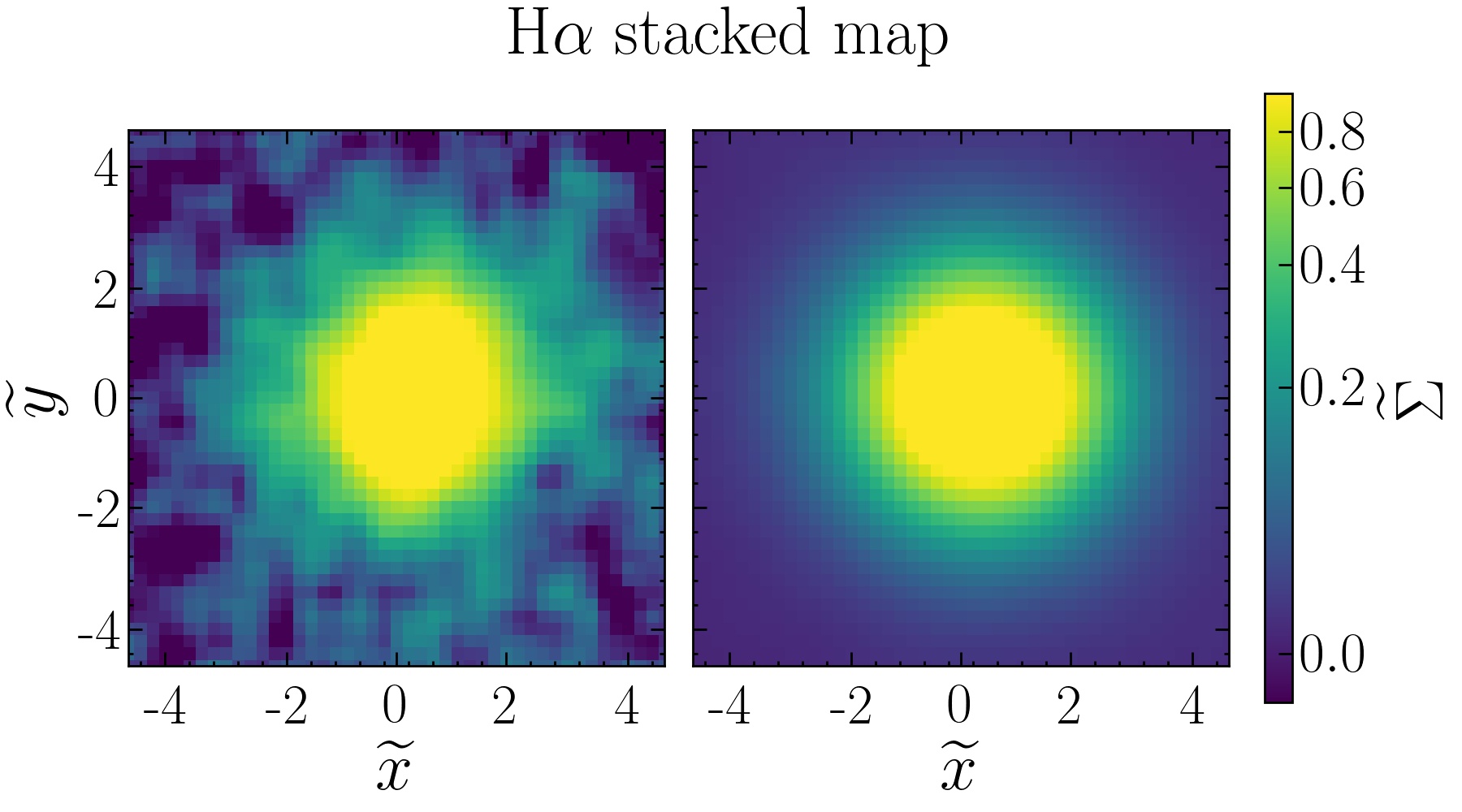}
\includegraphics[width=0.48\textwidth]{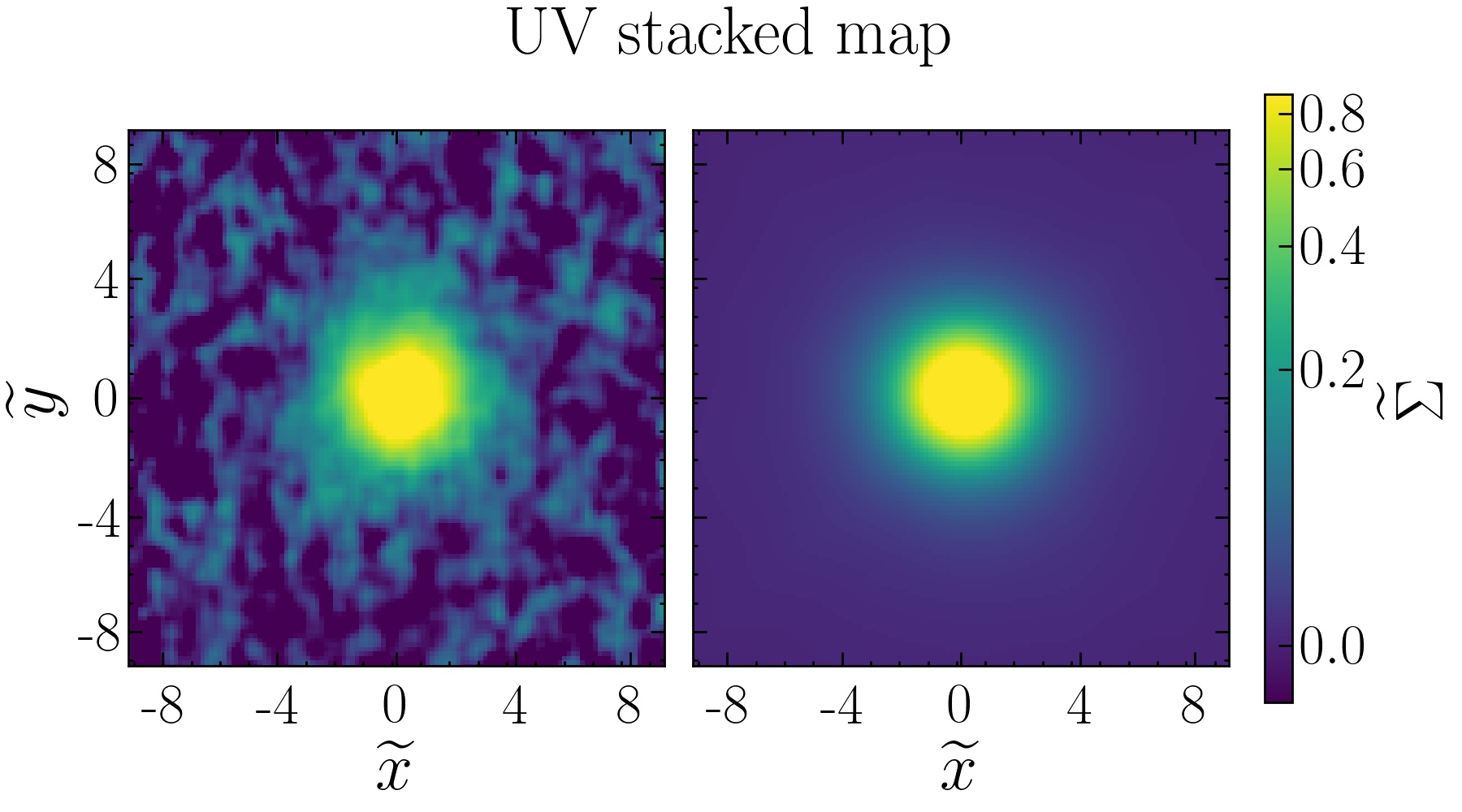}
\caption{Maps of the H$\alpha$ (top panels) and UV (bottom panels) stacked clumps (left panels) and best-fitting models (right panels) obtained with two 2D Moffat (parameters in Table \ref{tab:moffat2d}), residuals. The axes and the fluxes are normalized to scale quantities as described in Sect. \ref{sec:stacking}.}
\label{fig:stack_moffat}
\end{figure}

\begin{figure}[t!]
\centering
\includegraphics[width=0.49\textwidth]{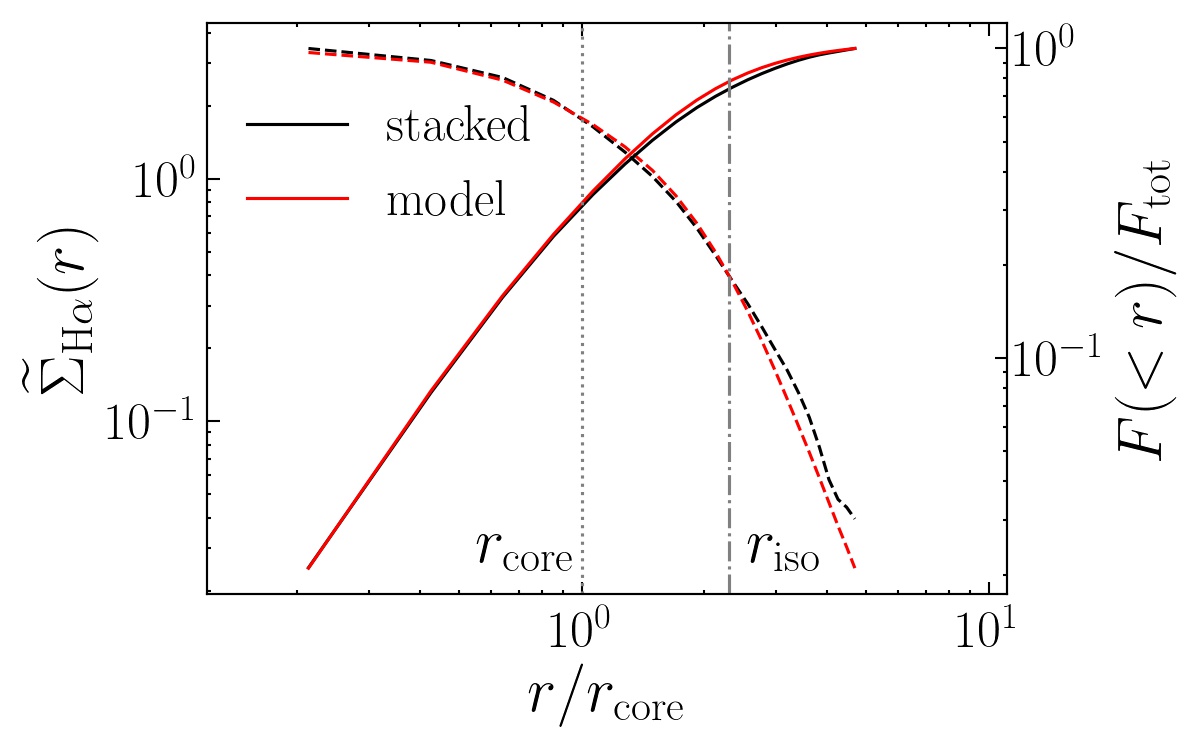}
\includegraphics[width=0.49\textwidth]{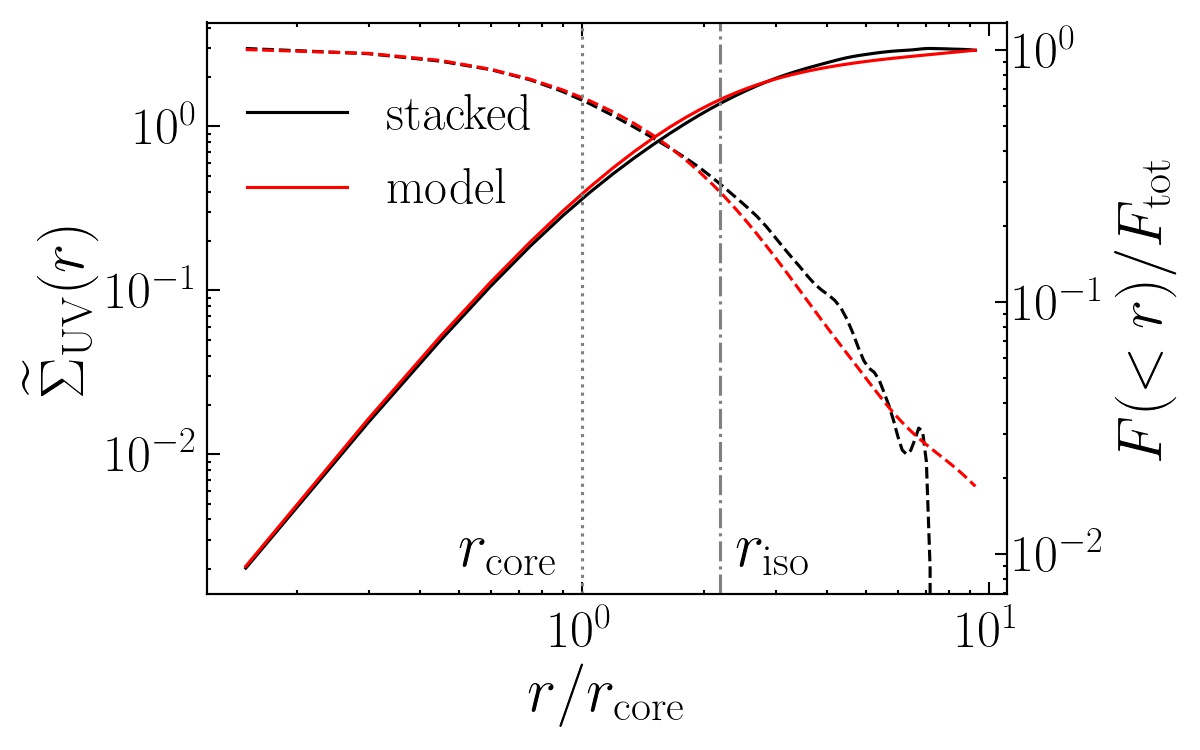}
\caption{Comparison between the stacked profile (black lines) and the best-fitting model (red lines) both in case of H$\alpha$ (top panel) and UV (bottom panel). The dashed lines show the surface brightness (left $y-$scale), the solid lines show the cumulative normalized flux (right $y-$scale). Vertical grey dotted and dash-dotted lines are the core and isophotal radii of the stacked profile.}
\label{fig:stacked_vs_model}
\end{figure}

In Table \ref{tab:moffat2d} we list the best-fitting values of the morphological parameters of the Moffat functions, already re-normalized where needed. The stacked clumps and the best-fitting models are shown both for H$\alpha$ and UV in Fig. \ref{fig:stack_moffat} as 2D maps. In these maps the axes and the flux $\widetilde{F}_\mathrm{stack}$ are re-normalized for the average core radius and total flux of the clumps used for the stacking, respectively, while the model flux $\widetilde{F}_\mathrm{model}$ is normalized for the total flux of the best-fitting model.
In Fig. \ref{fig:stacked_vs_model}, we compare the surface density and the cumulative flux of the stacked clump and the best-fitting model. As one can notice, the profiles are in pretty good agreement within the isophotal radius, which is the constrained interval within which all our clumps are selected and defined, confirming that the models gives good descriptions of the stacked profiles both in case of H$\alpha$ and UV.
Thanks to this procedure, we retrieved the average intrinsic shape of the a typical H$\alpha$ or UV clump, that can be re-scaled to physical units according to our purposes.

\FloatBarrier


\section{Modelling the Spectral Energy Distribution}\label{sec:sed}
The SED of a clump is comprised of many factors, primarily its age and mass. The former mainly drives the shape of the SED (i.e. the relative flux of the photometric bands with respect to each other). Typical stellar populations younger than 10 Myr and 200 Myr are characterized by strong H$\alpha$-line emission and UV continuum, respectively. A clump with age in between 10 and 200 Myr, for example, will still be very bright in the F275W filter, but will have faint or no emission coming from H$\alpha$. If the clump is even older than 200 Myr, then, it will get progressively faint in the blue filters of our set.

Once the shape of the SED has been determined, it must be scaled to physical units by a factor which depends on the mass-to-light ratio, which is a function of the age of the clump of which we want to create a mock image. The physical fluxes are then used to scale the normalized flux of the stacked clump defined in Sect. \ref{sec:stacking}.

In the following Sections, we are going to show the procedures we followed to generate the SED of a mock clump, starting from its shape and then moving to its $M/L$. As already done to define the intrinsic shape of the clumps (Sect. \ref{sec:stacking}), the procedure is followed independently for H$\alpha$ and UV clumps. Furthermore, in this case we have also treated the clumps to different spatial categories (tail, extraplanar, disk) separately.

\subsection{Shape of the Spectral Energy Distribution}\label{sec:shape_sed}
\begin{figure*}[t!]
\includegraphics[width=0.48\textwidth]{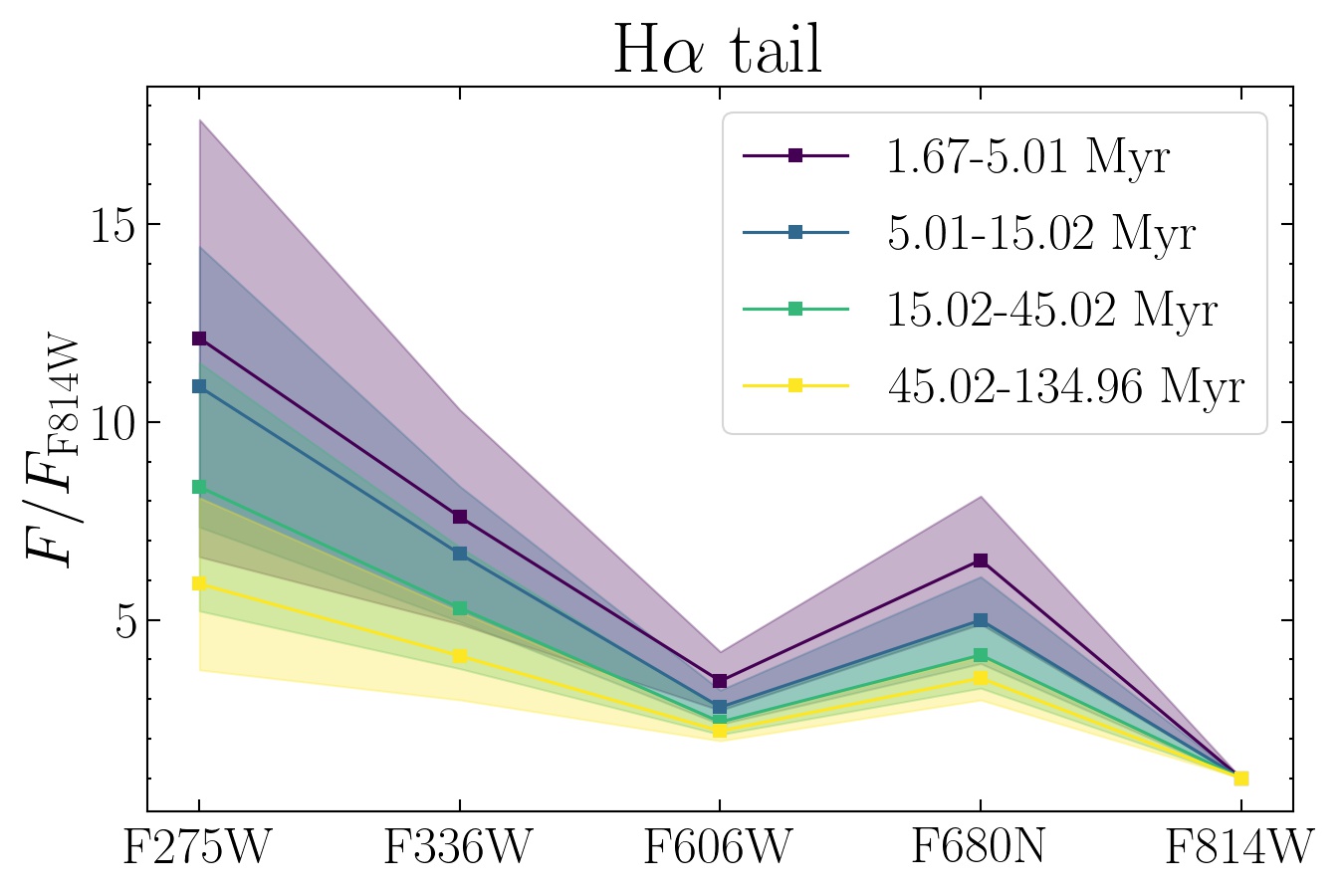}
\includegraphics[width=0.48\textwidth]{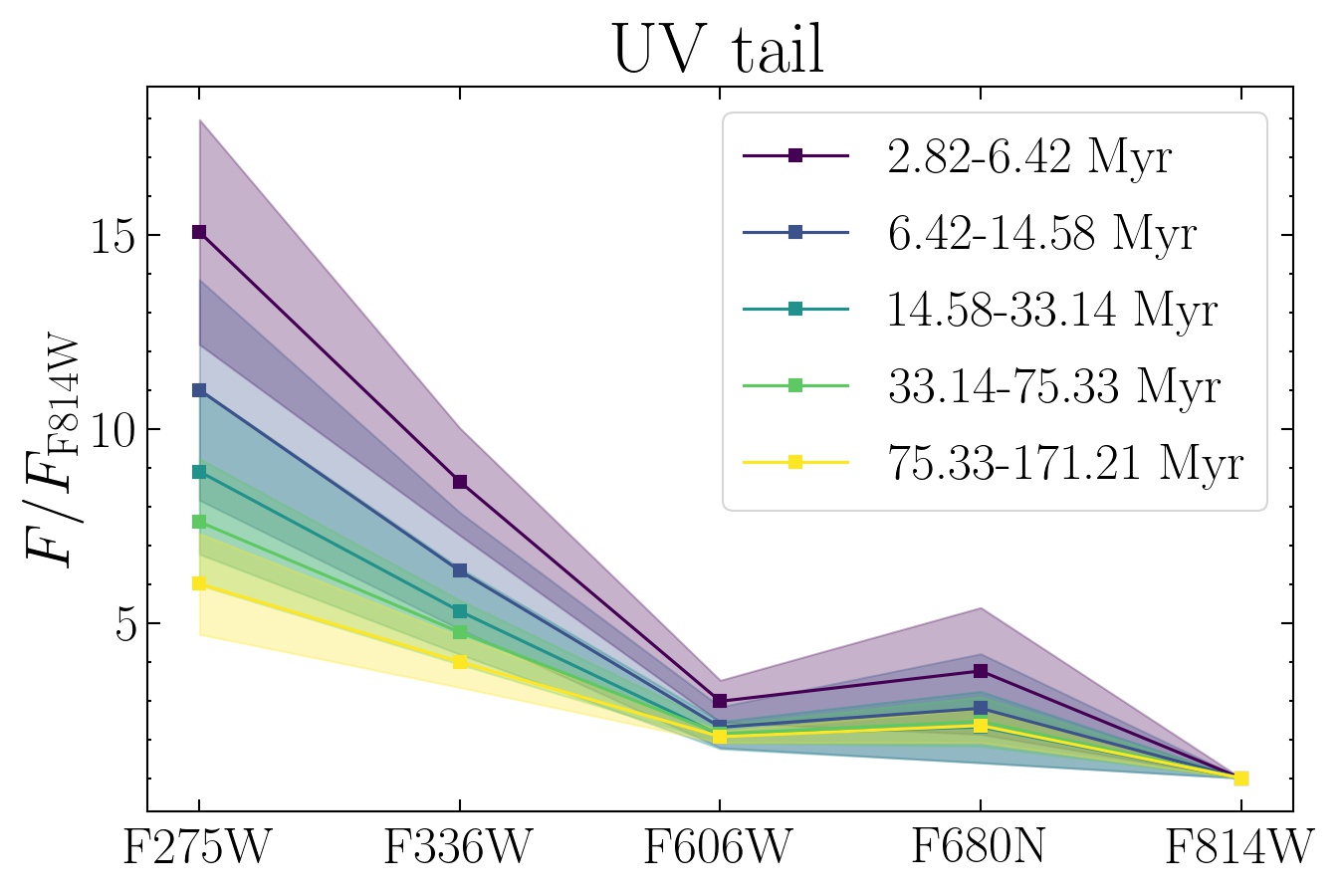}\vspace{-0.2cm}\\
\includegraphics[width=0.48\textwidth]{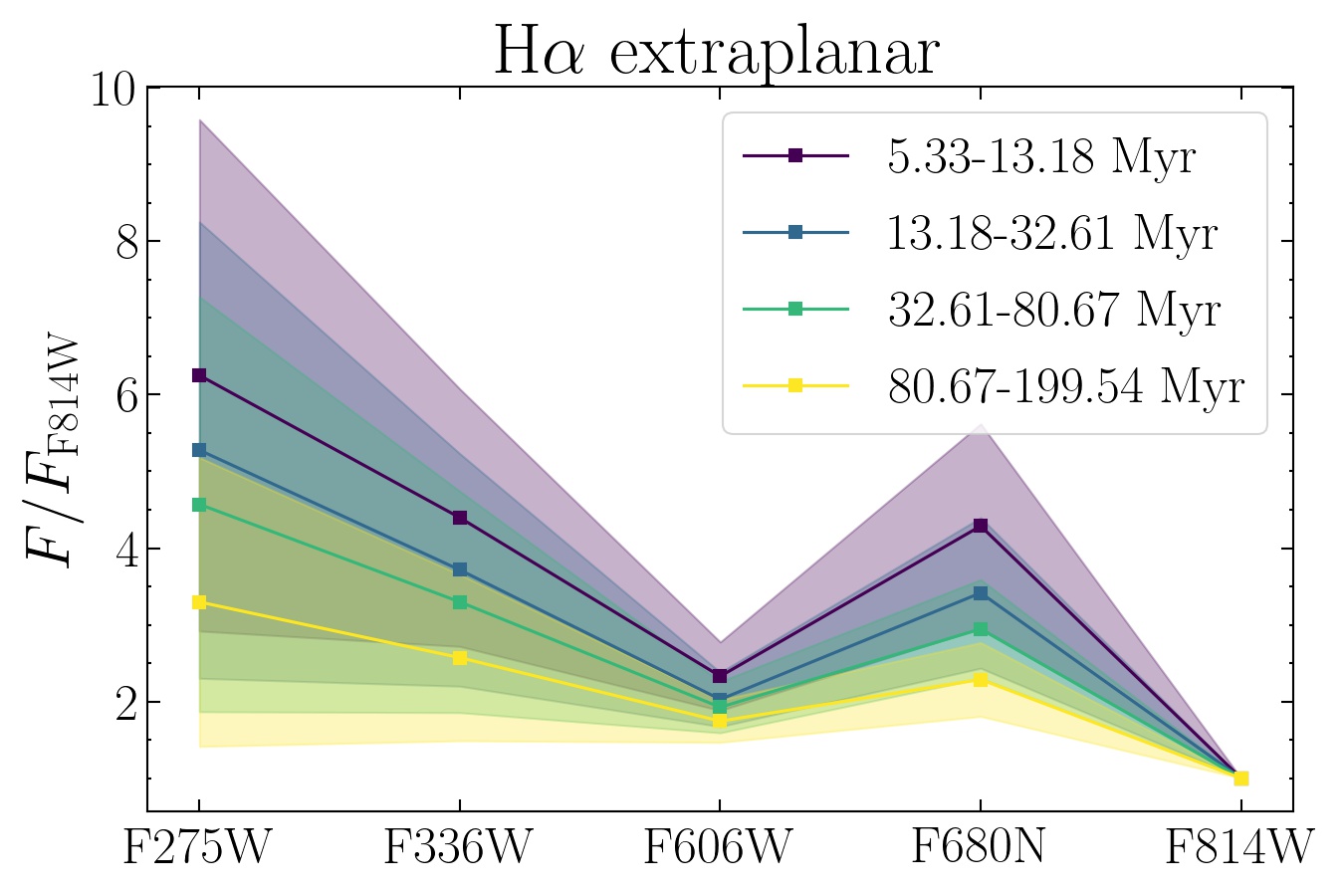}
\includegraphics[width=0.48\textwidth]{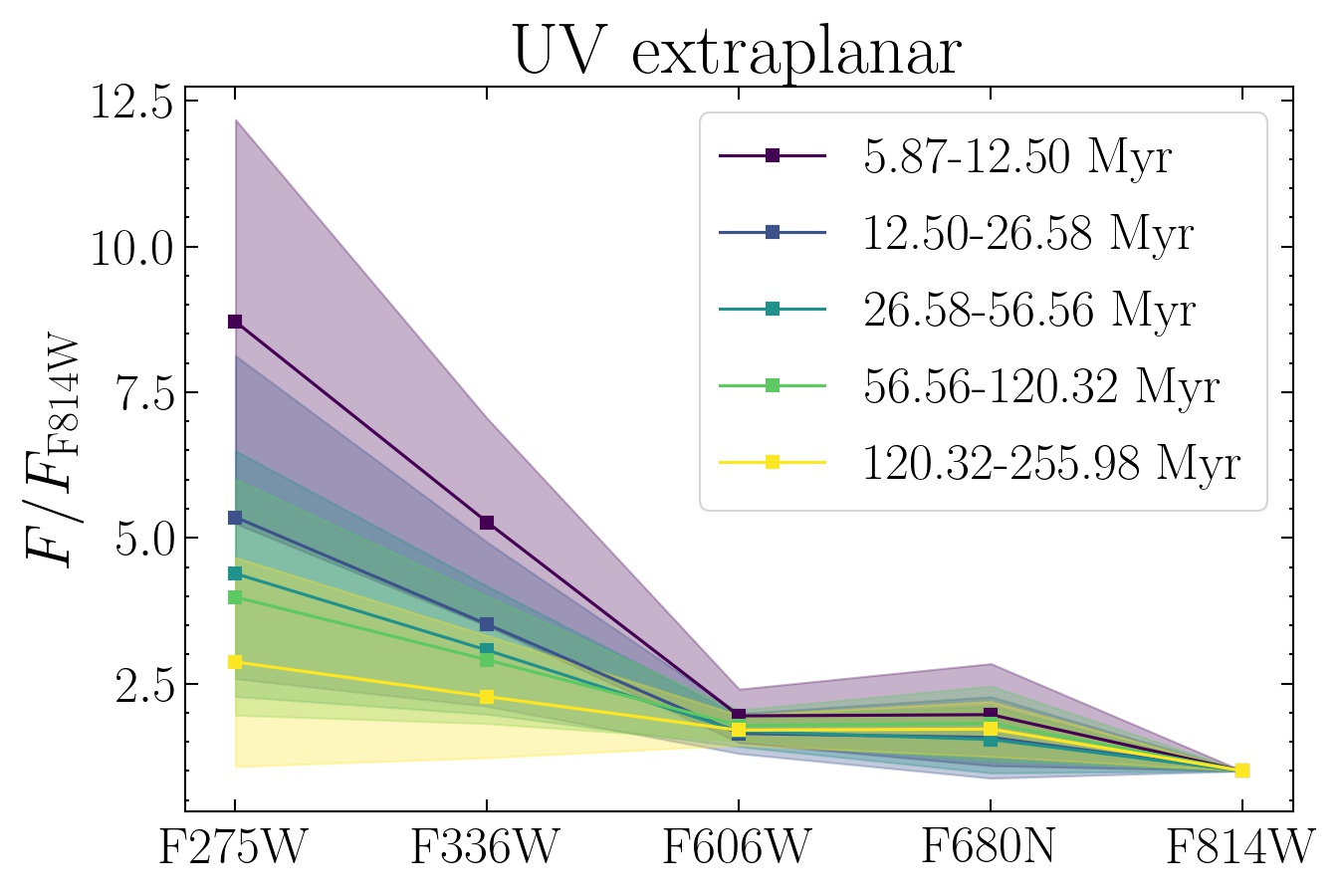}
\caption{SED distribution of tail (top row) and extraplanar (bottom row) clumps, normalized for the flux in the F814W filter. H$\alpha$ clumps are shown in the left panels and UV clumps in the right panels.
The clumps are divided in age bins (from purple to yellow). Solid lines are the median SEDs in each age bins, while the shaded area covers the $1\sigma$ interval. Details about how these normalized SEDs are obtained are given in Sect. \ref{sec:shape_sed}.}
\label{fig:sed_norm}
\end{figure*}

The aim is to generate a library of normalized SEDs of H$\alpha$ and UV clumps in different spatial categories as a function of the stellar mass-weighted age. Therefore, for each selection filter and spatial category, we selected all the leaf clumps for which \textsc{Bagpipes} gives reliable estimates of stellar mass and mass-weighted age (Sect. \ref{sec:method_mass}), and binned them by age in logarithmic scale (either 4 or 5 bins according to available number of clumps, in order to have enough statistics in each bin).
Then the SED of each clump is normalized for the flux in the F814W filter, to make the SEDs of the clumps with different masses comparable.
For each age bin and photometric band, a flux distribution is built starting from the clumps normalized fluxes. This choice lets us assign to each age bin a median normalized flux with a scatter than can well reproduce the variations in the SEDs even among clumps of similar age.
As an example, in Fig. \ref{fig:sed_norm} we show the results of the normalized SEDs for H$\alpha$ and UV tail clumps. As one can notice, the normalization for the F814W flux forces that filter to be anchored at 1, while the shapes of the SEDs follow pretty well the expected behaviour for the H$\alpha$-line emission and the near UV continuum, both getting fainter and fainter with the stellar age.

\subsubsection{Clumps mass-to-light ratios}\label{sec:m_to_l}
\begin{figure*}[t!]
\centering
\includegraphics[width=0.49\textwidth]{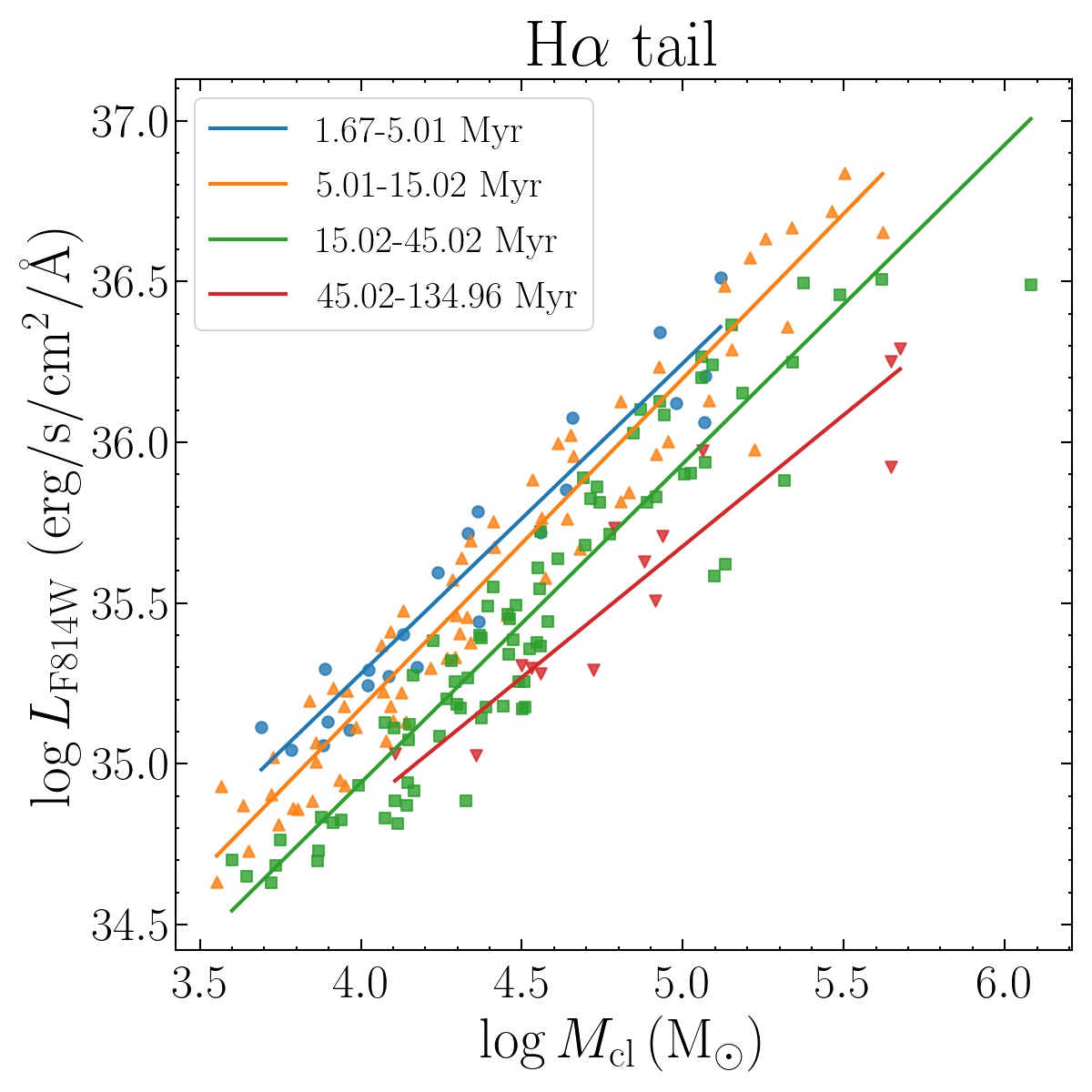}
\includegraphics[width=0.49\textwidth]{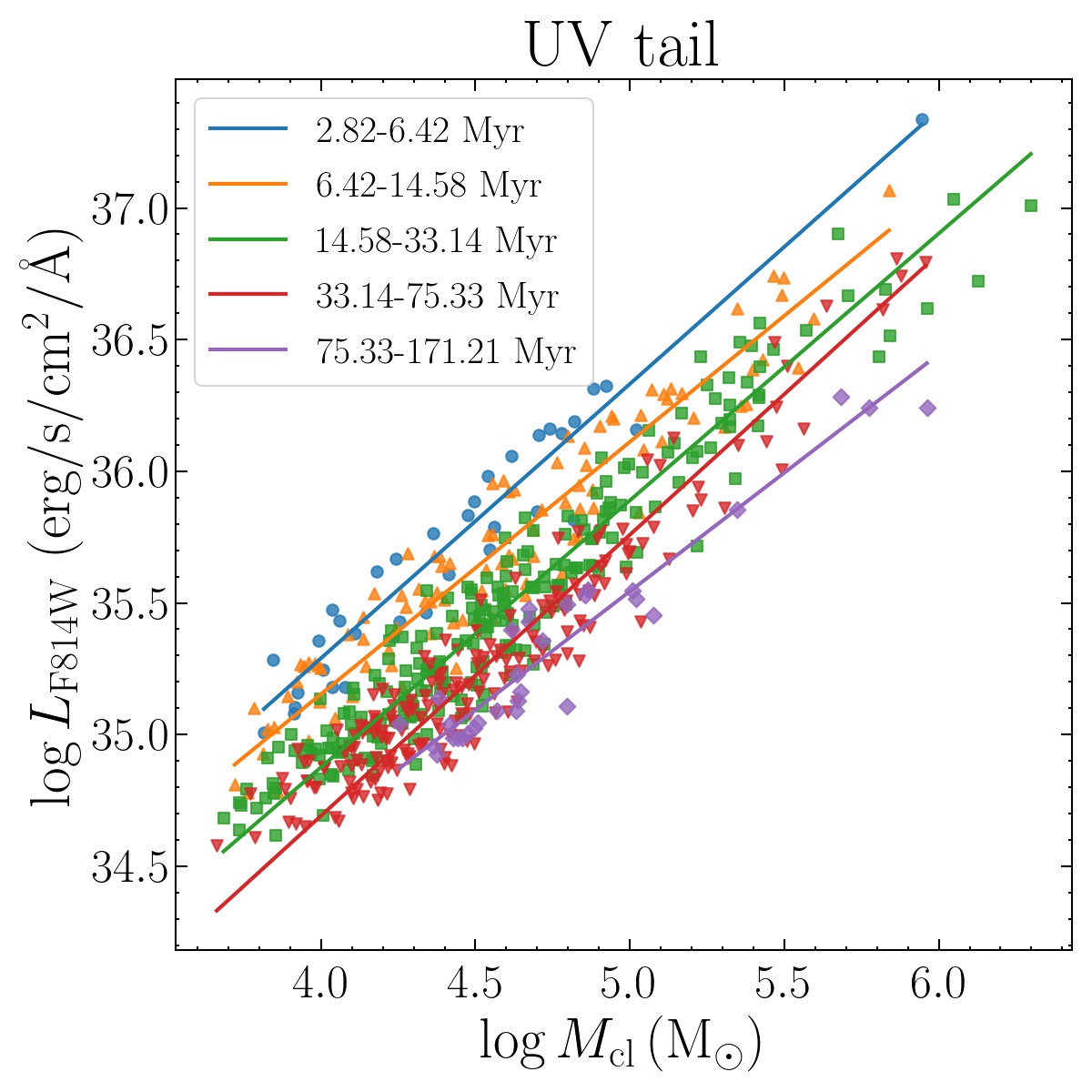}\\
\includegraphics[width=0.49\textwidth]{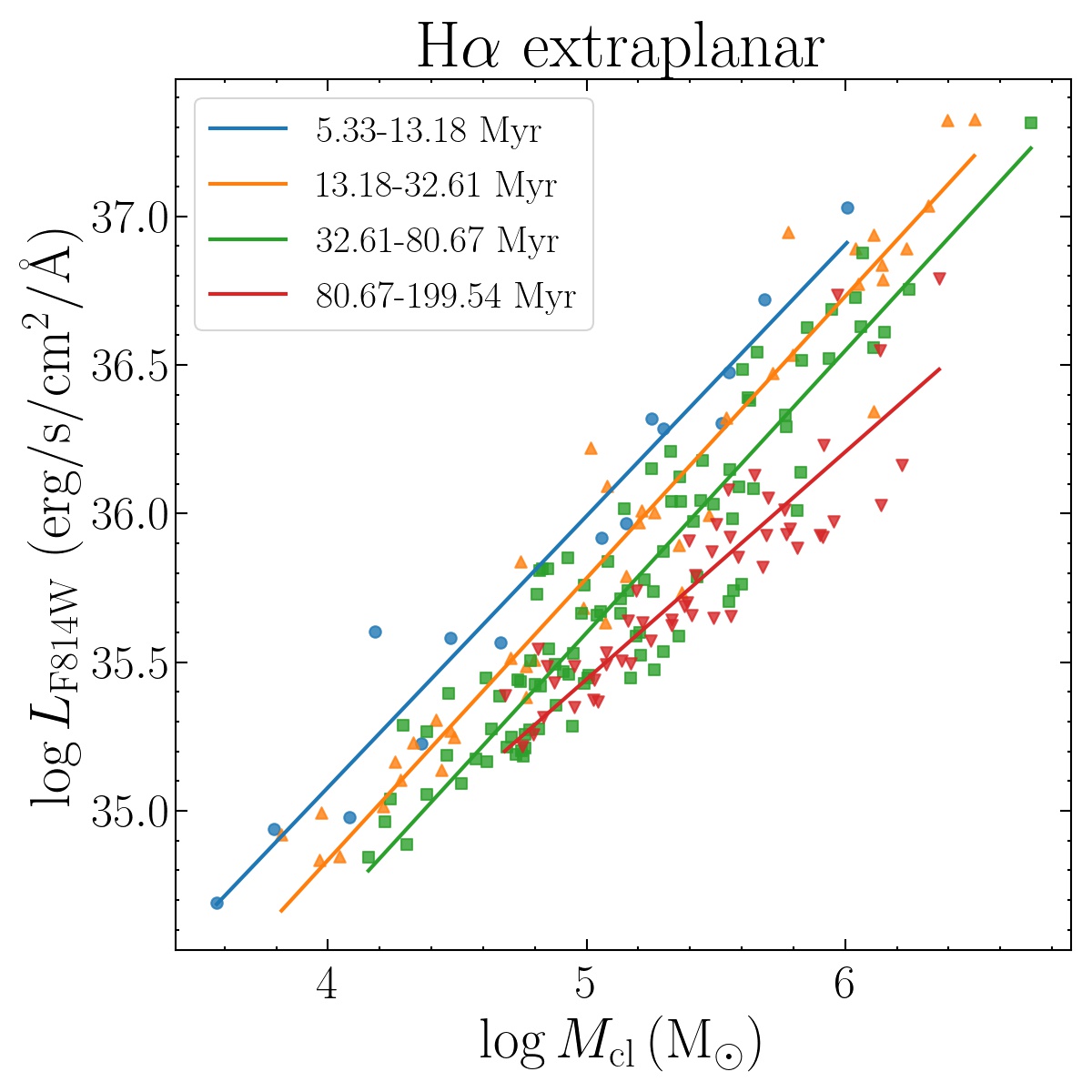}
\includegraphics[width=0.49\textwidth]{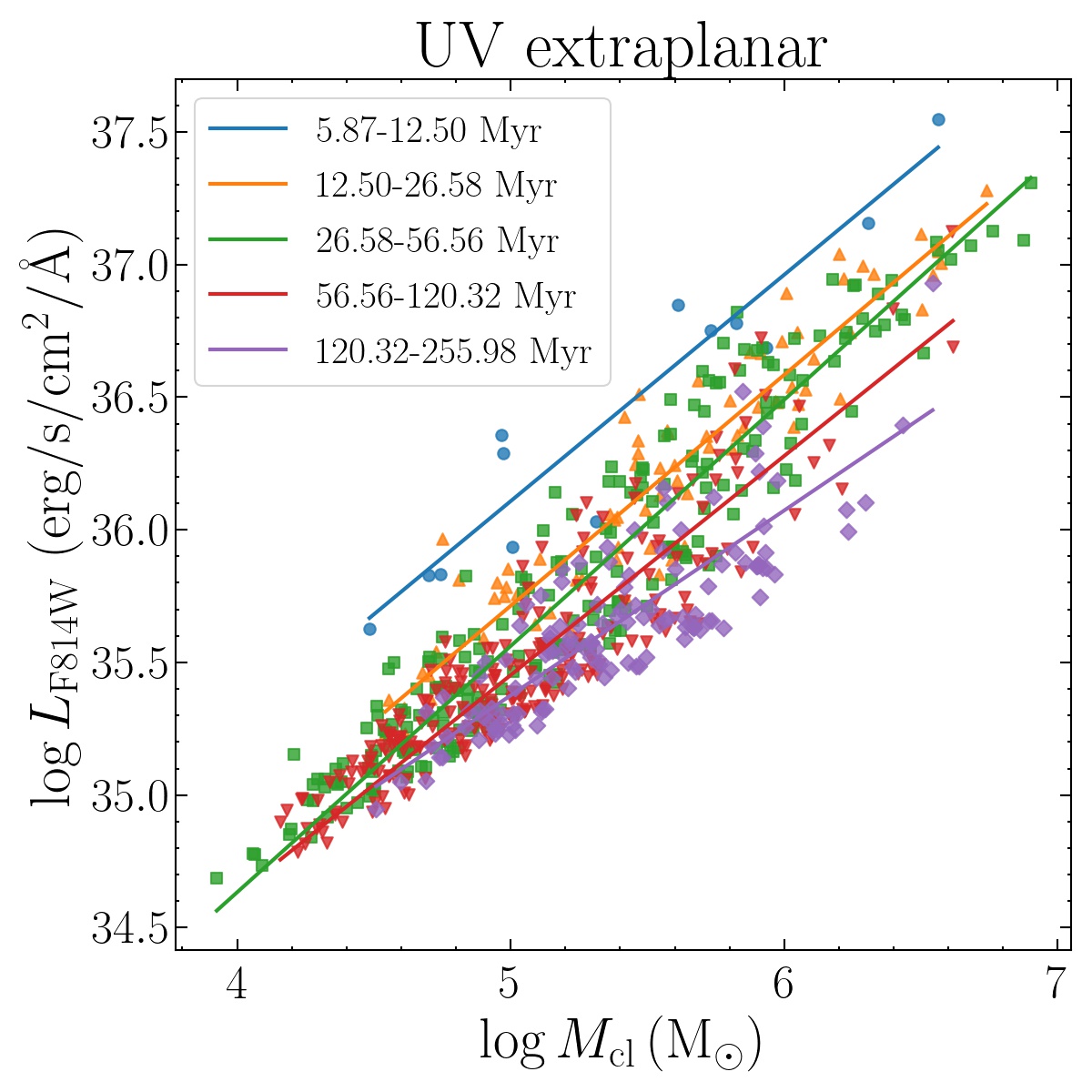}
\caption{Clumps $\log L_\mathrm{F814W}-\log M_\mathrm{cl}$ plane (top row for tail clumps, bottom row for extraplanar clumps, H$\alpha$ clumps in the left panels, UV clumps in the right panels). Clumps are divided in age bins as described in Sect. \ref{sec:shape_sed} and plotted with different colours and shapes (see the legend). For each age bin, a linear fit to the clumps $\log L_\mathrm{F814W}-\log M_\mathrm{cl}$ is performed and plotted as solid line of the corresponding bin.}
\label{fig:m_to_l}
\end{figure*}

We define the mass-to-light ratio as the ratio between the clump stellar mass $M_\mathrm{cl}$ and the F814W luminosity $L_\mathrm{F814W}$. $M_\mathrm{cl}/L_\mathrm{F814W}$ is then computed for each age bin by performing a linear fit in the $\log M_\mathrm{cl}-\log L_\mathrm{F814W}$ plane. The linear fit is performed using the Bayesian fitting tool \textsc{linmix}\footnote{\url{https://linmix.readthedocs.io/en/latest/}} \citep{Kelly2007}, including also the intrinsic scatter to the linear relation. As an example, the results concerning tail clumps are shown in Fig. \ref{fig:m_to_l}, where the H$\alpha$ and UV samples are also divided per age bins. As expected, the F814W at a given mass is larger for young clumps (i.e. small $M_\mathrm{cl}/L_\mathrm{F814W}$).
That is, for a given clump mass and age, it is possible to assign the F814W luminosity, which is used to scale the normalized SED defined by its age (Sect. \ref{sec:shape_sed}) to physical units.

\FloatBarrier

\section{Radius-mass relation}\label{sec:size_mass}
In order to assign a physical radius to the intrinsic profile obtained in Sect. \ref{sec:stacking} for any mock clump of a given mass, we fitted the core radius-mass relation $\rcorecorr-M_\mathrm{cl}$ with the Bayesian orthogonal linear fitting procedure firstly used in \cite{Posti2018} and described in \cite{Bacchini2019}, which lets us introduce an intrinsic scatter in the clumps radius at a given mass. The fit is performed only on the clumps which are spatially resolved, defined as the clumps with $\rcorecorr$ larger than two full-width half maximum of the \textit{HST} PSF \citep{Giunchi2023a}.
The log-likelihood is the same as in Eq. \ref{eq:logL_inout}, where $N_\mathrm{cl}$ is the number of resolved clumps that are fitted, $d_i$ is the distance between a given data point $(\log\widetilde{M}_\mathrm{cl},\log\widetilde{r}_\mathrm{core,corr})$ and the model $(\log\widetilde{M}_\mathrm{cl},m\cdot\log \widetilde{M}_\mathrm{cl}+\widetilde{q})$.
The data points $(\log\widetilde{M}_\mathrm{cl},\log\widetilde{r}_\mathrm{core,corr})$ are obtained by shifting the physical quantities $(\log M_\mathrm{cl},\log\rcorecorr)$ for their median values to remove the covariance between $m$ and $q$ (which becomes $\widetilde{q}$ after the shift).
In this case the total scatter is defined as $\sigma_\mathrm{tot}^2=\sigma_\perp^2+\sigma_{M_i,\perp}^2+\sigma_{r_i,\perp}^2$, where $\sigma_\perp$ is the orthogonal intrinsic scatter, while $\sigma_{M_i,\perp}=\sigma_{M_i}\sin\theta$ and $\sigma_{r_i,\perp}=\sigma_{r_i}\cos\theta$ are the projections of the $\log\widetilde{M}_\mathrm{cl}$ and $\log\widetilde{r}_\mathrm{core,corr}$ uncertainties $\sigma_{M_i}$ and $\sigma_{r_i}$, respectively ($\theta=\arctan m$).

\begin{figure*}[t!]
\includegraphics[width=0.48\textwidth]{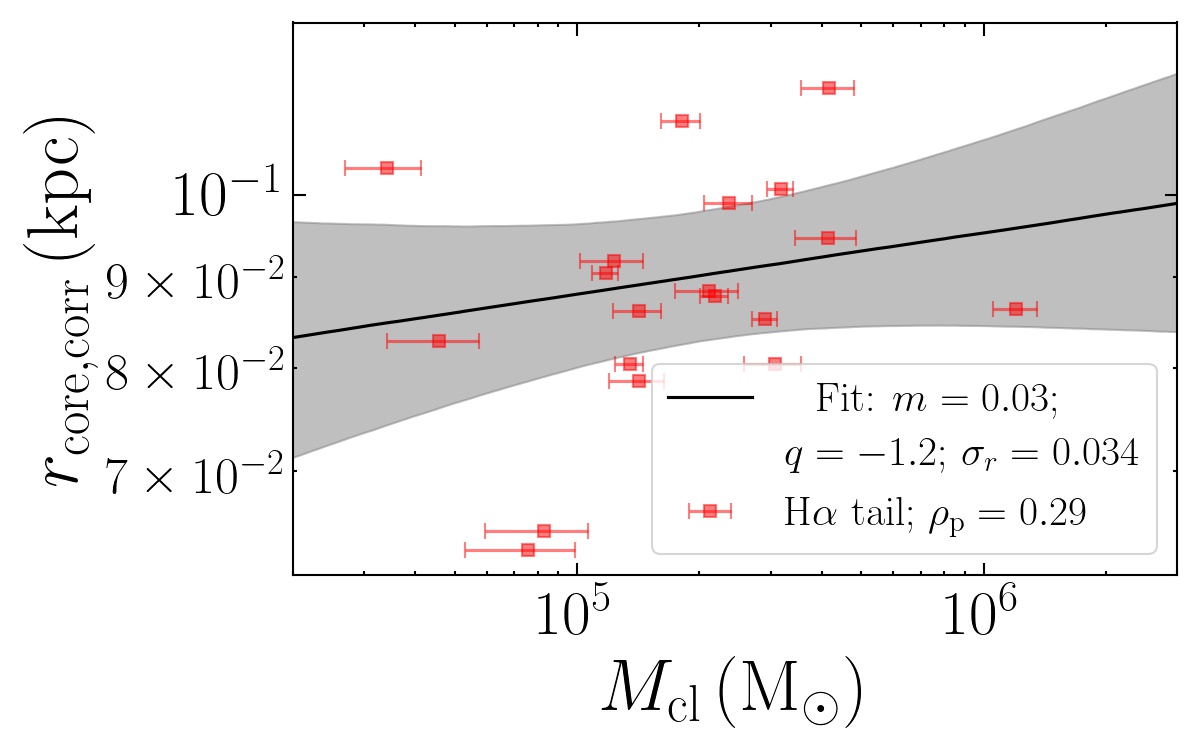}
\includegraphics[width=0.48\textwidth]{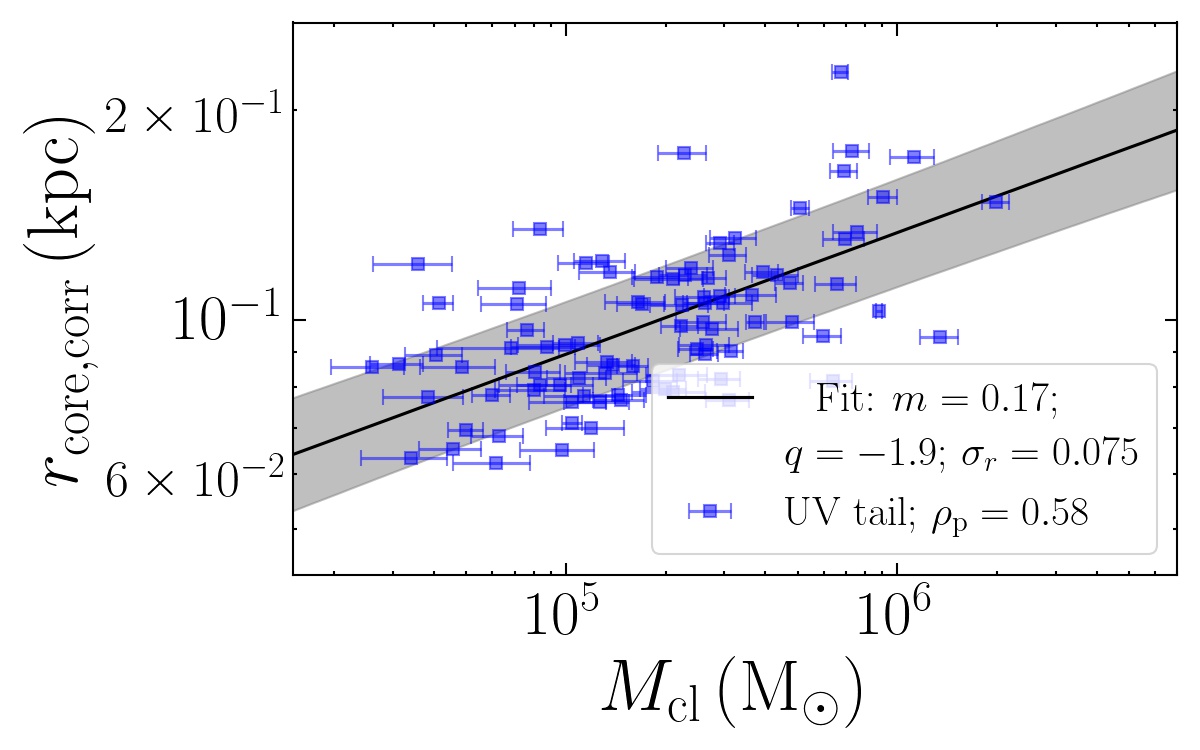}\vspace{-0.2cm}\\
\includegraphics[width=0.48\textwidth]{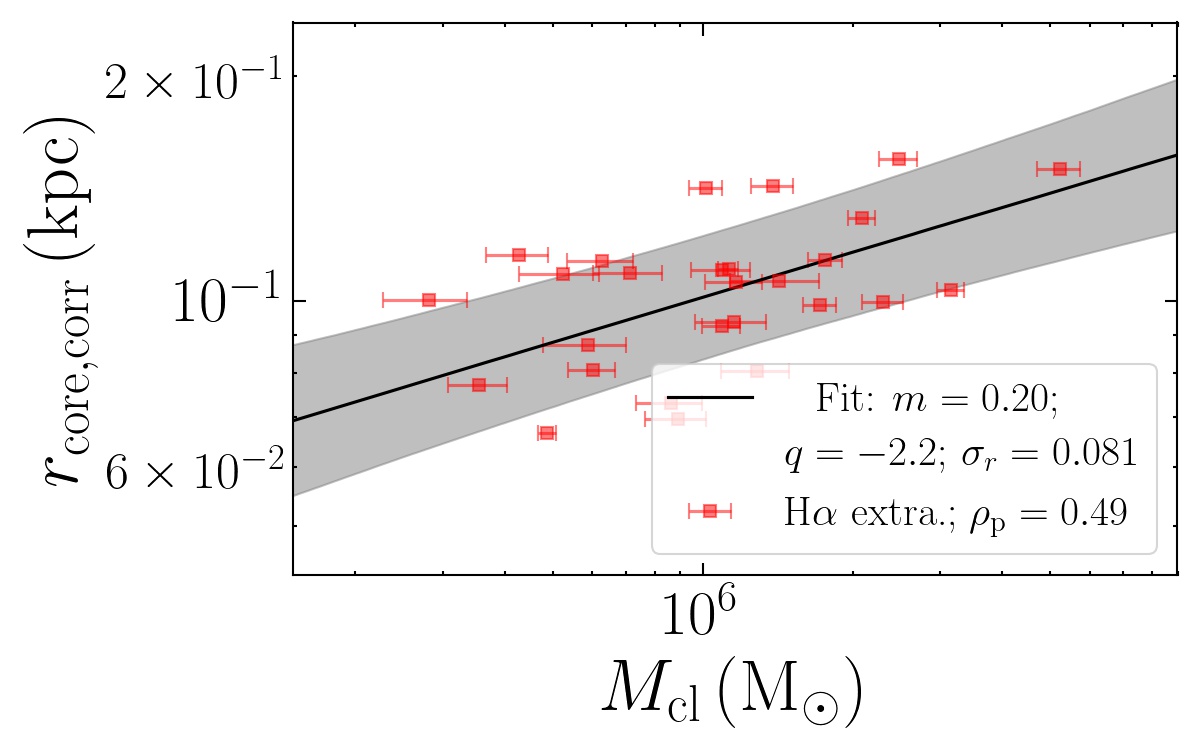}
\includegraphics[width=0.48\textwidth]{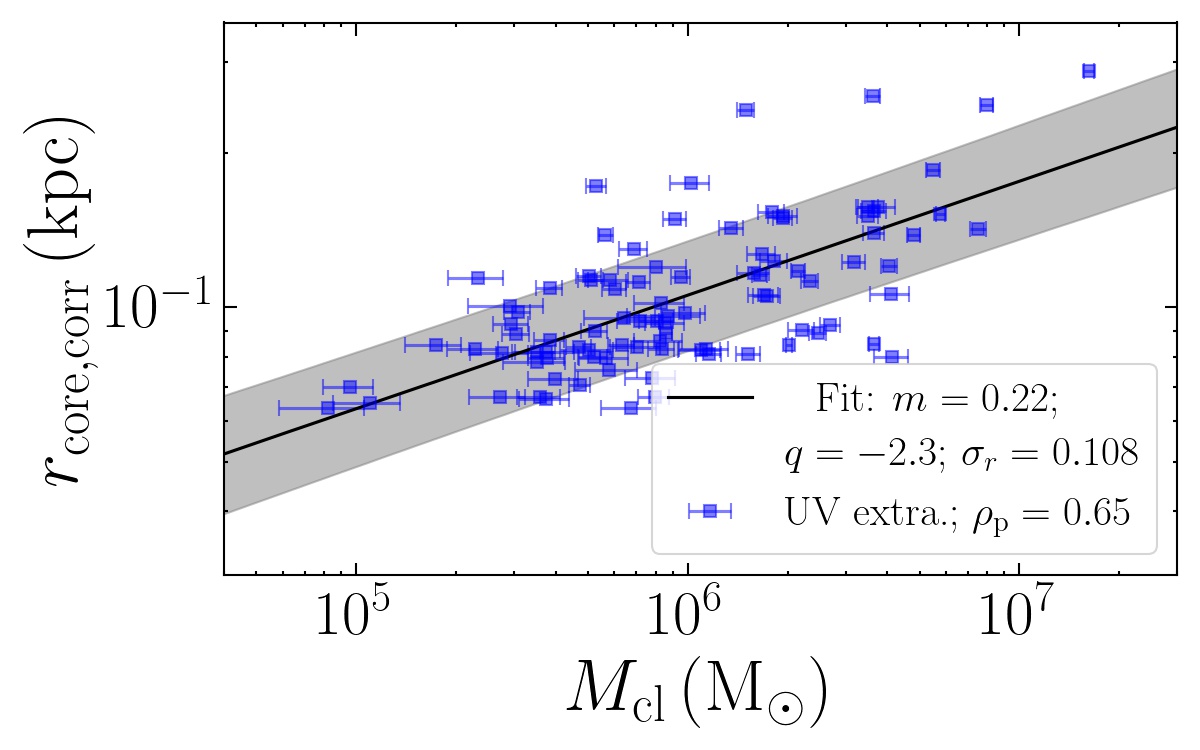}
\caption{Core radius-stellar mass relation for tail (top row) and extraplanar (bottom row) resolved clumps (H$\alpha$ in the left panels and UV in the right panels, respectively). Each clump is plotted as a red (for H$\alpha$) or blue (for UV) square with stellar mass uncertainties. The best-fitting linear relation (black solid line) is obtained as described in Sect. \ref{sec:size_mass}, and the grey area is the $1\sigma$ interval of the fit. In the legend, the best-fitting parameters are reported, together with the Pearson coefficient and the P-value.}
\label{fig:size_mass}
\end{figure*}

We considered the median of the parameters distributions as the best-fitting parameters and are listed in Table \ref{tab:size_mass} in the form $(m,q,\sigma_r)$, which are the parameters needed to infer the clumps size at a given mass. In Fig. \ref{fig:size_mass} the size-mass log-plane is shown; we note good correlations for tail and extraplanar UV clumps (with Pearson's coefficients larger than $0.5$), while the correlations for the H$\alpha$ clumps are quite weak (especially in the tail, for which the Pearson's coefficient is $0.29$).
However, we stress that the main purpose of this fitting procedure is to assign a realistic radius to a clump of given stellar mass: in this sense, the presence of a weak correlation is not a problem, since the large scatter is modelled by introducing the intrinsic scatter.

\begin{table}[t!]
\fontsize{10pt}{10pt}\selectfont
\setlength{\tabcolsep}{4pt}
\renewcommand{\arraystretch}{1.4} 
\centering
\caption{Best-fitting parameters of the radius-mass relation, modelled as described in Sect. \ref{sec:size_mass}.}
\begin{tabular}{l|c|c|c|c}
\bottomrule
\bottomrule
 & \multicolumn{2}{|c}{H$\alpha$} & \multicolumn{2}{|c}{UV}\\\hline
Parameter & Tail & Extra. & Tail & Extra.\\\hline
$m/10^{-1}$ & $0.3\pm 0.6$ & $2.0\pm0.7$ & $1.7\pm0.3$ & $2.2\pm0.3$\\
$q$ & $-1.2\pm 0.3$ & $-2.2\pm0.4$ & $-1.9\pm0.1$ & $-2.3\pm0.2$\\
$\sigma_r/10^{-2}$ & $3.4\pm1.8$ & $8.1\pm1.7$ & $7.5\pm0.9$ & $11\pm1$\\
\hline
\end{tabular}
\tablefoot{From left to right: the list of free parameters (slope $m$, intercept $q$ and radius intrinsic scatter $\sigma_r$), the results for H$\alpha$ tail clumps, H$\alpha$ extraplanar clumps, UV tail clumps and UV extraplanar clumps.}
\label{tab:size_mass}
\end{table}

\FloatBarrier

\section{Mock clumps generation}\label{sec:mock_clumps}
In the previous Sections, we have defined all the quantities necessary to generate mock images in all the \textit{HST} filters for a clump with given mass, age and axial ratio.
We have generated a set of $1000$ H$\alpha$ clumps and $1000$ UV clumps for each galaxy, for which the input parameters listed above are defined as follows:

\begin{enumerate}
    \item{The clump stellar mass $M_\mathrm{cl}$ is extracted from a flat distribution ranging from the minimum to the maximum clump stellar mass derived from the real clumps. Thanks to this choice, all masses are uniformly sampled when studying the mass completeness and discrepancy (Sect. \ref{sec:mass_function}).}
    \item{Age and axial ratio are extracted from the corresponding distribution of real clumps. In this way, we can explore combinations of mass-age-shape that are realistic and weighted by their frequency in the observed sample. We point out that, since the age and axial ratio distributions are taken from observations, they may suffer of incompleteness as well. For a fixed mass, for example, an old clump is fainter in the detection band than a young one. However, a systematic study of the completeness of the age distribution is beyond the scope of this work, it would increase the dimensionality of the problem and would require the study of $N^2$ mock clumps, with $N$ the number of mock clumps studied in this work. Therefore we opted for adopting the observed distribution, in order to be as close as possible to the clump sample, even with no further correction.}
\end{enumerate}

Once the intrinsic properties of the clumps are defined, we can derive the observables:

\begin{enumerate}
    \item{By means of the size-mass relation, at a given mass we can derive the value of $\rcorecorr$ (Sect. \ref{sec:size_mass}), which is converted to angular scale given the redshift of the galaxy. The intrinsic scatter of the linear relation is taken into account with a Gaussian extraction. The inferred core radius is used to assign a physical angular scale to the clump intrinsic profile (Sect. \ref{sec:stacking}), which is then convoluted for the \textit{HST} PSF.}
    \item{The clump image is stretched for the extracted axial ratio and rotated by a position angle uniformly extracted between 0 and $2\pi$.}
    \item{The shape of the SED of the clump is inferred from the age as follows. As described in Sect. \ref{sec:shape_sed}, for each age bin and photometric filter we derived a distribution of normalized fluxes; after uniformly extracting a random number between 0 and 1, we take the normalized flux in each filter at the percentile corresponding to the number extracted. By doing this, we am able to reproduce the variety of SED shapes present in each bin.}
    \item{Given the mass and the age, we derive $M_\mathrm{cl}/L_\mathrm{F814W}$ (Sect. \ref{sec:m_to_l}) and convert it to F814W flux, which letusscale the normalized SED previously determined to a physical quantity. Given the fluxes in the 5 photometric filters, we generate an image of the clump for each filter, by multiplying the clump intrinsic profile for the corresponding flux.}
    \item{The clump images of each filter are convoluted for the corresponding PSF curve.}
\end{enumerate}

\begin{figure*}[t!]
\includegraphics[width=0.99\textwidth]{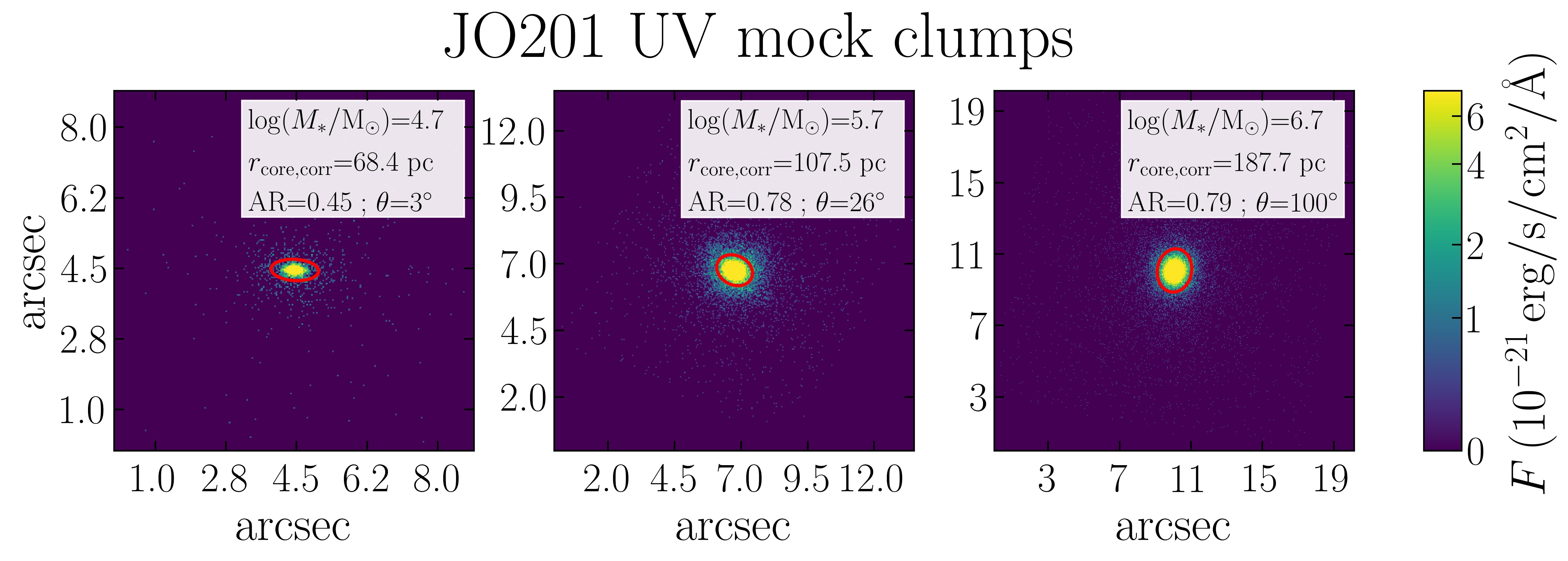}
\caption{Examples of F275W mock images of three UV clumps generated for JO201 as described in Sect. \ref{sec:mock_clumps}. The red ellipse is built using the semi-major axis, semi-minor axis and position angle of the clump. For each clump, the stellar mass, core radius, axis ratio and position angle are reported in the legend.}
\label{fig:mock_clumps}
\end{figure*}

That is, we created a library of 1000 mock H$\alpha$ clumps (and 1000 mock UV) for each galaxy, each of which constituted by 5 images.
As examples, we show in Fig. \ref{fig:mock_clumps} three F275W mock images of UV mock clumps generated to belong to JO201. The clumps have masses $5\times 10^4\,\msun$, $5\times 10^5\,\msun$ and $5\times 10^6\,\msun$. One can notice the increasing scale of the image, due to the increasing mass and size of the mock clumps, and the variety of elongations and position angles.

\FloatBarrier

\section{Re-detection of the mock clumps}\label{sec:re_obs}
The aim of these mock observations is to quantify the effects of background noise and contamination by background and foreground sources on the detection of clumps in the tails of the jellyfish galaxies of our sample. The classical way to mimic the same conditions affecting the galaxy real images consists of adding the mock clumps images to the real images themselves, perform the same procedure adopted to detect the real clumps and check whether mock clumps are re-detected and what their output properties are. The way the mock clumps are added to the image changes whether we are studying tail or extraplanar clumps and it is based on the need not to change the crowding properties of the corresponding region of the studied galaxy. The two methods are listed below.

\begin{figure*}[t!]
\includegraphics[width=0.99\textwidth]{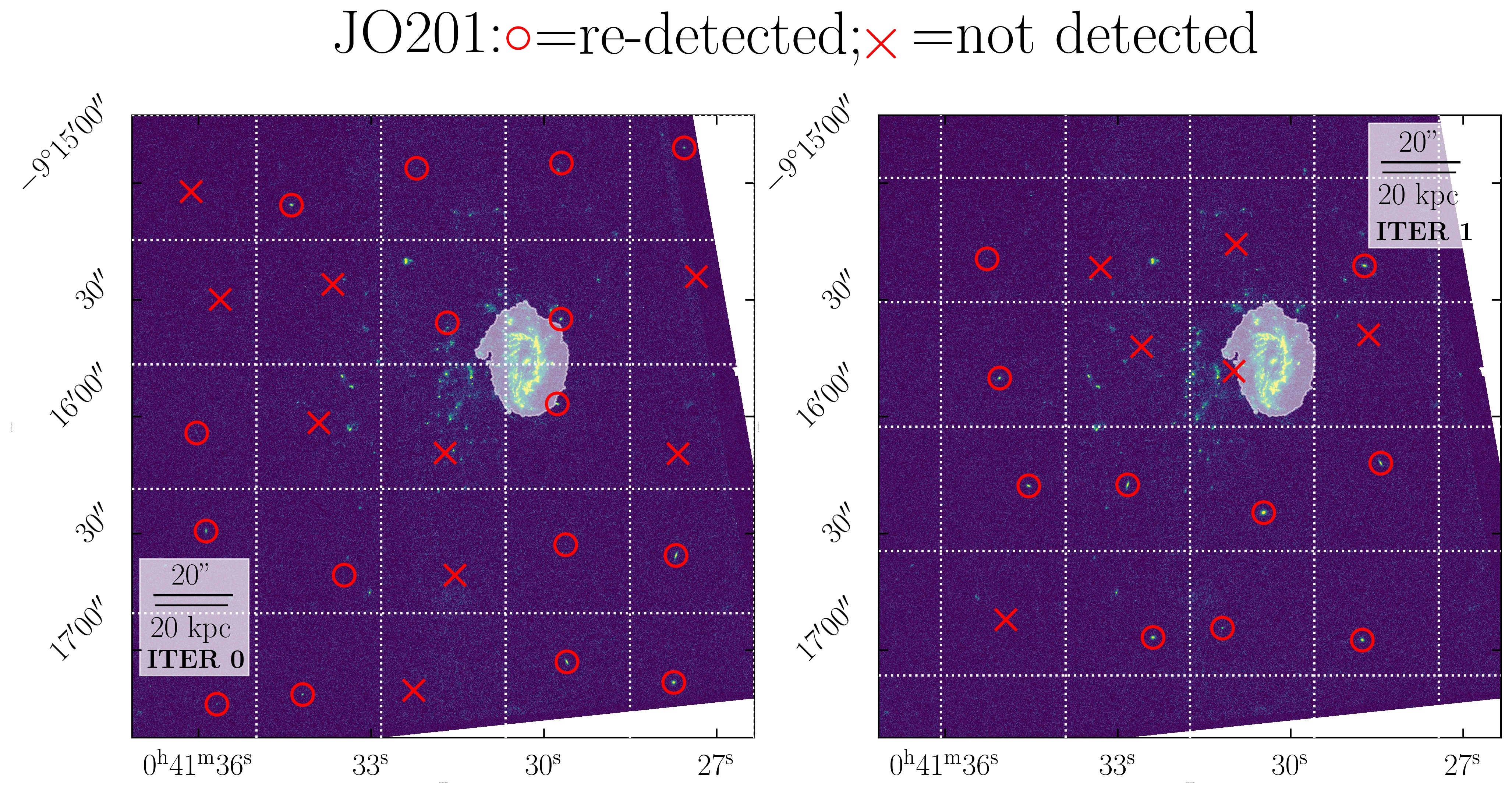}
\caption{First two F275W mock images generated for JO201 following the prescription in Sect. \ref{sec:re_obs}. The white dotted lines define the grid in which the FOV is divided, while the white shaded area is the optical disk (Sect. \ref{sec:spatcat}). The mock clumps that are added to the real images are inside a red circle if they got re-detected, otherwise they are marked by a red cross.}
\label{fig:JO201_grid}
\end{figure*}

\textit{Mock clumps addition in the tails.} In order to evenly cover the area of the image where the clump detection was run, the field of view of each galaxy is divided in a grid of $N_\mathrm{cell}$ squared cells (700 pixels in size for JO175, JO206 and JW100, 800 for JO201, JO204, JW39). Then a number of mock clumps equal to the number of cells is randomly extracted by the corresponding library (Sect. \ref{sec:mock_clumps}) and added to each cell. 
The centre of the clump is randomly extracted within a square region of size 350/400 pixels (half of the cell size), centred in the corresponding cell.
This is done in order to fully explore the background conditions inside a certain cell, still avoiding to place the mock clump at the border of the cell and therefore to overlap it with other adjacent mock clumps. Furthermore, the centre of the mock clump can not be inside the galactic optical disk.
For the subsequent mock observation, the cell grid is shifted both along RA and Dec by 350/400 pixels, again to exhaustively explore the whole galactic FOV. As examples, in Fig. \ref{fig:JO201_grid} we show the first two mock observations for JO201, where one can notice where the mock clumps are added, how the FOV is divided in cells and how the grid is shifted in the second mock observation.
In order to have a statistically robust estimate of the completeness and input-output discrepancy in the whole mass interval, we need to run at least 1000 clumps per galaxy, which result in $N_\mathrm{mock}=1000/N_\mathrm{cell}$ mock observations (rounded up).

\textit{Mock clumps addition in the extraplanar region.} The extraplanar region is located in the outskirts of the galactic stellar disk (as described in Sect. \ref{sec:spatcat}). Being this region highly irregular in all the six galaxies of the sample, we can not apply the same aforementioned procedure used for the tail clumps. That is, we opted for a random extraction of the position of the clumps within the extraplanar area. The number of added mock clumps per iteration in a given galaxy is chosen to be equal to the Poissonian uncertainty on the number of real extraplanar clumps detected in that galaxy $N_\mathrm{extra,g}$, which is $N_\mathrm{extra,g}^{0.5}$. This choice allows us to add clumps without changing the crowding conditions of the real sample of clumps. The total number of simulated mock clumps is about $10^4$ in H$\alpha$ and $1.4\times 10^4$ in UV.
\\
\par
Once the mock image is ready, we re-ran the clumps detection (Sect. \ref{sec:clumps}). From now on, we will call this catalogue the new catalogue, while the one with real clumps is the original catalogue.
The new catalogue is matched with the original one to find re-detected real clumps: if the centre of an original clump is within a circle centred in the centre of a new clump and radius equal to the new clump core radius \citep{Giunchi2023a}, and their fluxes are comparable within a factor 2, then the new clump is flagged as a re-detected original clump and removed by the new catalogue.
The remaining new clumps are matched with the mock clumps that were added to the image, and if the centre of a mock clump is within the area of a new clump, the new clump is flagged as a re-detected mock clump. Mock clumps not matched to any new clump are flagged as non-detected.
Finally, the mass and mass-weighted age of the re-detected mock clumps are derived with \textsc{Bagpipes} in the same way described in Sect. \ref{sec:method_mass}.
After having followed these steps, out of 6125 H$\alpha$ and 6125 UV clumps, we re-detected and computed stellar masses for 3416 and 2982, respectively.

To summarize, at the end of this process we can study the fraction of lost mock clumps as a function of their input mass while, for those that are re-detected, we can compare the input mass with the one resulting from \textsc{Bagpipes}.

\FloatBarrier

\begin{figure*}[t!]
\section{Completeness and mass estimation bias}\label{sec:compl_bias_plots}
\centering
\resizebox{0.94\textwidth}{!}{\includegraphics[height=1cm]{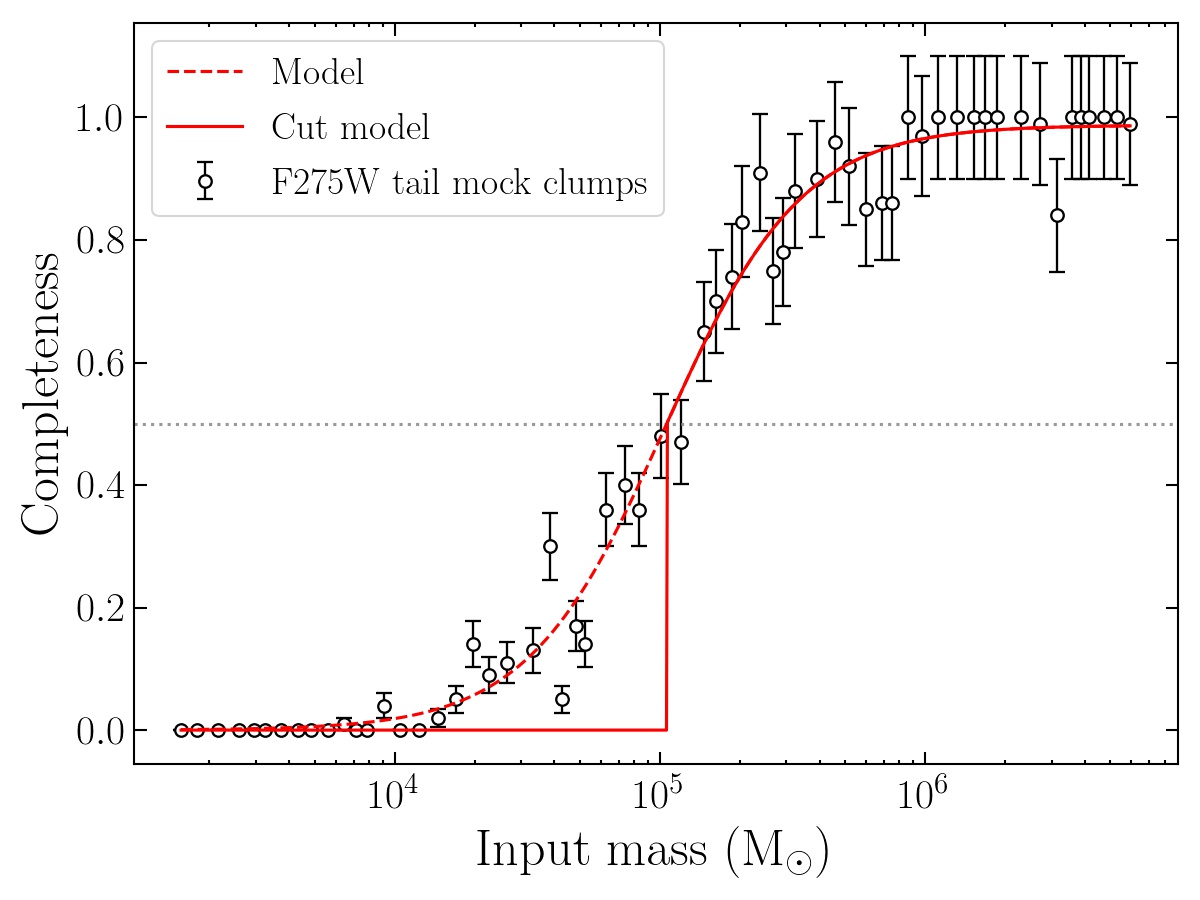}
    \includegraphics[height=1cm]{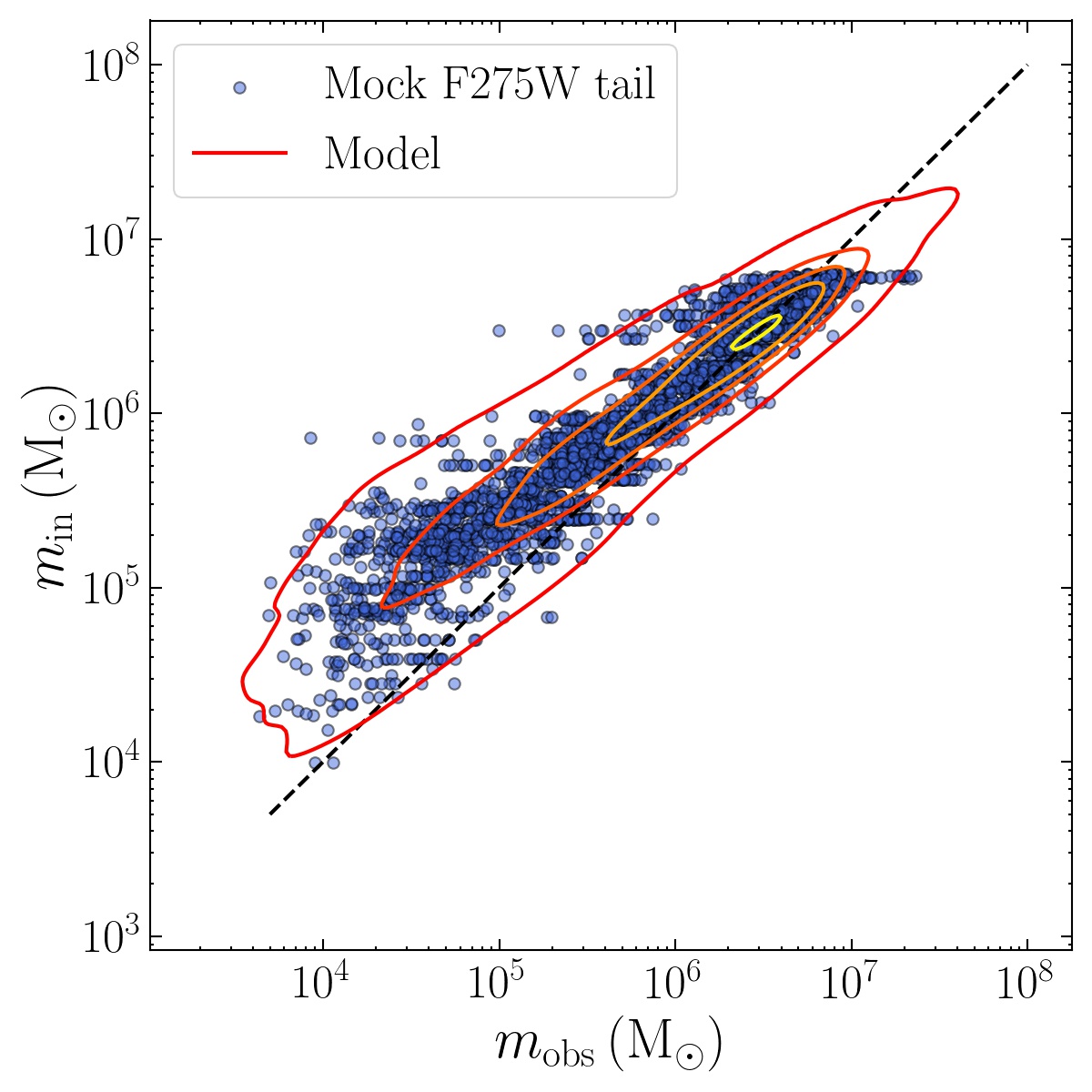}}\\
\vspace{-0.15cm}
    \resizebox{0.94\textwidth}{!}{\includegraphics[height=1cm]{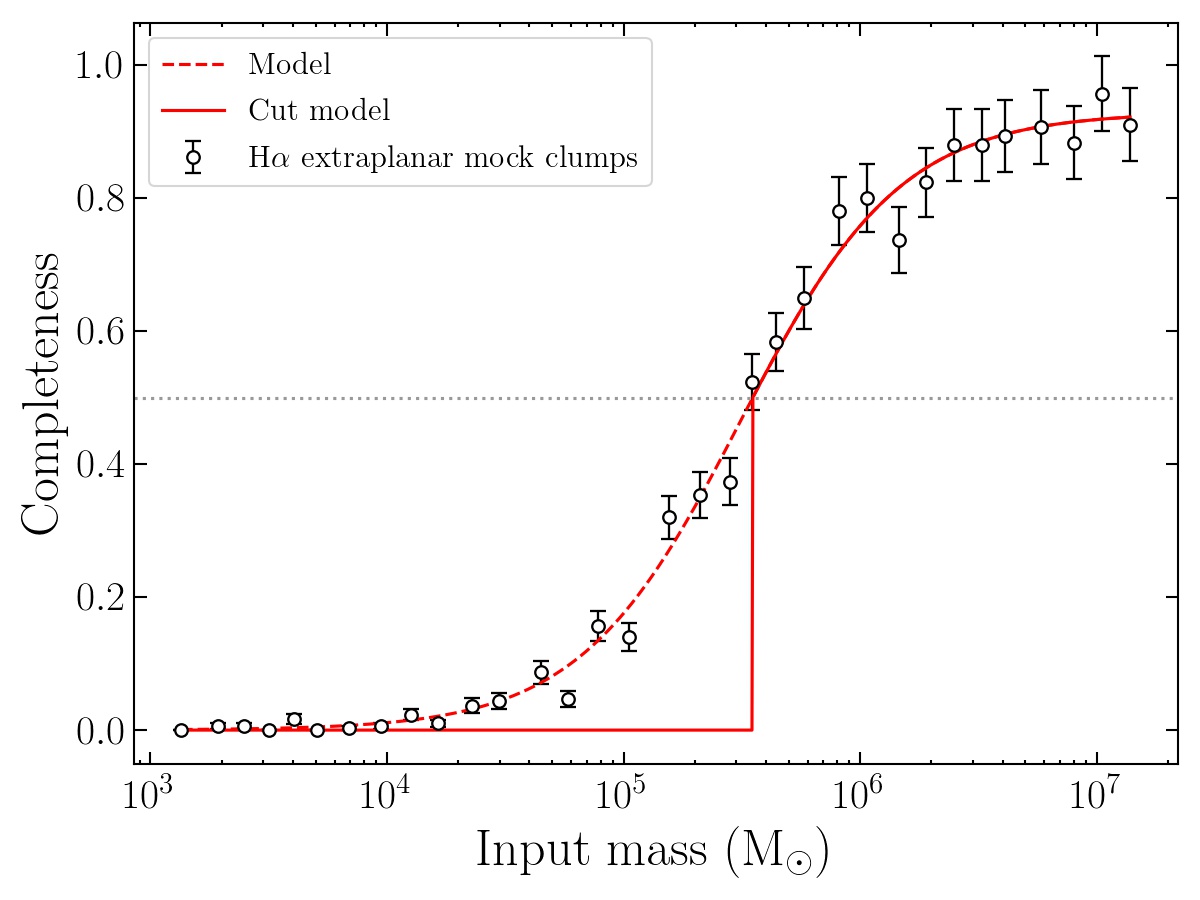}
    \includegraphics[height=1cm]{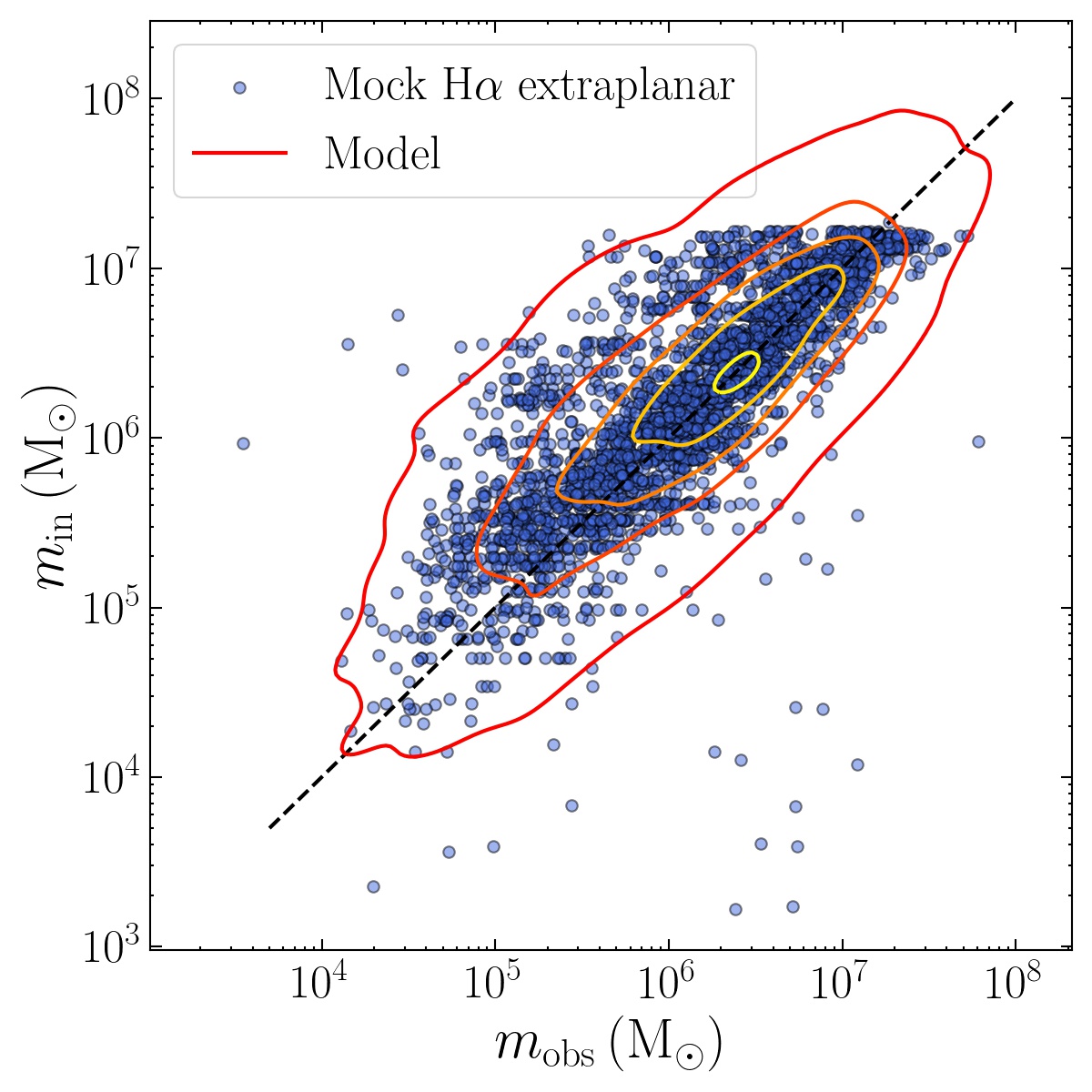}}\\\vspace{-0.15cm}
    \resizebox{0.94\textwidth}{!}{\includegraphics[height=1cm]{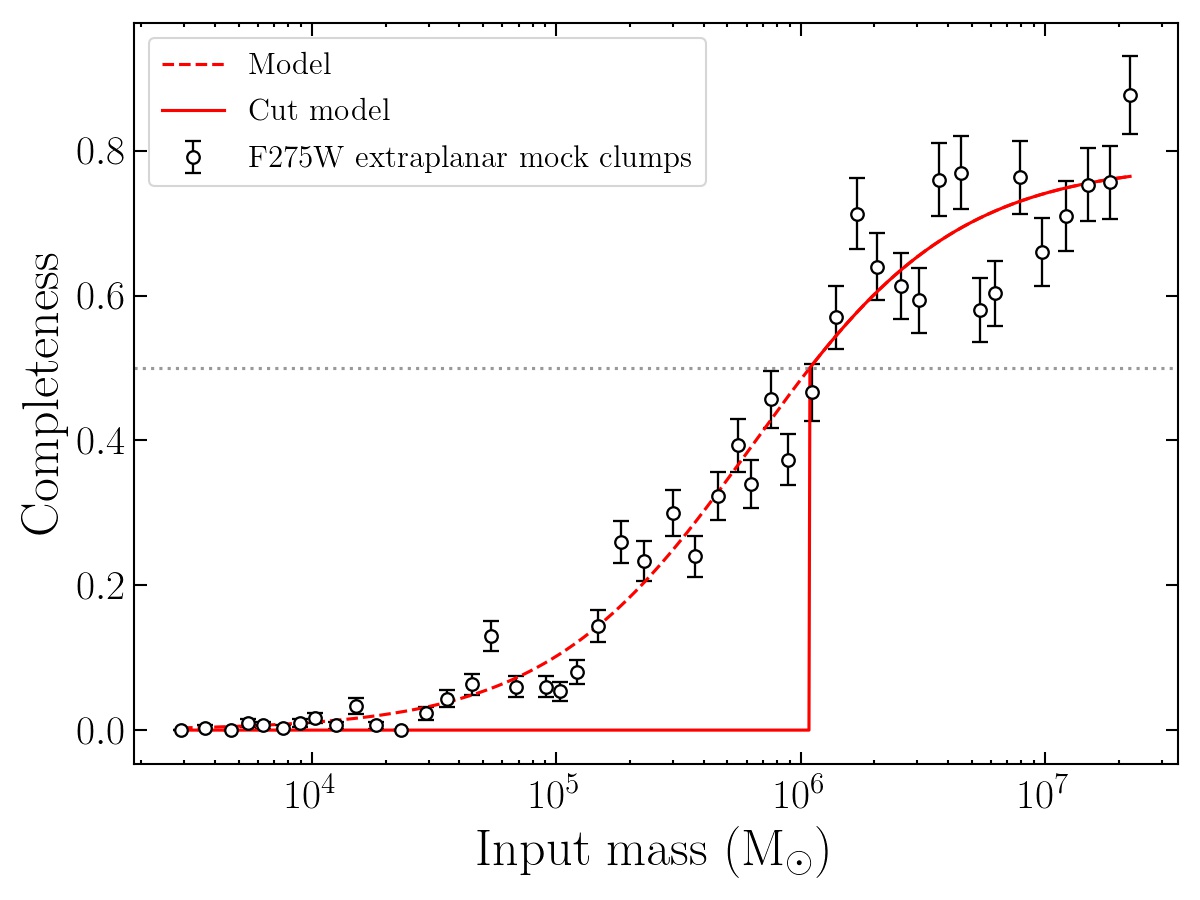}
    \includegraphics[height=1cm]{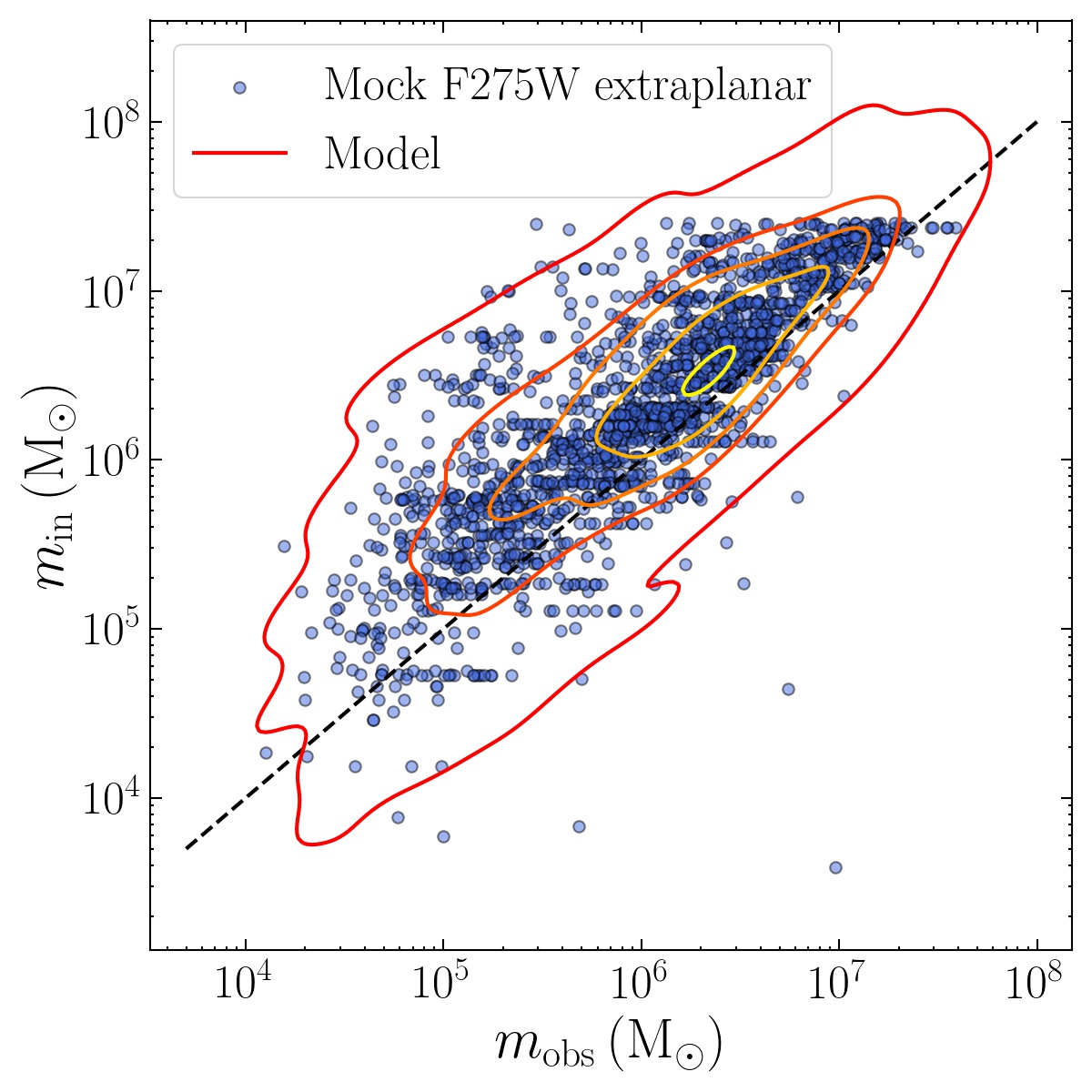}}\vspace{-0.2cm}
\caption{Plots of the completeness (left column) and mass estimation bias (right column), as in Fig. \ref{fig:ha_tail_params}, but for UV tail (top row), H$\alpha$ extraplanar (middle row) and UV extraplanar (bottom row) clumps. The parameters are estimated as in Sects. \ref{sec:mass_completeness} and \ref{sec:in_out_mass}, listed in Table \ref{tab:completeness_discrepancy} and the plots shown in Fig. \ref{fig:uv_tail_params}.}
\label{fig:uv_tail_params}
\end{figure*}

\FloatBarrier

\section{Testing the mass function fitting method}\label{sec:mock}
In order to test the reliability of the fitting method, we generated two samples of mock clumps from a distribution of equation

\begin{equation}\label{eq:mf_compl}
    \mathrm{MF}_C(M_*)=\mathrm{MF}(M_*)\times C(M_*),
\end{equation}

\noindent where MF is the intrinsic mass function and $C$ is the completeness function (\ref{sec:mass_completeness}), for which we assumed the same parameters derived for the tail H$\alpha$ clumps.
We tested two different functional forms for the mass function, which are the most used and debated in literature to fit observations: sample A is extracted from a power law mass function $\mathrm{MF}\propto M_*^{-\alpha}$, with slope $\alpha=2$; sample B is extracted from a Schechter function $\mathrm{MF}\propto M_*^{-\alpha}\,e^{-M_*/M_c}$, with slope $\alpha=2$ and cut-off mass $\log M_c=10^{5.5}\,\mathrm{\msun}$.

\begin{figure*}[t!]
\centering
\includegraphics[width=0.25\textwidth]{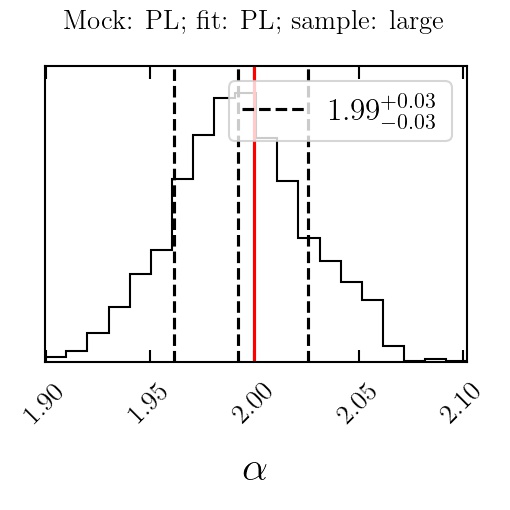}\hspace{-0.3cm}
\includegraphics[width=0.25\textwidth]{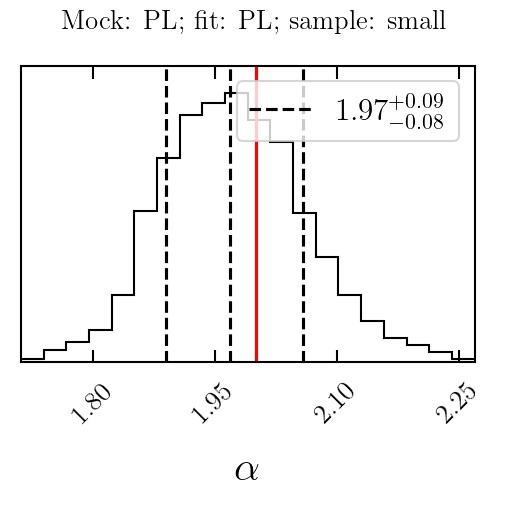}\hspace{-0.3cm}
\includegraphics[width=0.25\textwidth]{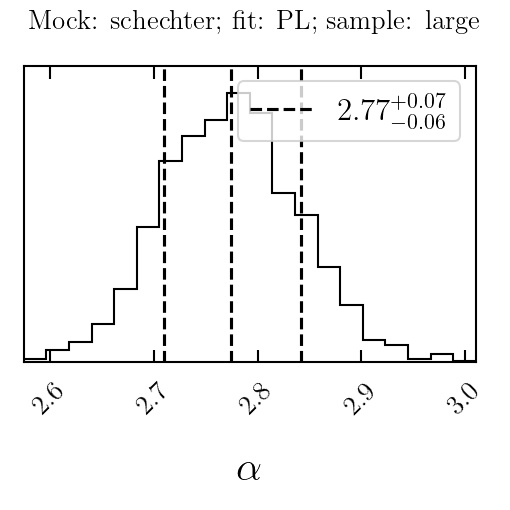}\hspace{-0.3cm}
\includegraphics[width=0.25\textwidth]{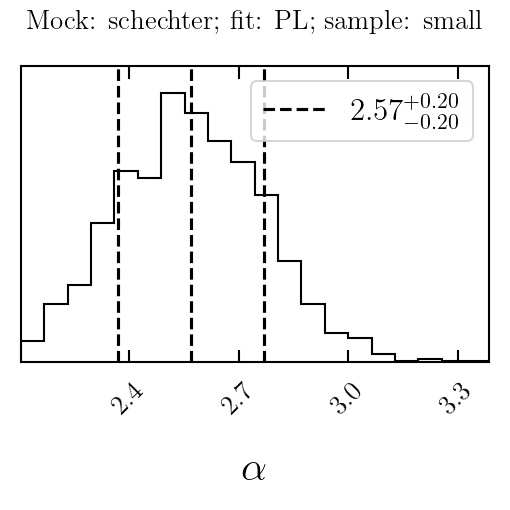}
\caption{Histogram of the best-fitting slope $\alpha$ distribution, obtained fitting a power law mass function to a sample of mock clumps extracted either from a power law (first two panels) or a Schechter (last two panels) function. The plots to the left refer to cases in which the fitted sample was large ($\sim 1000$ objects), while for those to the right the sample was small ($\sim 100$ objects).
When present (because within the interval of explored values), the red line marks the input slope assumed for both the power law and the Schechter mock distributions ($\alpha=2$). The values in the legend of each histogram are the median of each distribution for the corresponding parameter, with uncertainties given by the 84th percentile minus the median and the median minus the 16th percentile (upper and lower uncertainty, respectively). The same values are also reported as black dashed vertical lines in the histograms.}
\label{fig:mock_hist}
\end{figure*}

\begin{figure*}[t!]
\includegraphics[width=0.47\textwidth]{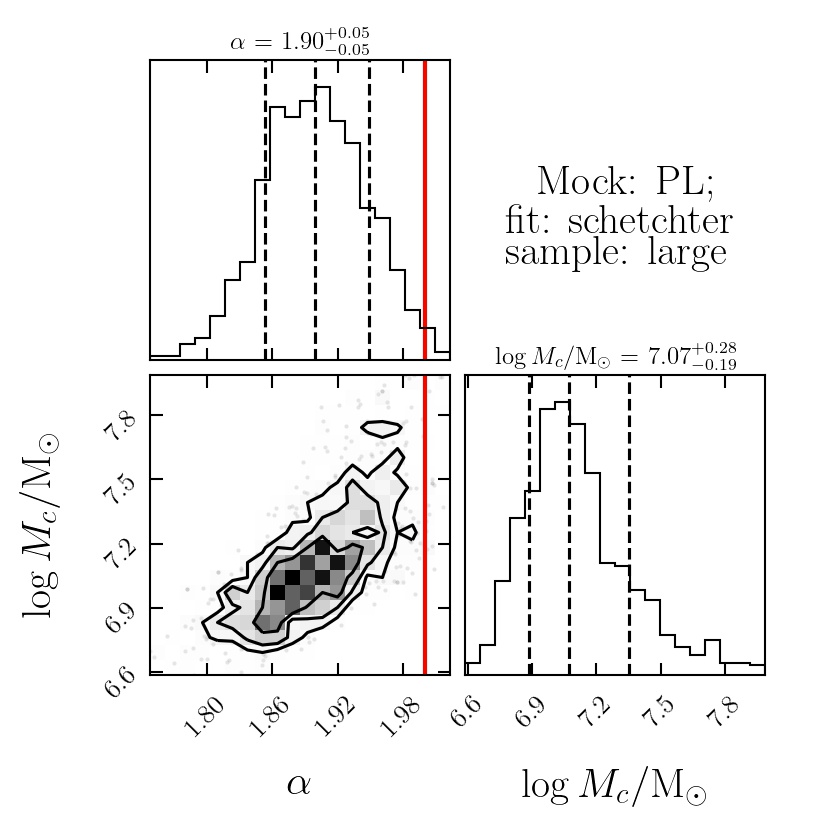}\hspace{-0.7cm}
\includegraphics[width=0.47\textwidth]{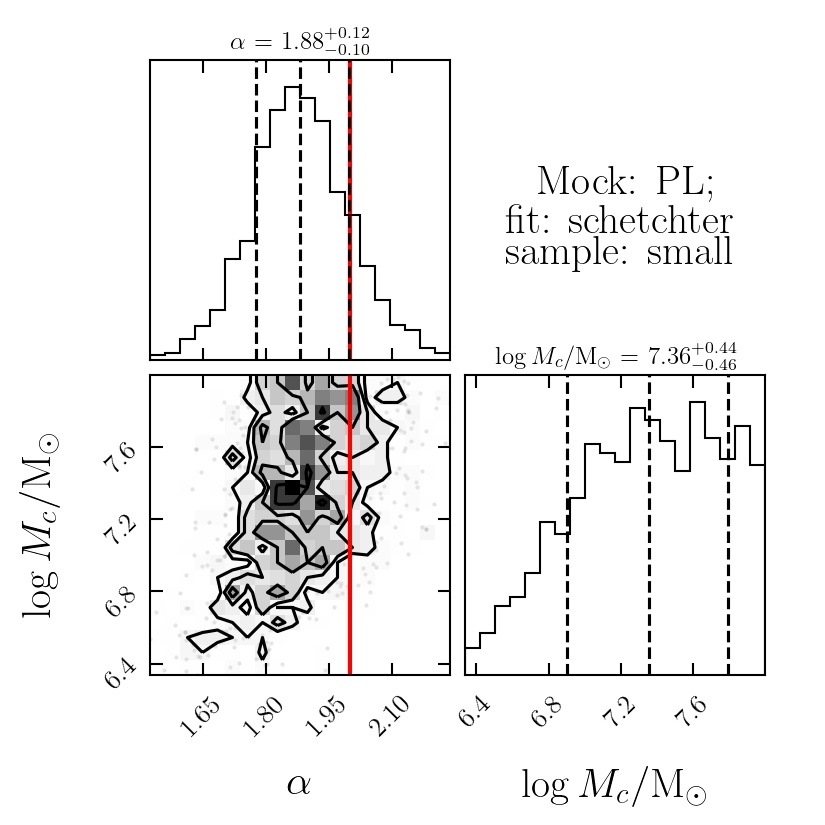}\vspace{-0.25cm}\\
\includegraphics[width=0.47\textwidth]{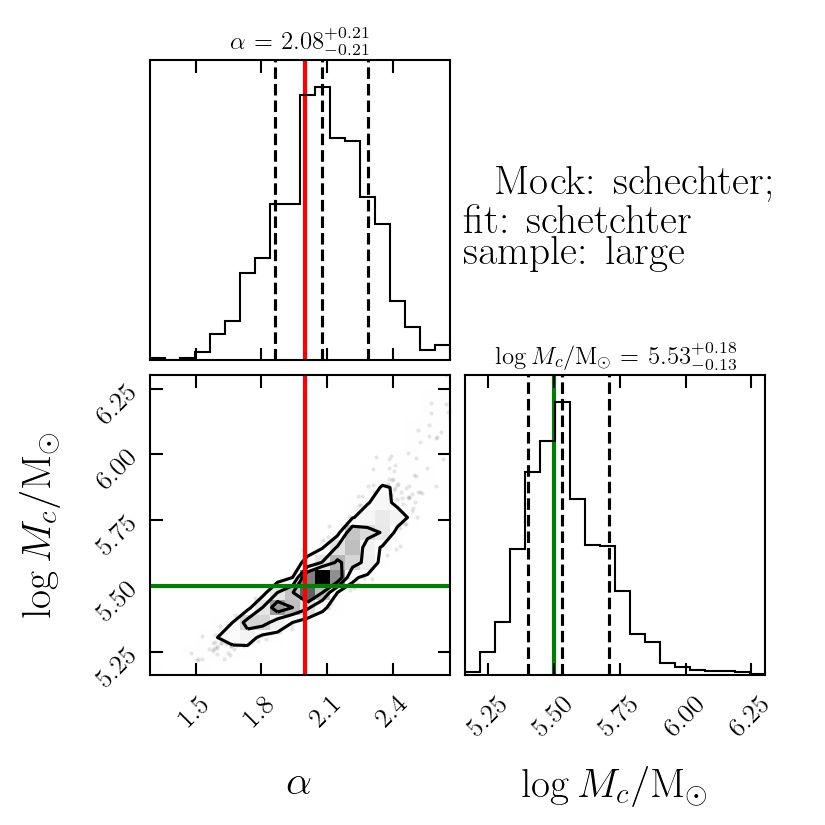}\hspace{-0.7cm}
\includegraphics[width=0.47\textwidth]{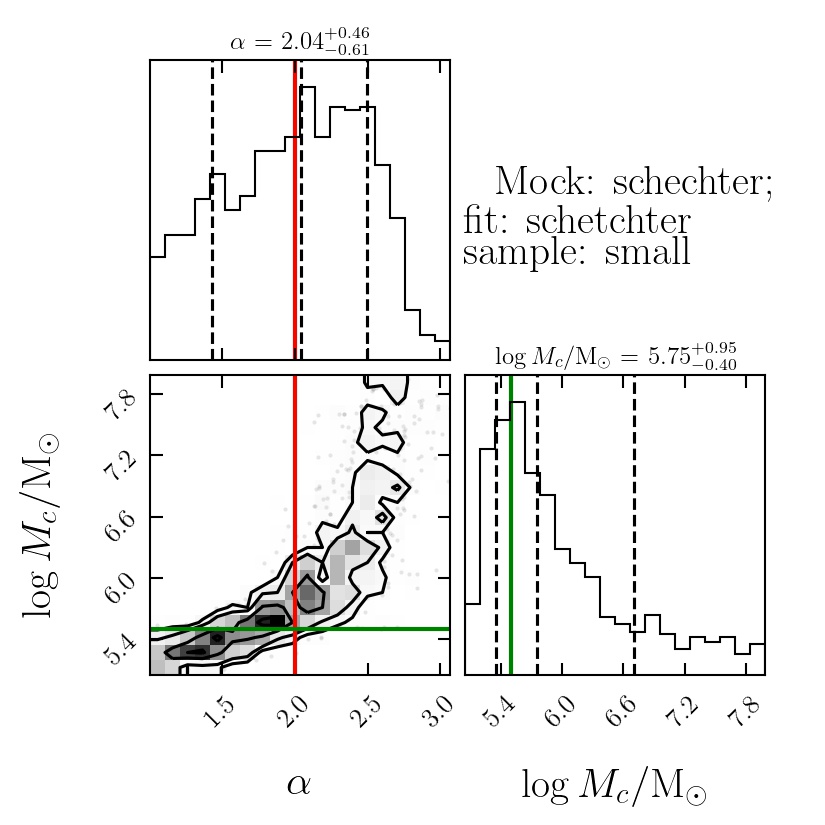}
\caption{Corner plots of the parameter space $(\alpha,\log M_c)$ (the slope and the cut-off mass, respectively), obtained fitting a Schechter mass function to a sample of mock clumps extracted either from a power law (top row) or a Schechter (bottom row) function. The plots to the left refer to cases in which the fitted sample was large ($\sim 1000$ objects), while for those to the right the sample was small ($\sim 100$ objects).
The red line marks the input slope of both the power law and the Schechter mock distributions ($\alpha=2$). The green line marks the cut-off mass ($\log M_c/\mathrm{\msun}$), which is present only in the case of the Schechter. Values on top of each histogram are the median of each distribution for the corresponding parameter, with uncertainties given by the 84th percentile minus the median and the median minus the 16th percentile (upper and lower uncertainty, respectively). The same values are also reported as black dashed vertical lines in the histograms.}
\label{fig:mock_corners}
\end{figure*}

In testing the likelihood of Eq. \ref{eq:logL_mf}, we fitted both samples either assuming a power-law and a Schechter mass function, and varying the sample size from $\sim 1000$ to $\sim 100$ clumps (with the latter case resembling our typical conditions). The aim is to understand if the slope can be correctly retrieved and how the presence of the cut-off affects our results.
Applying a certain mass function to one of the two samples brings to the following results, as a function of the sample size:

\begin{enumerate}
    \item \textit{Fitting sample A with a power law}, first two panels of Fig. \ref{fig:mock_hist}: the fitting procedure correctly retrieves the input slope within $1\sigma$, with the sample size affecting just the uncertainty on the estimate.
    \item \textit{Fitting sample A with a Schechter}, top row of Fig. \ref{fig:mock_corners}: regardless of the sample size, the retrieved cut-off mass is larger than the most massive clump of the sample, correctly indicating that an exponential cut-off is absent within the fitted range of masses. For what concerns the slope, the value inferred by the model underestimates the intrinsic one (yet being consistent within $3\sigma$), with the sample size affecting just the uncertainties.
    \item \textit{Fitting sample B with a power law}, last two panels of Fig. \ref{fig:mock_hist}: as shown and discussed in previous works \citep{Gieles2009}, the presence of a cut-off leads to a slope steeper than the input one, since it is not taken into account by the chosen functional form chosen to fit the mass function. This happens regardless of the sample size, which affects the uncertainties.
    \item \textit{Fitting sample B with a Schechter}, bottom row of Fig. \ref{fig:mock_corners}: in this case the sample size highly affects the results. With a large available statistics (left panel), the fitting procedure correctly infers both the input slope and cut-off mass. When the sample is small (right panel), the cut-off mass is still well constrained, while the slope is not constrained and only an upper limit on it can be put.
\end{enumerate}

These tests suggest that fitting a Schechter function puts good constraints about the presence of the exponential cut-off (and therefore if the shape of the mass function is a power law of a Schechter function), independently on the size of the sample.
However, when the available statistics is small (about $100$ objects), the slope can not be constrained, especially if the intrinsic shape of the mass function is a Schechter function.
Typically, fitting a power law gives steeper values for the slope than the intrinsic one, if the real shape of the mass function is a Schechter, but can be the only method to put constraints on this parameter.

\end{appendix}

\end{document}